\documentclass[useAMS,usenatbib]{mn2e}

\usepackage{graphicx}
\usepackage{amsmath}

\title[Modification of the gamma-ray spectra in OVV blazars]
{Modification of the gamma-ray spectra by internal absorption in OVV blazars: 
The example cases of 3C~273 and 3C~279}
\author[J. Sitarek \& W. Bednarek]
{J. Sitarek$^{1,2}$ \& W. Bednarek$^{1}$\\
$^{1}$Department of Experimental Physics, University of \L \'od\'z,
ul. Pomorska 149/153, 90-236 \L \'od\'z, Poland; bednar@fizwe4.fic.uni.lodz.pl\\
$^{2}$Max-Planck-Insitut f\"ur Physik, D-80805 M\"unchen, Germany; jsitarek@mppmu.mpg.de}
\begin{document}

\date{Accepted . Received ; in original form }

\pagerange{\pageref{firstpage}--\pageref{lastpage}} \pubyear{2007}

\maketitle

\label{firstpage}

\begin{abstract}
Recent observations with the low threshold Cherenkov telescopes proved that sub-TeV $\gamma$-rays are able to arrive from active galaxies at relatively large distances in spite of expected severe absorption in the extragalactic background light (EBL). 
We calculate the $\gamma$-ray spectra at TeV energies from two example OVV quasars,
3C~273 and 3C~279, assuming that $\gamma$-rays are injected in the inner parts of the 
jets launched by the accretion disks. It is assumed that $\gamma$-rays 
in the broad energy range (from MeV up to TeV) are produced in these blazars with the power law spectrum with the spectral index as observed from these objects by the EGRET telescope at GeV energies. We take into account 
the internal absorption of these $\gamma$-rays by considering a few different models for the radiation field surrounding the jet. The classical picture of the relativistic blob in jet model for the injection of primary $\gamma$-rays is considered with the injection rate of $\gamma$-rays as observed by the EGRET telescope in the GeV energy range. 
The results of calculations are compared with positive detection and the upper limits on the sub-TeV $\gamma$-ray fluxes from these two sources. It is concluded that even with the Stecker EBL model, the level of the $\gamma$-ray emission from 3C~279 is close to the recent measurements in the sub-TeV $\gamma$-ray energies provided that the injected $\gamma$-ray spectrum extends from the GeV energies over the next two decades with this same spectral index. We also suggest that a few day time scale flare from 3C 273 can be detected by the MAGIC II stereo telescopes. 

\end{abstract}

\section{Introduction}

The TeV $\gamma$-rays are detected from compact sources (massive binary systems, active galactic nuclei) in which their absorption at the source has to be carefully investigated. In fact,
the escape conditions of $\gamma$-ray photons from the vicinity of the accretion disks
surrounding compact objects have been studied in the past in the context of the 
$\gamma$-ray production close to the accretion disk radiation starting with the first reports on the TeV $\gamma$-ray emission from the binary systems in the eighties and the GeV $\gamma$-ray emission from the active galaxies (see e.g.~early calculations of absorption in the disk radiation by Protheroe \& Stanev~(1987), Carraminana~(1992), Bednarek~(1993), Backer \& Kafatos~(1995), and in the quasi-isotropic radiation around the disk-jet by e.g.~
Blandford \& Levinson~(1995), Dermer \& Schlickeiser~(1994). The problem of $\gamma$-ray escape became again very attractive with the discovery of TeV $\gamma$-ray emission from active galaxies of the BL Lac type (see more recent calculations by e.g.~Protheroe \& Biermann~(1997), Bednarek~(1997), Donea \& Protheroe~(2003), Liu \& Bai~(2006), Reimer~(2007)). Thanks to the low energy threshold the MAGIC telescope, it is at present clear that also quasars of the OVV type at very large distances emit at least sub-TeV $\gamma$-rays (e.g. quasar 3C~279, Albert et al.~2008).

However, higher energy $\gamma$-rays 
(above $\sim 100$ GeV) should be efficiently absorbed in the soft radiation field filling the intergalactic space. Therefore observations of TeV $\gamma$-rays from the population of AGNs at different distances might be used to put the upper limits on the Extragalactic Background Light (EBL). The conclusions reached based on this method strongly depend on the assumptions on the production spectrum of the $\gamma$-rays at the source. Usually a simple power law is applied with the spectral index which in the extreme case can be constrained by the likely radiation mechanisms of particles accelerated at AGNs (for the most recent analysis see e.g.~Aharonian et al.~(2006a), Mazin \& Raue~(2007)). However, if the internal absorption inside the source is important, the shape of the emerging $\gamma$-ray spectrum can be quite complex, i.e. the spectrum can become significantly flatter (see e.g. Fig. 9 in~Bednarek~1997 and Aharonian et al.~2008) than considered their limiting values $\propto E_\gamma^{-1.5}$ or even $\propto E_\gamma^{-2/3}$. Therefore, derived upper limits on EBL are in fact model dependent, strongly depending on the details of the production of $\gamma$-rays inside the source. 

In this paper we calculate the opacities for $\gamma$-ray photons inside the central regions of AGNs in the most general case, i.e. for arbitrary angles of injection of $\gamma$-rays in respect to the jet axis and also for arbitrary inclination of the jet in respect to the plane of the accretion disk. The opacities for the $\gamma$-rays propagating in the spherically symmetric Broad Line Region are also taken into account. Therefore, we can obtain the three-dimensional $\gamma$-spheres around the central engine which can be specially useful in the case of highly inclined jets towards the observer, e.g. as it might occur in the case of strongly variable TeV $\gamma$-ray radio galaxy M~87 (Aharonian et al.~2006b, Aciari et al.~2008). 

Moreover, we show how the intrinsic spectra of $\gamma$-rays
emitted from AGNs should be modified in the case of two example EGRET blazars, 
3C~273 (Hermsen et al.~1993, von Montigny et al.~1993, Lichti et al.~1995) and 3C~279 (Kniffen et al.~1993, Wehrle et al.~1998, Hartman et al.~2001). The second source has been  also recently detected by MAGIC telescope (Albert et al.~2008), being at present the most distant sub-TeV $\gamma$-ray source. 
Based on the new observations of 3C~279 in the sub-TeV energies, we estimate the location of the production region inside the jet assuming that: (1) the $\gamma$-ray flare observed by the MAGIC telescope was comparable to the highest activity state observed by the EGRET telescope at GeV energies; (2) the production spectrum of $\gamma$-rays at the source extends from GeV energies up to TeV energies without a break. The effects of possible absorption of TeV $\gamma$-rays during propagation in the EBL are taken into account by applying its extreme estimates derived by~Primack et al.~(2005) and~Stecker et al.~(2006).

\section{The model for soft radiation inside AGN}

Up to now a few different scenarios have been considered in which absorption of the 
$\gamma$-rays in collisions with the soft photon field surrounding the inner jets of AGNs
is expected to be important. The soft radiation can be produced inside the jet (synchrotron origin), in the accretion disk (thermal origin), in the broad line region (BLR, disk scattered emission and the emission lines), or in the large equatorial torus (thermal infrared emission) (see references mentioned in the Introduction). 
Absorption in the synchrotron radiation field is mostly important in the case of BL Lac objects, where accretion disk radiation field is rather weak and the $\gamma$-ray emission regions are very compact (estimated from observations of very short variability time scales). 
Since radiation from torus is mainly in the IR band, it would contribute mainly to the absorption of higher energy gamma rays. In this paper we consider only the absorption of $\gamma$-rays in the accretion disk and in the BLR.
These two radiation fields seems to be the most important since the very short variability time scale of TeV $\gamma$-ray emission (of the order of a few minutes observed recently in the case of Mrk 501 (Albert et al.~2007) and PKS 2155-304 (Aharonian et al.~2007)) strongly suggest that the emission regions in blazars are located within the inner part of the jet, i.e. relatively close to the accretion disk.

\begin{figure}
\includegraphics[width=0.45\textwidth]{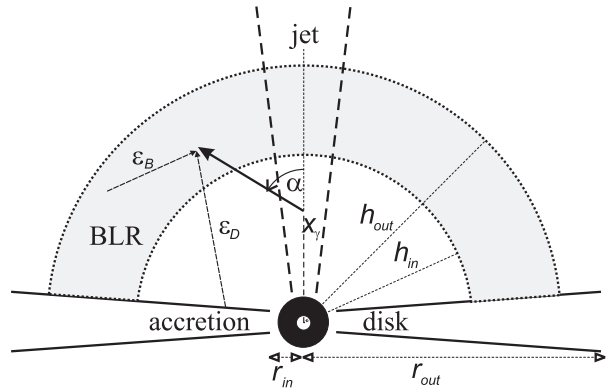}
\caption{Schematic picture of the geometry of inner part of the AGN. Gamma-rays are injected within the jet, which propagates in this specific case perpendicularly to the disk surface, at a distance $x_{\gamma}$ from the base of the jet, and at the angle $\alpha$ to the jet axis. These $\gamma$-rays interact with soft radiation produced inside the accretion disk (photons marked by $\varepsilon_{\rm D}$, disk defined by the inner radius $r_{\rm in}$ and outer radius $r_{\rm out}$) and with the photons produced (or scattered) in the broad line emission region (shaded area) marked by $\varepsilon_{\rm B}$. The scattering centers of the BLR are distributed at distances between $h_{\rm in}$ and $h_{\rm out}$ from the central black hole. }
\label{fig1}
\end{figure}

We assume the classical geometry for the emission region inside the inner part of the 
AGN as shown in Fig.~\ref{fig1}. The accretion disk extends around the black hole 
within the range of distances defined by the inner $r_{in}$ and the outer $r_{out}$ disk radius. In is usually considered that the jet, launched in the inner part of the disk, propagates perpendicularly to the disk plane. Although, in general some specific inclination of the jet to the disk axis, due to e.g. accretion from molecular clouds whose orbits have 
different inclinations,  can be also considered. The inner engine is surrounded by 
a spherically symmetric Broad Line Region (BLR) extending from $h_{in}$ to $h_{out}$. 
A part of the disk radiation is absorbed and reprocessed in the BLR either in the form 
of emission lines or in the form of continuous radiation. Gamma-rays are produced in the 
jet at a specific range of distances from the black hole (between $x_{\gamma}^{\rm min}$ 
to $x_{\gamma}^{\rm max}$), which can be determined from the observed variability time 
scales of high energy flares. They are injected at the angle $\alpha$ in respect to 
the direction of the jet. Below, we describe in a more detail the parameters which define 
the applied radiation field.

\subsection{The accretion disk radiation}

We investigate a few different models for the soft radiation field produced by the accretion disk around the black hole in the center of AGN, since it is not clear what is in fact the radiation field created by the accretion disks (see e.g. review by Koratkar \& Blaes~1999).
In the simplest case (our model I), we adopt the optically thick and geometrically thin Shakura \& Sunyaev~(SS, 1973) disk model.
The emission of the accretion disk is treated as a black body with a power law temperature dependence on the distance, $r$, from the black hole, $T=T_{in}(r/r_{in})^{-3/4}$, where $T_{\rm in}$ is the temperature at the inner disk radius $r_{\rm in}$. For large disks (i.e. $r_{out} \gg r_{in}$), the total disk luminosity can be calculated from, 
\begin{equation}
L_{\rm d}=4\pi\sigma_{\mathrm{\rm SB}}r_{\rm in}^2 T_{\rm in}^4, 
\label{eq:Ldisk}
\end{equation} 
where $\sigma_{\mathrm{SB}}$ is the Stefan-Boltzmann constant. In the case of some quasars the disk emission, its luminosity and the location of the peak in the spectrum, is clearly observed (e.g. 3C~273). Then, the above formula can be used to estimate the inner radius of the accretion disk. 

The observed luminosity of the accretion disk in 3C 273 is $L_{disk}=2\times 10^{46} \mathrm{erg\, s^{-1}}$, and the temperature of the inner part of the disk is estimated as $T_{\rm in}=2.6\times 10^4$ K (Malkan \& Sargent~1982, Malkan~1983). Paltani \& T\"urler~(2005), using the reverberation method, give the estimation of black hole mass $M\approx7 \times 10^9 M_\odot$ and the size of the BLR: $R_{BLR}=986^{+21}_{-37}$ days.
Based on observed luminosity and location of the peak emission, the inner radius of the disk can be fixed on $r_{\rm in}=7.8\times 10^{15} \mathrm{cm}$. 
This value is consistent with independent estimate of the gravitational radius of the black hole, $r_{\rm g} = GM/c^2\approx 10^{15}\mathrm{\rm cm}$ from its estimated mass. The outer radius of the accretion disk has been fixed on $r_{\rm out}=10^4 r_{\rm in}$. 

We also consider the second OVV type blazar, 3C~279. For this source we apply the disk luminosity $L_{\rm disk}=2\times 10^{45}\mathrm{\rm erg\, s^{-1}}$ and its inner temperature $T_{\rm in}=20000$ K (Pian et al.~1999).  The inner radius of the accretion disk in 3C~279 is estimated on $r_{\rm in}=4.2 \times 10^{15} \mathrm{\rm cm}$.

The second model (model II) for the disk radiation is suggested by the observations of the strong bump in the optical-UV energy band in 3C~273. It shows clear power law tail at soft X-rays which seems to smoothly continue after the UV peak. It can be interpreted as due to the emission from the hot disk corona above the surface of the optically thick, geometrically thin accretion disk (see Staubert et al.~1992, Leach et al.~1995). We add such additional component to the thermal emission of the optically thick, geometrically thin disk. This soft X-ray tail is described by a simple power law with the differential spectral index $\sim 2.7$ (Kriss et al.~1999), meeting the peak in the thermal emission. Therefore, in the model II we add to the thermal SS type disk emission also emission from the disk corona with the power law spectrum mentioned above. 

We also consider the third model (model III) for the radiation of the accretion disk in which the inner disk temperature is significantly larger than predicted by classical SS model. This might happen in the case of the accretion disk around a 
rapidly rotating black hole (the Kerr black hole). In such a case the gravitational energy release per unit area of the disk increases faster with the disk radius (Page \& Thorne~1974) resulting in the increase of the inner disk temperature by a factor of $\sim 3$.
Also the slim disks, which are expected in the case of the high accretion rates on the lower mass black hole (see Szuszkiewicz et al.~1996), can produce disk spectra extending to 
higher energies resulting in higher effective temperatures of the disk radiation than in the case of SS thin disks.
Note that interpretation of the UV results from 3C~273 by such hotter accretion disks can not be excluded since the peak of emitted radiation falls into the UV region which is unobservable due to interstellar absorption. Therefore, in our model III we also consider the accretion disks with the inner temperatures by a factor of 3 larger than in the case of the classical Shakura-Sunyaev disk model but with similar temperature profile.
Note that 3 times higher temperature at the inner disk radius  requires simultaneous decrease of inner disk radius $r_{\rm in}$ by a factor of $3^{4/3} \approx 4.3$ in order to satisfy condition on the total power emitted by the accretion disk.

\subsection{The Broad Line Region radiation}

The quasi-spherical region around the central black hole, in which Broad Emission Lines are produced (the BLR), consists of a large number of clouds. In these clouds the disk UV radiation is absorbed and re-emitted isotropically in the form of emission lines. For the calculations of line emission from the BLR we follow Liu \&\ Bai (2006), who calculate the emissivity $j(r)$ of 35 lines (in the wavelength range of $\lambda=1 - 6.5\times 10^3 \mathrm{\AA}$).
The photon density of this re-processed radiation is calculated by integration along an arbitrary direction through the BLR. However, in contrary to Liu \&\ Bai, we consider only the upper hemisphere of the BLR (in respect to the accretion disk), since the lower hemisphere is in fact obscured by the accretion disk. Due to the presence of free electrons in the BLR, a part of the disk radiation is isotropized in the Compton Scattering process.
Since this scattering process occurs in the Thomson regime, the angular distribution of scattered photons is roughly isotropic and their spectrum is conserved. 
Therefore, the energy spectrum of the continuous part of BLR emission is the same as the black body spectrum integrated all over the disk. 
The BLR continuous spectrum, obtained in the above way, has a long low energy tail, which is important in calculating the absorption of gamma rays above $\approx 1 TeV$. 

Total luminosity of the line, $L_{\rm l}$, and the continuous, $L_{\rm c}$, part of BLR spectrum is normalized to the disk luminosity by introducing two parameters, i.e. the conversion efficiency of the disk luminosity to the BLR line luminosity, $\eta_{\rm l}$, and to the continuous luminosity $\eta_{\rm c}$. Therefore, $L_{\rm l}=\eta_{\rm l} L_{\rm d}$ and $L_{\rm c}=\eta_{\rm c} L_{\rm d}$. Note, that the BLR emission shows much larger level of isotropy than the direct disk emission. Therefore, their maximum contribution to the absorption of high energy $\gamma$-rays is shifted to lower energies.

The dimension of the BLR in 3C 273 is estimated on $h_{\rm out} = 500 r_{in}$ (with agreement with reverberation mapping, Paltani \& T\"urler~2005). The inner radius of the BLR is difficult to estimate by any arguments. We apply the value of $h_{in}=100 r_{in}$.
In the example calculations shown below, it is assumed that $\eta_{\rm l} = \eta_{\rm c} =0.01$. Bear in mind, that optical depths for $\gamma$-rays interacting with the BLR radiation scale linearly with appropriate $\eta$ coefficient. The size of BLR in 3C~279 is rescaled from the size of 3C~273 according to the general prescription $h_{\rm out}\propto L_{\rm D}^{0.7}$ (Kaspi et. al. 2000). Therefore, we assume that the clouds of the BLR in 3C~279 are confined between 20 and 190 $r_{in}$.

\begin{figure*}

\includegraphics[width=0.33\textwidth, height=0.18\textheight, trim= 0 35 0 0,clip]{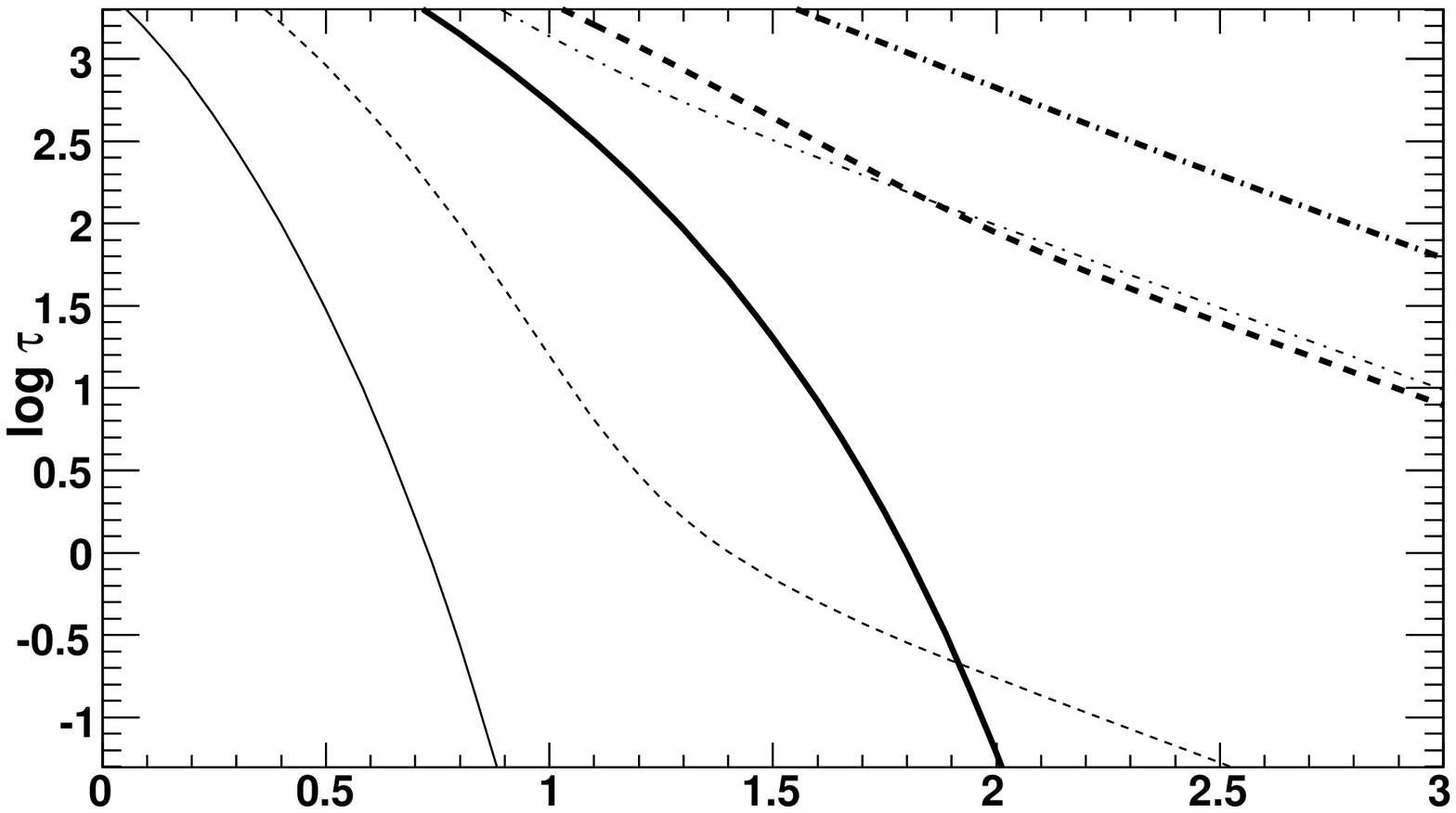}
\includegraphics[width=0.33\textwidth, height=0.18\textheight, trim= 5 35 0 0,clip]{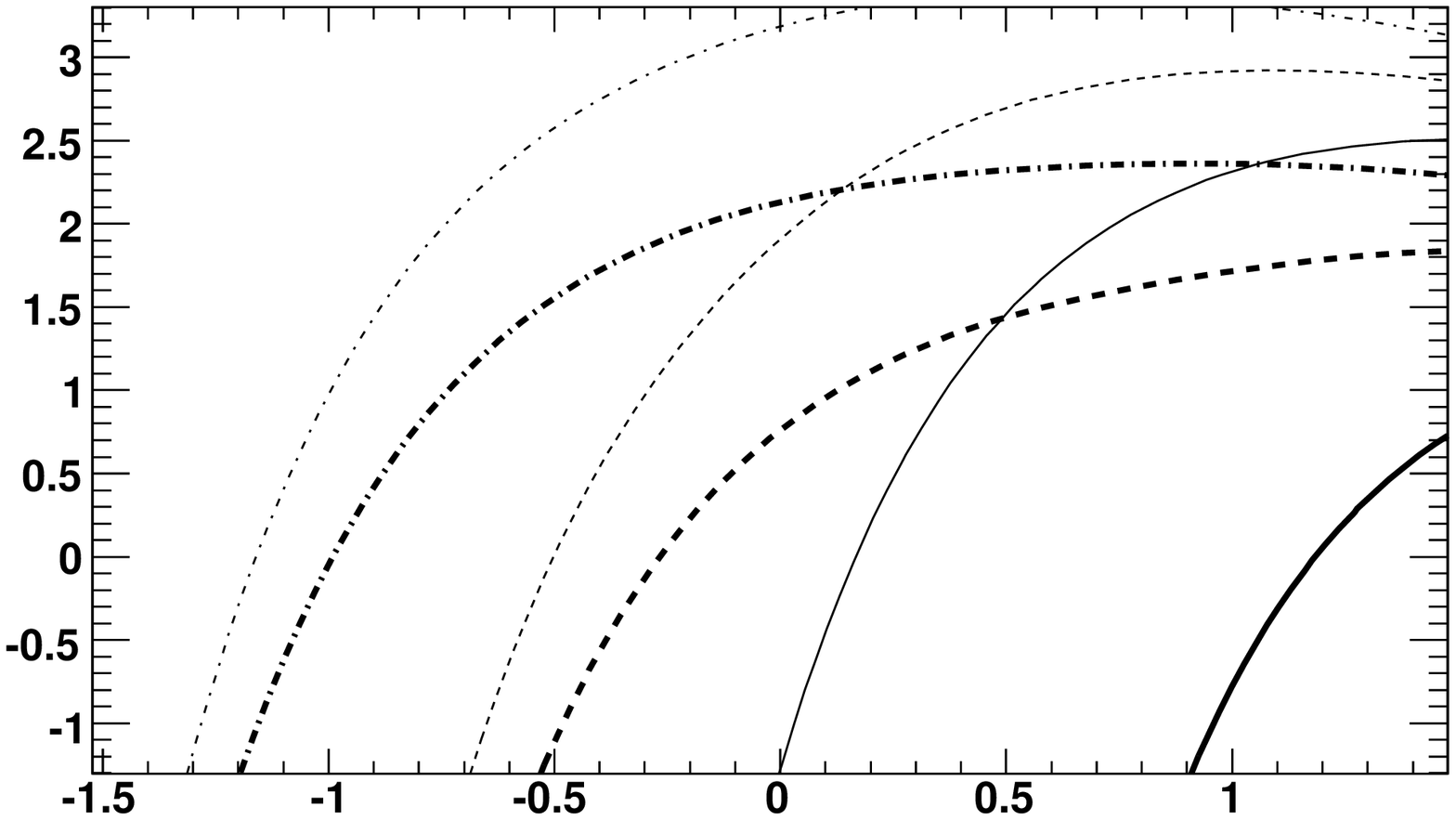}
\includegraphics[width=0.33\textwidth, height=0.18\textheight, trim= 5 35 0 0,clip]{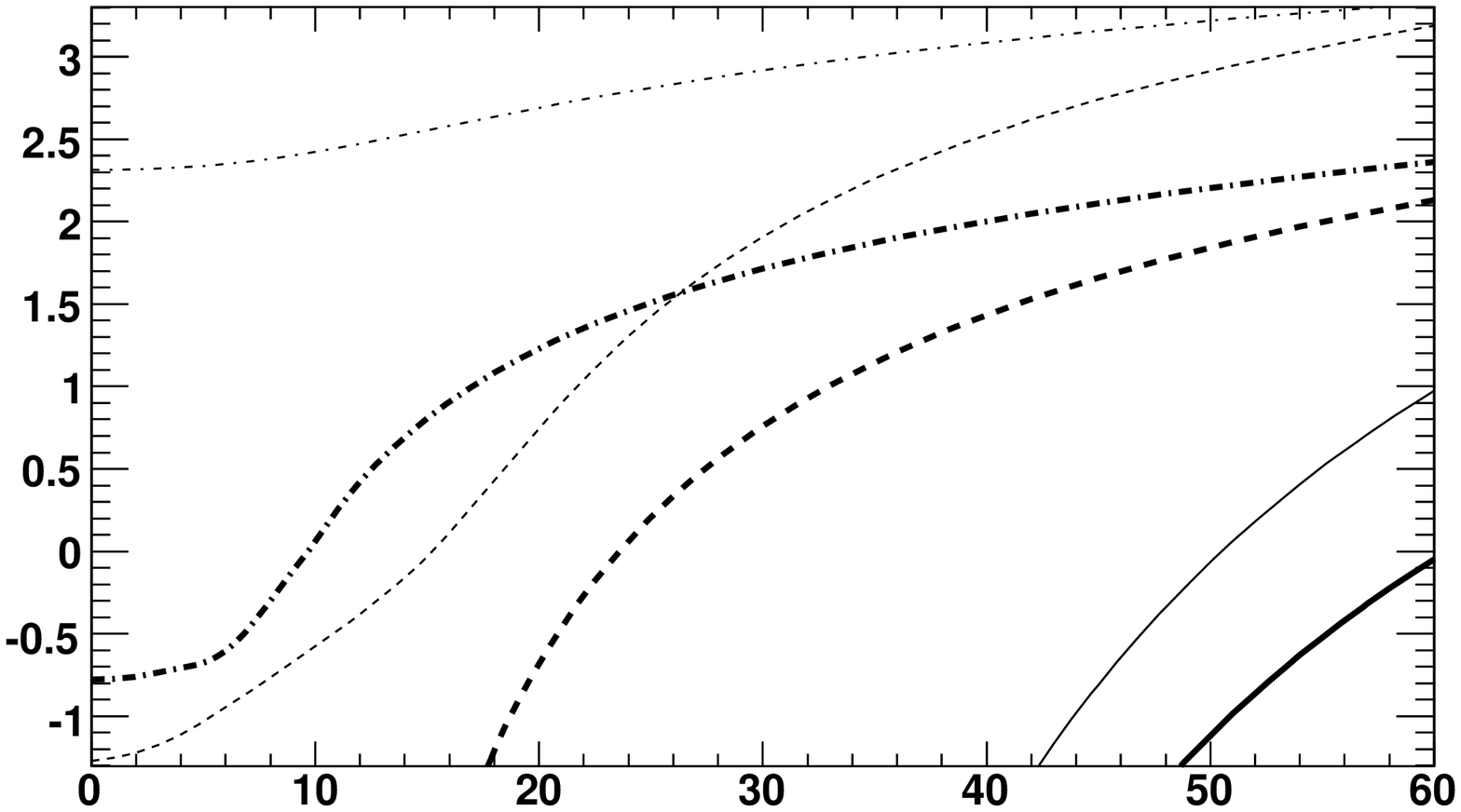}

\includegraphics[width=0.33\textwidth, height=0.18\textheight, trim= 0 35 0 0,clip]{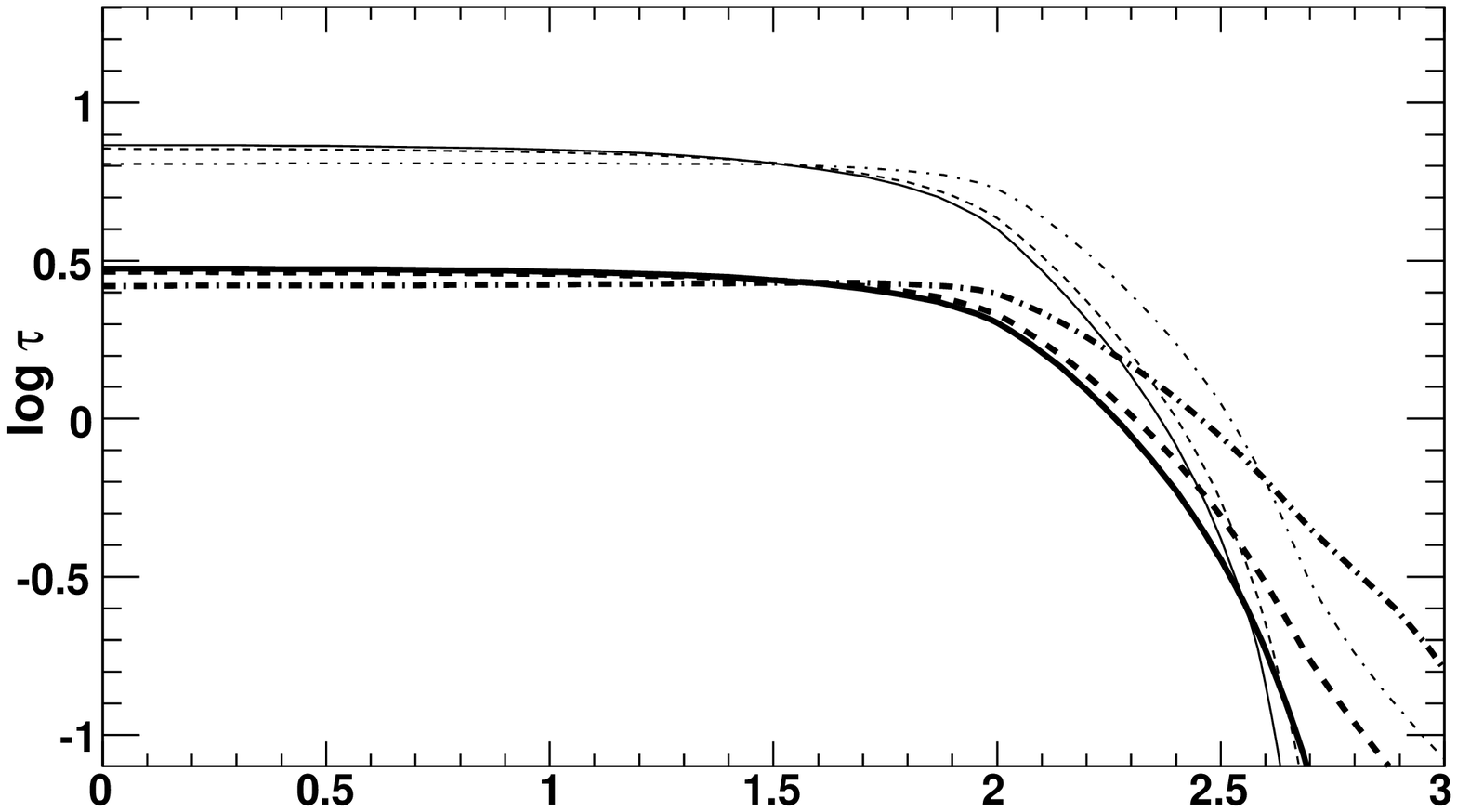}
\includegraphics[width=0.33\textwidth, height=0.18\textheight, trim= 5 35 0 0,clip]{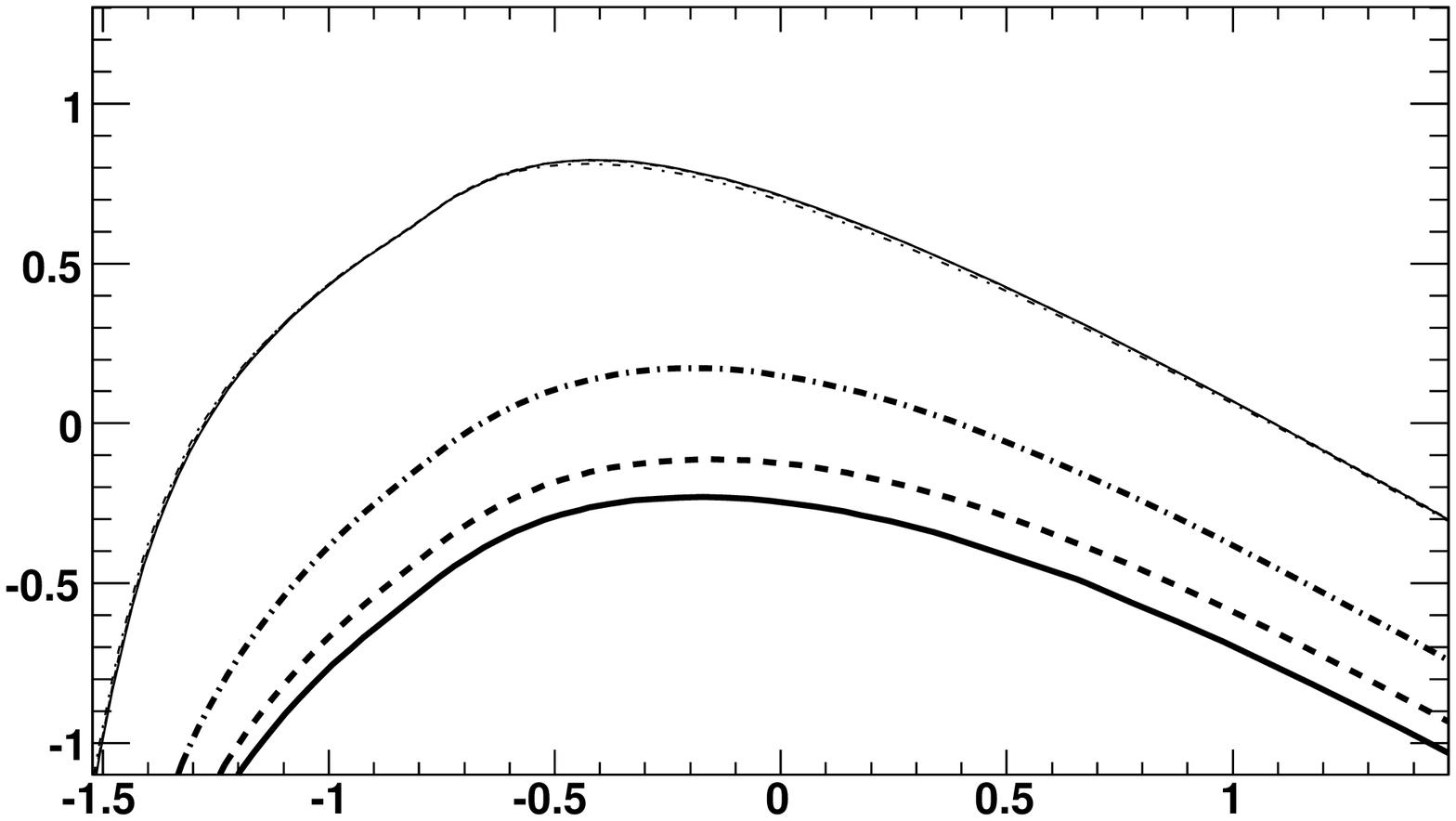}
\includegraphics[width=0.33\textwidth, height=0.18\textheight, trim= 5 35 0 0,clip]{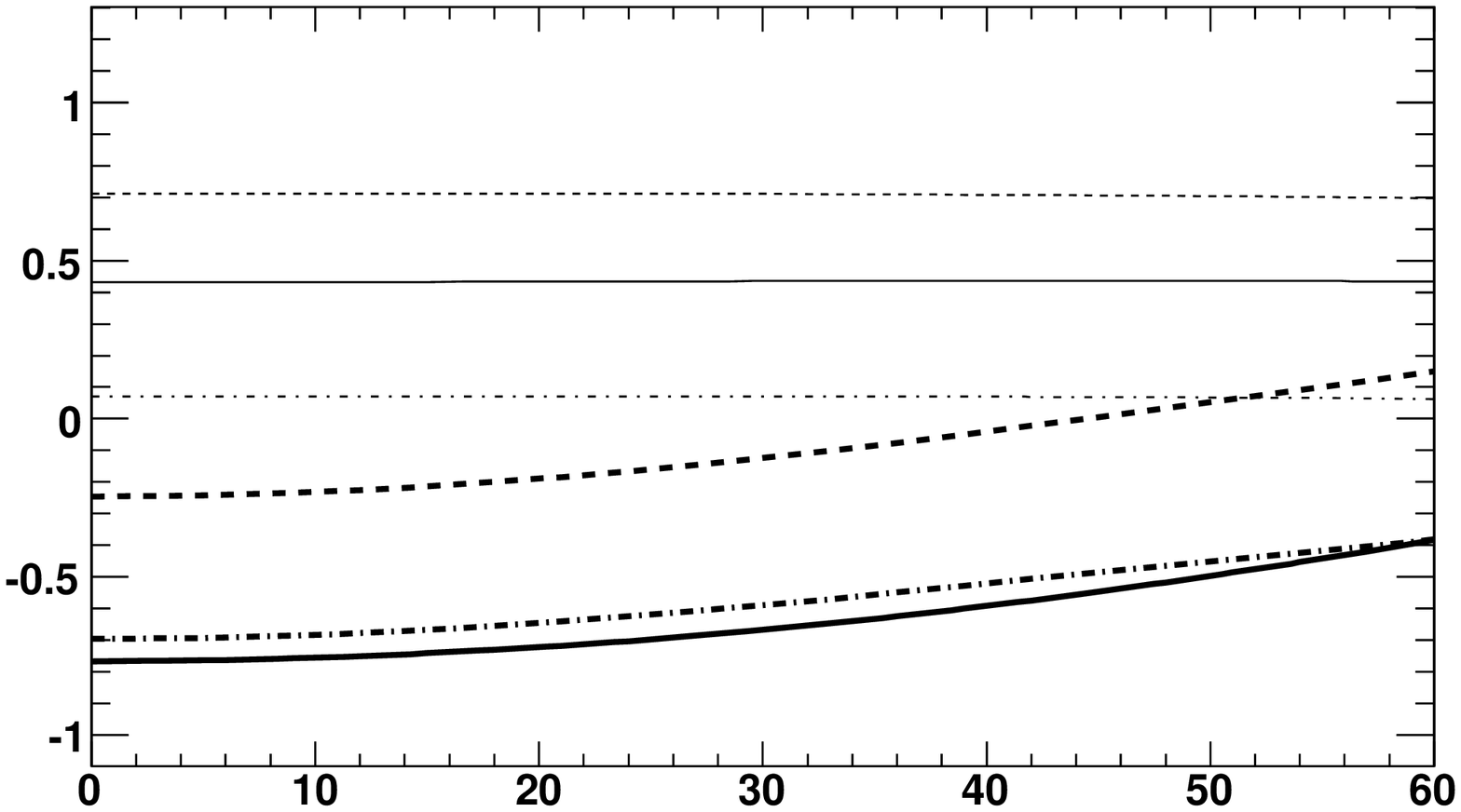}

\includegraphics[width=0.33\textwidth, height=0.18\textheight, trim= 0 0 0 0,clip]{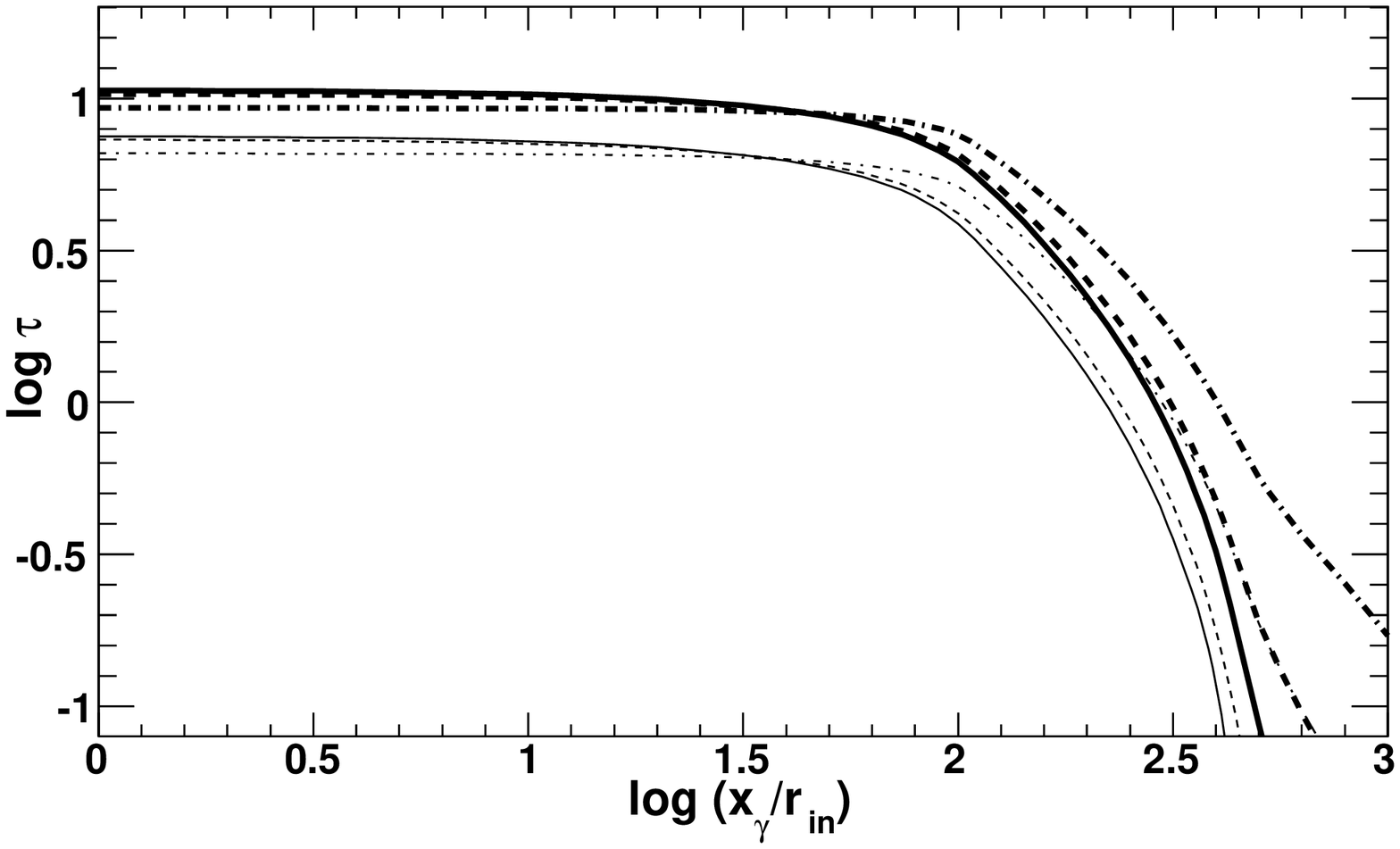}
\includegraphics[width=0.33\textwidth, height=0.18\textheight, trim= 5 0 0 0,clip]{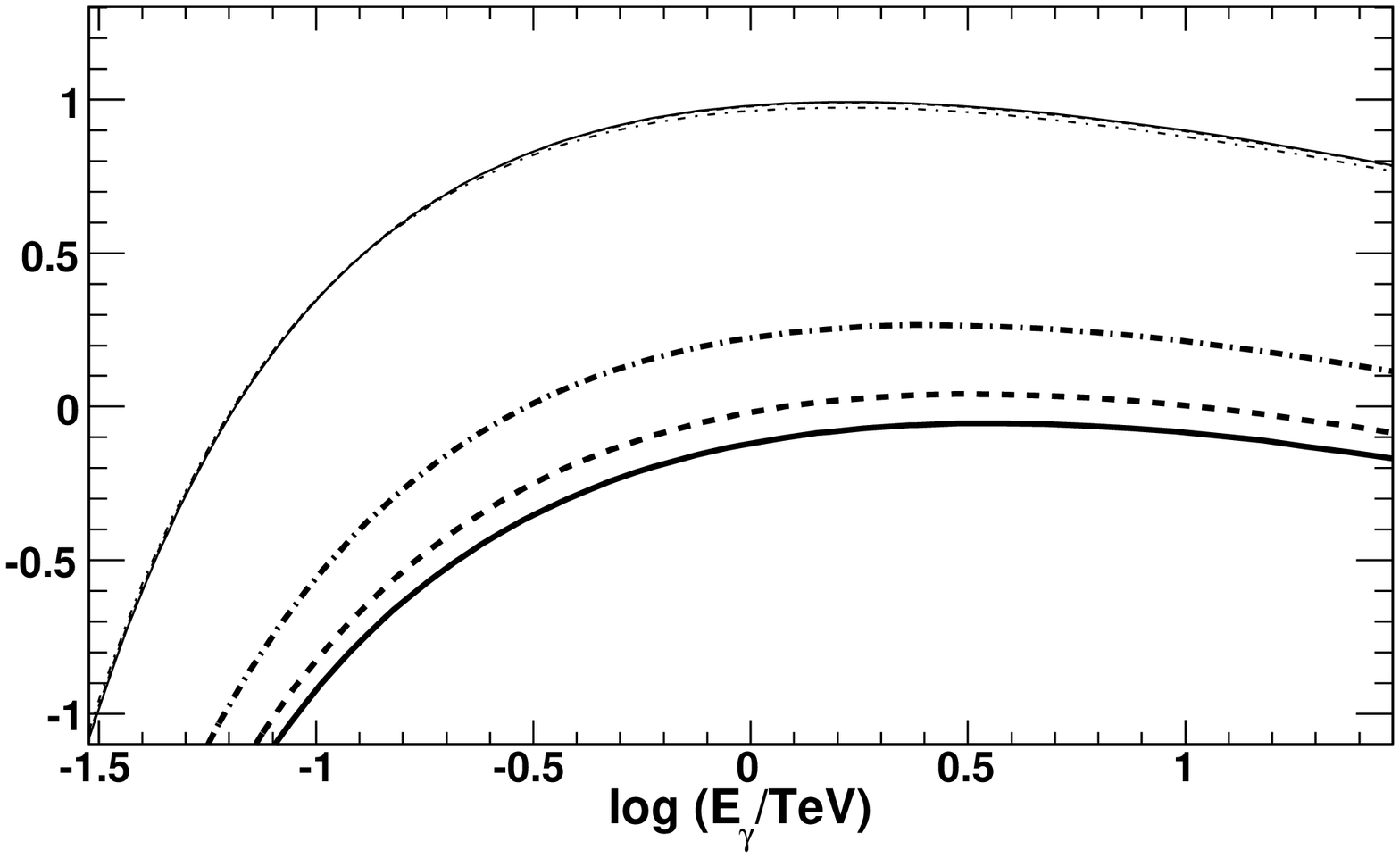}
\includegraphics[width=0.33\textwidth, height=0.18\textheight, trim= 5 0 0 0,clip]{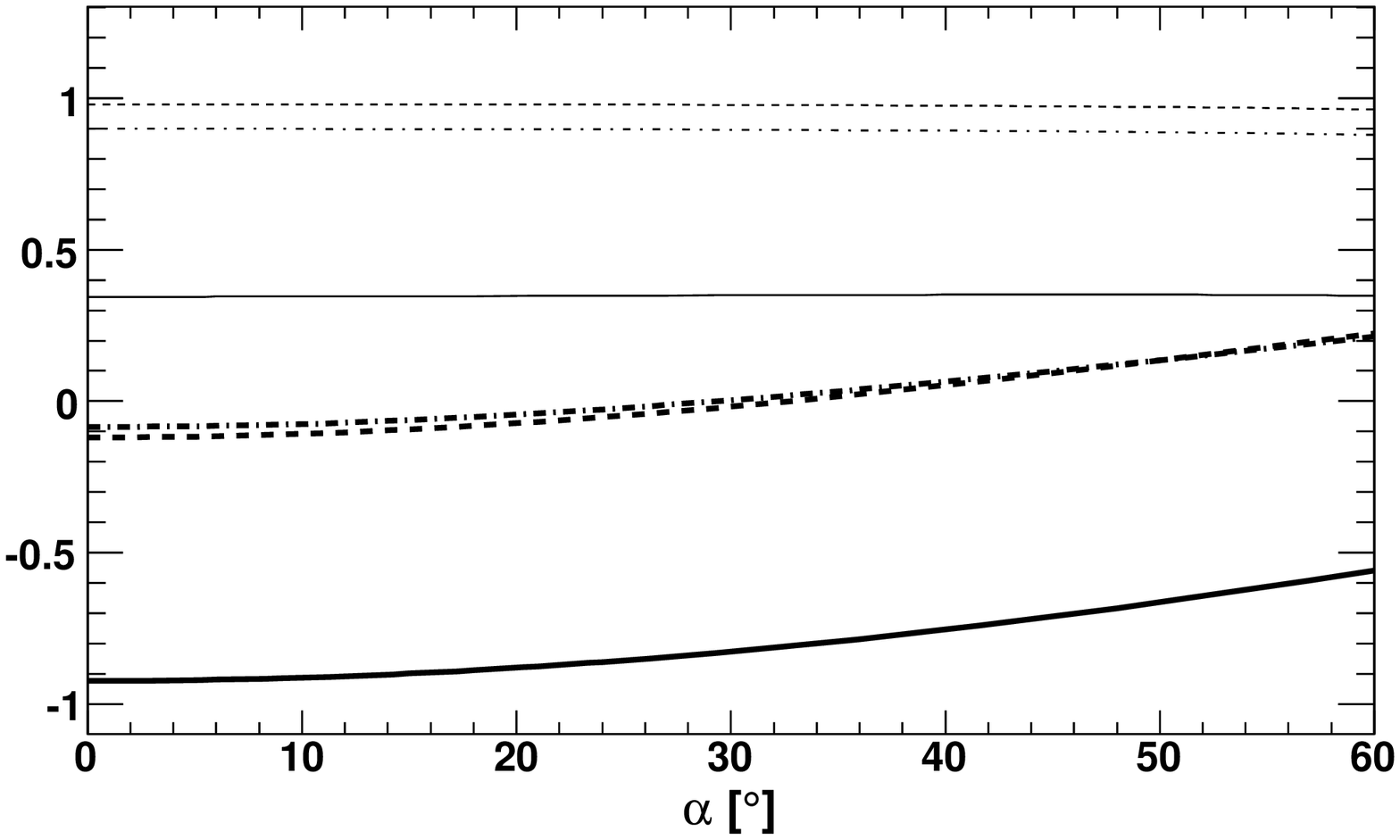}
\caption {Optical depths for gamma-ray photons are shown for the example parameters of 3C~273 for various radiation fields: disk radiation (upper panel), lines (middle) and continuous emission of BLR (lower). Dependence on the  
injection distance, $x_\gamma$, are shown on the left figures for two energies of the $\gamma$-ray photons: $E_\gamma$=0.3 TeV (thin curves), 3 TeV (thick curves) and different injection angles $\alpha=0^\circ$ (solid), $30^\circ$ (dashed), and $60^\circ$ (dot-dashed), respectively. Dependence on the $\gamma$-ray energy is shown on 
the middle figures for different injection distances from the base of the jet: 
$x_\gamma=30\, r_{in}$  (thin curves), $300\, r_{in}$ (thick) and different injection 
angles: $\alpha=0^\circ$ (solid curves), $30^\circ$ (dashed), and $60^\circ$ (solid).
Dependence on the injection angle $\alpha$ is shown on the right figures for two injection distances $x_\gamma=30\, r_{in}$  (thin curves), 
$300\, r_{in}$ (thick curves) and different energies of $\gamma$-ray photons 
$E_\gamma=0.1$ TeV (solid curves), 1 TeV (dashed), and 10 TeV (dot-dashed).}
\label{fig2}
\end{figure*}

\section{Optical depths for gamma-rays}

We calculate the optical depth for $\gamma$-rays in the radiation field inside the central parts of the AGNs according to standard prescription,
\begin{equation}
\tau=\int_\ell dl \int d\epsilon d\Omega n(l, \epsilon, \Omega) \sigma_{\gamma\gamma}(\epsilon,\theta)(1-\cos\theta), \label{eq:tau}
\end{equation}
where $n(l, \epsilon, \Omega)$ is the differential number density of soft photons with energy $\epsilon$ which arrive inside the solid angle $\Omega$ to instantaneous location of the $\gamma$-ray photon at the propagation distance $l$ (soft photons arrive either from the accretion disk, or its corona, or the BLR), $\sigma_{\gamma\gamma}$ is the pair production cross section, and $\theta$ is the angle between the momentum vectors of the gamma-ray and soft photon. $\ell$ denotes the path along propagation direction of the gamma-ray photon in the soft radiation field. Note that in the most general situation, $\gamma$-rays injected at an arbitrary place above the accretion disk and at an arbitrary direction, the calculations are not straightforward since considered radiation field is highly anisotropic (see Bednarek~1993). 

As a first step, we calculate the optical depths for the case of $\gamma$-ray blazar 3C~273 in which clear disk emission is observed in the optical-UV energy range. We also perform such calculations to another OVV blazar, 3C~279, which has been recently discovered in sub-TeV $\gamma$-ray energies by the MAGIC telescope (Albert et al.~2008).

\subsection{Absorption in the disk radiation}

We investigate the optical depths for $\gamma$-rays as a function of their injection place, their injection angle and their energies for the radiation field defined by the parameters characteristic for 3C~273 and for the simple Shakura Sunyaev temperature profile (see upper Figs.~\ref{fig2}). Our calculations show that the absorption effects of the TeV $\gamma$-rays can be very important 
in the case of this source. The basic features of these results, e.g. strong dependence on the distance from the accretion disk and the injection angle of the $\gamma$-rays,
can be naturally understood. Note that, the number of photons emitted from the specific ring on the accretion disk, with the radius $r$, scales as $N\sim r T^3\sim r^{-5/4}$. This means that most of the disk radiation comes from its inner part
of the disk. Therefore, radiation field seen by $\gamma$-rays is highly anisotropic.
In the case of small propagation angles $\alpha$, the collision angle, $\theta$, between directions of $\gamma$-ray and soft photon, is also small. Therefore, $e^\pm$ pair production processes is strongly suppressed due to higher energy threshold and geometric factor $(1-\cos\theta)$ enclosed in Eq.~\ref{eq:tau}.

\begin{figure}
\includegraphics[width=0.49\textwidth, trim= 15 21 41 30,clip]{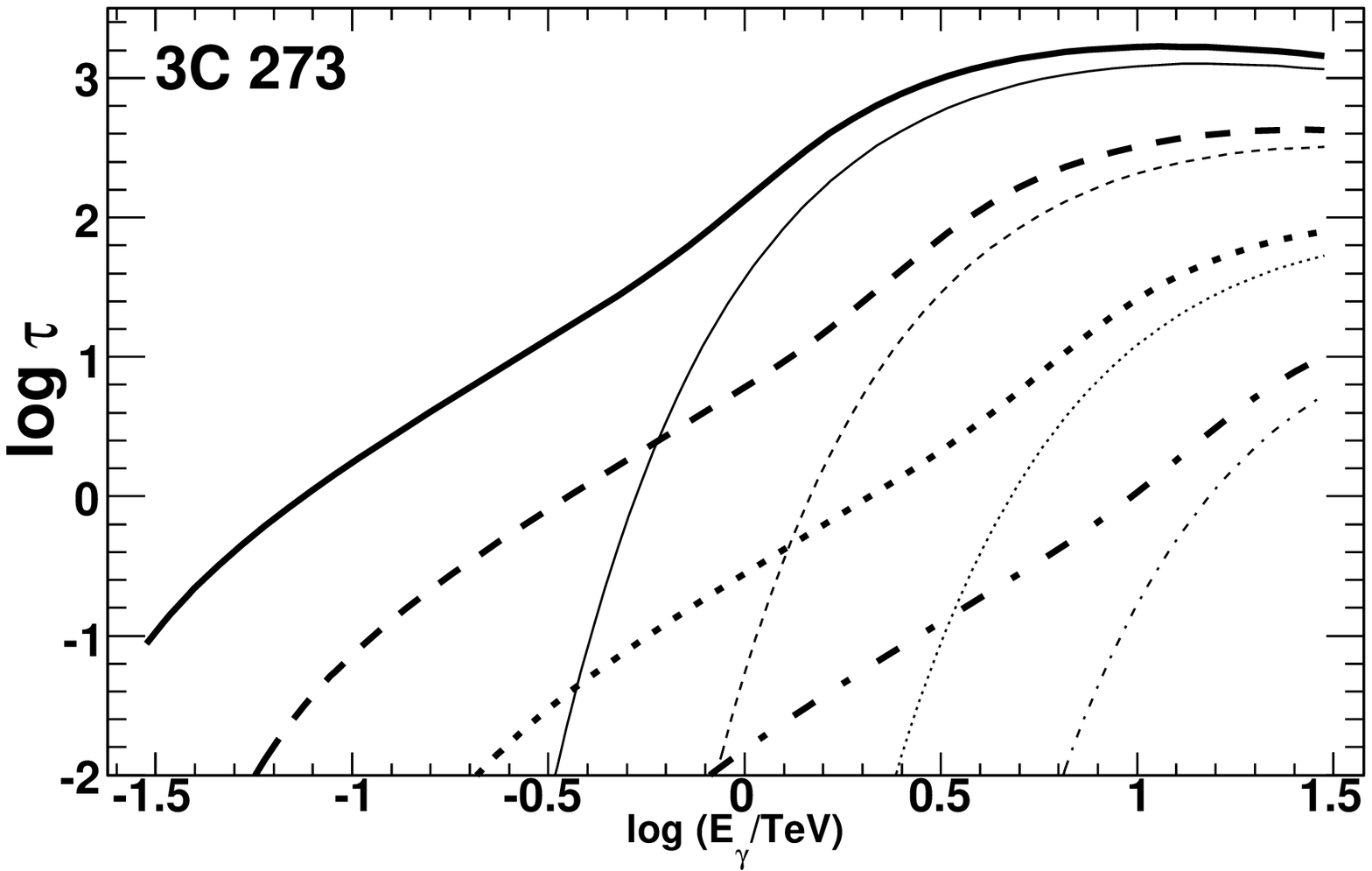}
\includegraphics[width=0.49\textwidth, trim= 15 0 41 30,clip]{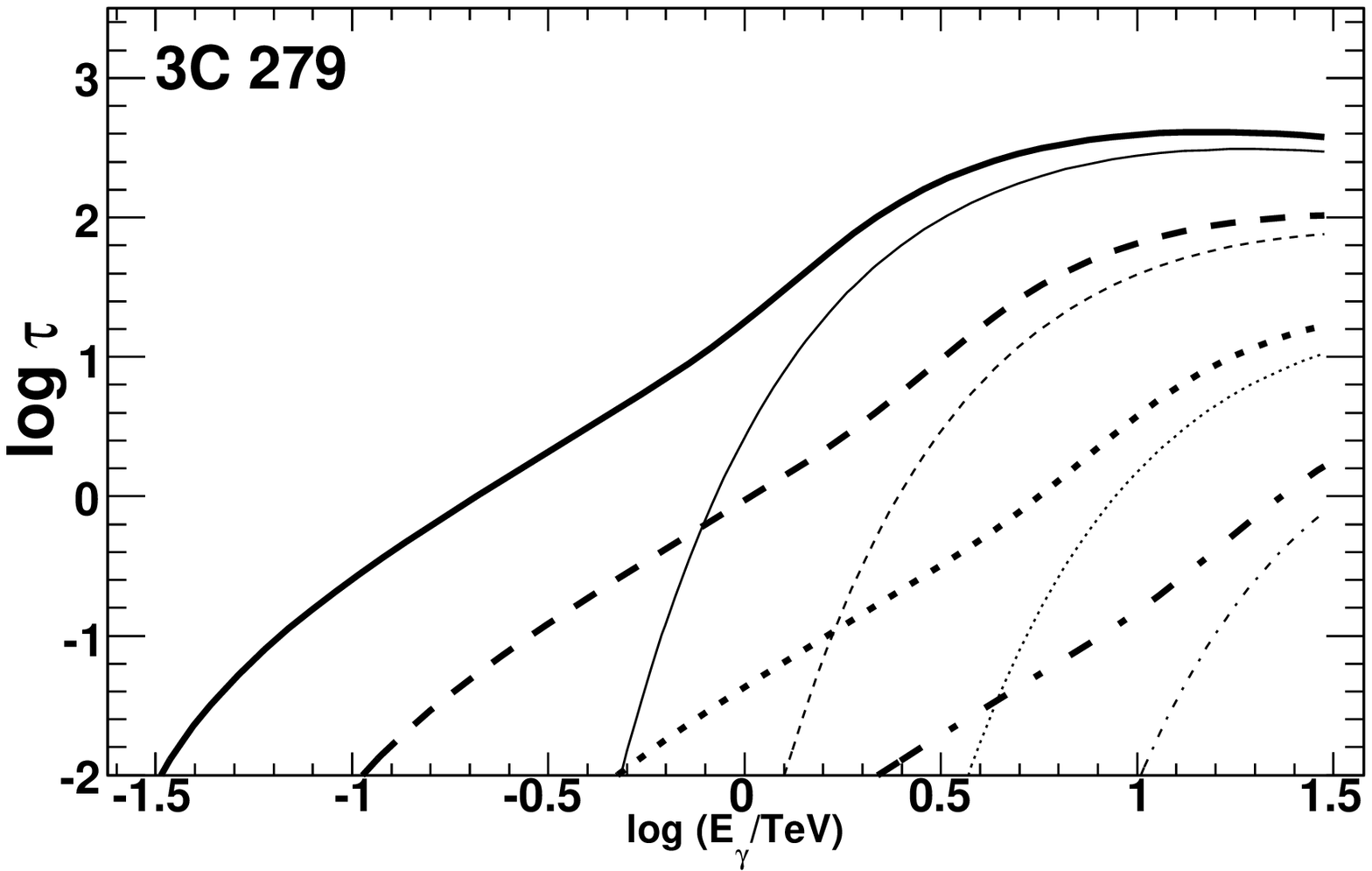}
\caption{Comparison of the optical depths for $\gamma$-rays in the radiation field of the Shakura Sunyaev accretion disk model (model I, thin curves) and this same accretion disk with the additional hot corona (model II, thick) for 3C~273 (upper figure) and 3C~279 (bottom). The optical depths are shown for $\gamma$-rays injected at the angle $\alpha = 0^{\rm o}$ and different distances from the base of the accretion disk: $x_\gamma = 10 r_{in}$ (solid curves), $x_\gamma = 30 r_{in}$ (dashed), $x_\gamma = 100 r_{in}$ (dotted) and $x_\gamma = 300 r_{in}$ (dot-dashed). The radiation field produced inside the accretion disk corona has a differential spectrum of a power low type ($E^{-2.7}$) and extends 
over whole disk. 
}\label{fig3}
\end{figure}

Dependence of the optical depths on the injection place is the strongest for $\alpha\approx 0^{\rm o}$ (see solid curves in the upper- left Fig.~2).
For $\gamma$-rays injected at $\alpha = 0^{\rm o}$ and relatively close to the black hole, a large part of soft photons from the accretion disk interacts with the $\gamma$-ray photon at large angle $\theta$ resulting in their quite efficient absorption. 
Dependence of the optical depths on the distance from the accretion disk is weaker in the case of higher energy $\gamma$-rays, when the threshold on $e^\pm$ pair production is not so important (compare the slope of the thin and thick curves in the upper-left Fig.~\ref{fig2}). Note how fast the absorption of $\gamma$-rays in disk radiation increases for higher values of the angle $\alpha$. The reason is that the most bright inner part of the disk is better seen and the interaction angle, $\theta$, between photons is large. 

The radiation field of the accretion disk considered in the case of 3C~273 seems to be one of the strongest in comparison to other blazars observed in the GeV $\gamma$-rays by the EGRET telescope. This might be the reason that the spectral index of the GeV emission from 3C~273 is rather steep in respect to the spectral indexes of other $\gamma$-ray OVV blazars, e.g. 3C~279. Due to larger optical depths, the primary TeV $\gamma$-ray photons can develop more efficient cascades in the 
radiation field surrounding the central engines of blazars. Such IC $e^\pm$ pair cascade in the anisotropic radiation is very complicated. Therefore, detailed calculations of the emerging cascade $\gamma$-ray spectra emerging from the optically thick central regions of AGNs (with highly anisotropic radiation fields) will be discussed in the future paper. 

We also compare the optical depths for $\gamma$-rays in the case of the presence of the hot
disk corona above the optically thick disk (model II). It is clear that the disk corona can significantly contribute to the optical depths at lower energies of $\gamma$-rays due to larger energies of soft photons in respect to photons coming from the thin disk alone (see Fig.~\ref{fig3}). This additional absorption in the radiation field of the disk corona can have an effect on the correct normalization of the calculated $\gamma$-ray spectra from 3C 273 and 3C 279 to their spectrum observed by the EGRET telescope in the GeV energies.

\subsection{Absorption in the BLR radiation}

It is assumed that emission of specific clouds in the BLR is isotropic. 
Thus, $\gamma$-ray photons collide with the large number of soft BLR photons at relatively large interaction angles $\theta$. Therefore, their absorption in BLR
radiation can be more important than in the radiation coming directly from the accretion 
disk even for relatively small re-radiation factors $\eta_{\rm l}$ and $\eta_{\rm c}$. 

The optical depths are roughly constant for the injection places which are inside the BLR, i.e. $x_\gamma < h_{in}$ (note that the thin curves in the right middle and lower Figs.~\ref{fig2} are nearly horizontal). However, the optical depths fall rapidly with increasing $x_\gamma$ inside the BLR $h_{\rm in} < x_\gamma < h_{\rm out}$. 
This is due to the fact that $\gamma$-rays, injected close to the black hole ($x_\gamma \ll h_{\rm in}$), see very similar BLR radiation field which is nearly independent on the $\gamma$-ray injection angle. Because a part of the BLR is obscured by the accretion disk, the values of the optical depths slightly decrease with increasing injection angle $\alpha$. On the other hand, for $\gamma$-rays injected inside the BLR ($x_\gamma < h_{\rm in}$), their propagation distance through the BLR increases with the injection angle $\alpha$. Therefore, the optical depths also increase with the angle $\alpha$. 

We assumed for simplicity that $\eta_c=\eta_l$, i.e. parts of the power of the accretion disk which is re-processed in the line and continuous radiation inside the BLR are comparable. Nevertheless, there is an important difference in the dependence on the optical depths for $\gamma$-rays with their energy in these two radiation fields. In the case of line emission from the BLR, all soft photons are concentrated in a small energy range. Therefore, absorption is the strongest for $\gamma$-rays with energies $E\sim \lambda m^2 c^3/h\approx 0.3$ TeV.
On the other hand, the continuous part of BLR radiation is composed of all wavelengths corresponding to the spectrum radiated by the accretion disk. This spectral distribution has a long low energy tail which is dominant target for the higher energy $\gamma$-rays.

\begin{figure}
\centering
\includegraphics[width=0.235\textwidth, height=0.235\textwidth, trim= 1 32 11 5,clip]{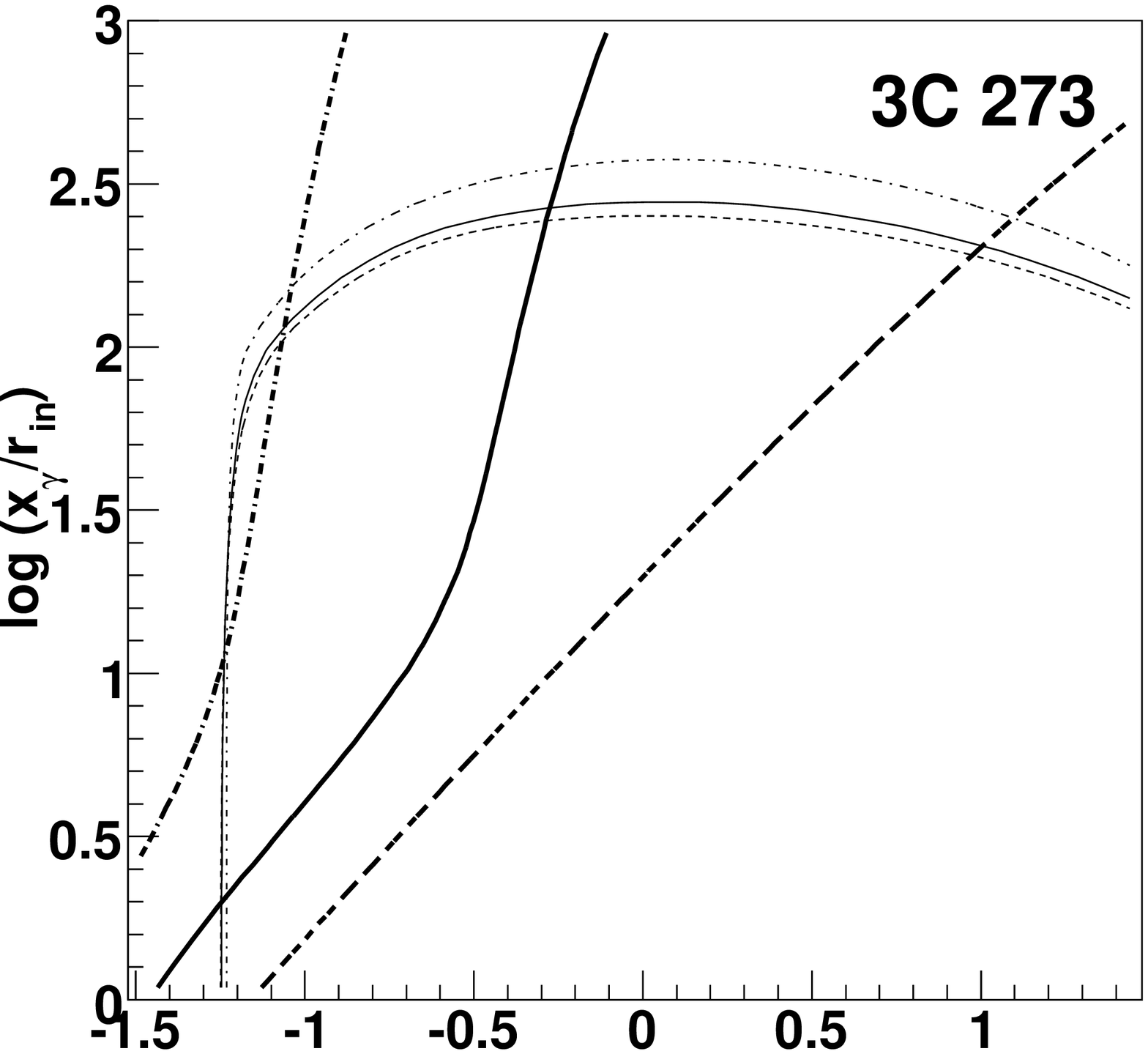}
\includegraphics[width=0.235\textwidth, height=0.235\textwidth, trim=26 32  0 4,clip]{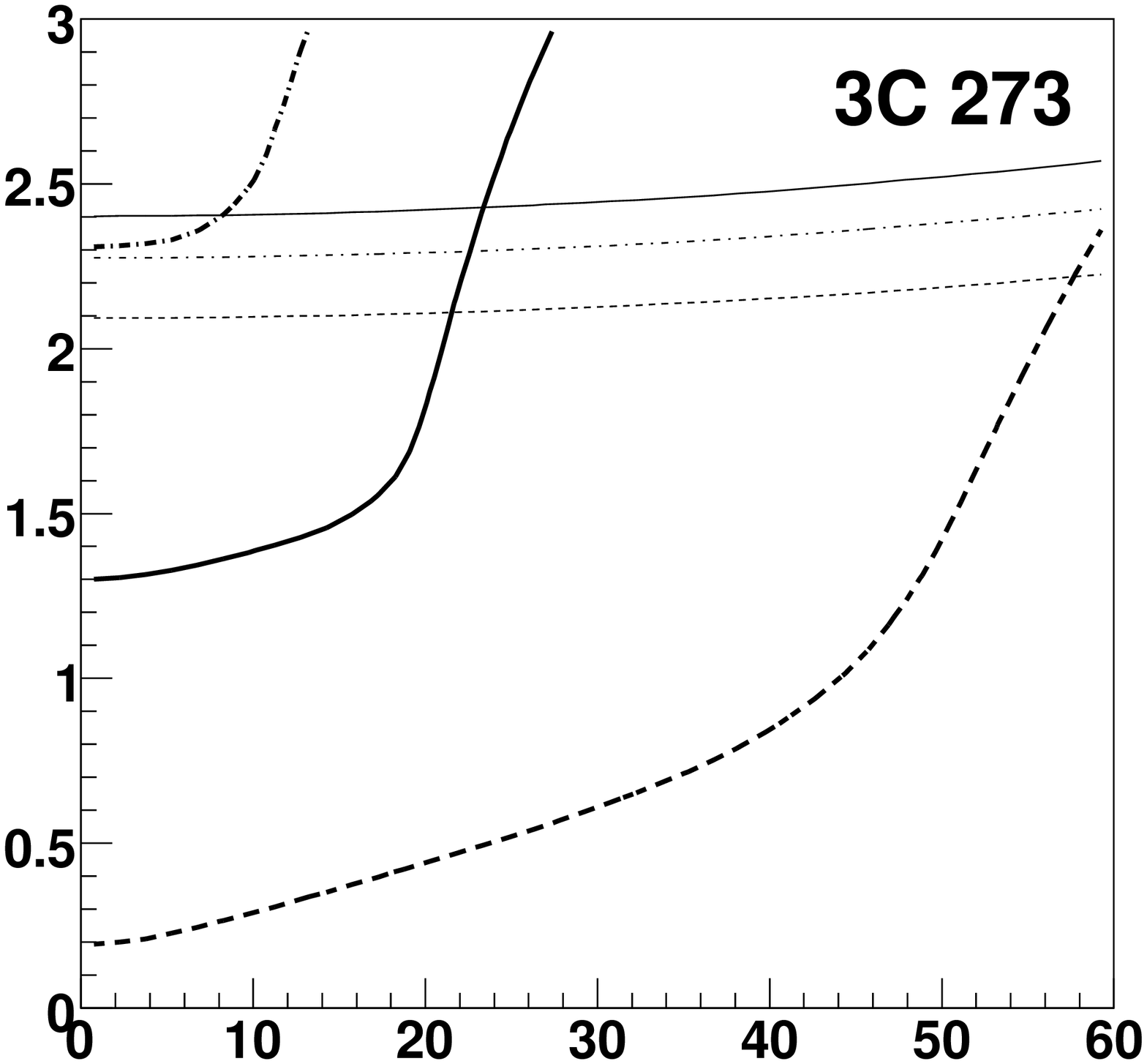}
\includegraphics[width=0.235\textwidth, height=0.235\textwidth, trim= 1 0  11 4.5,clip]{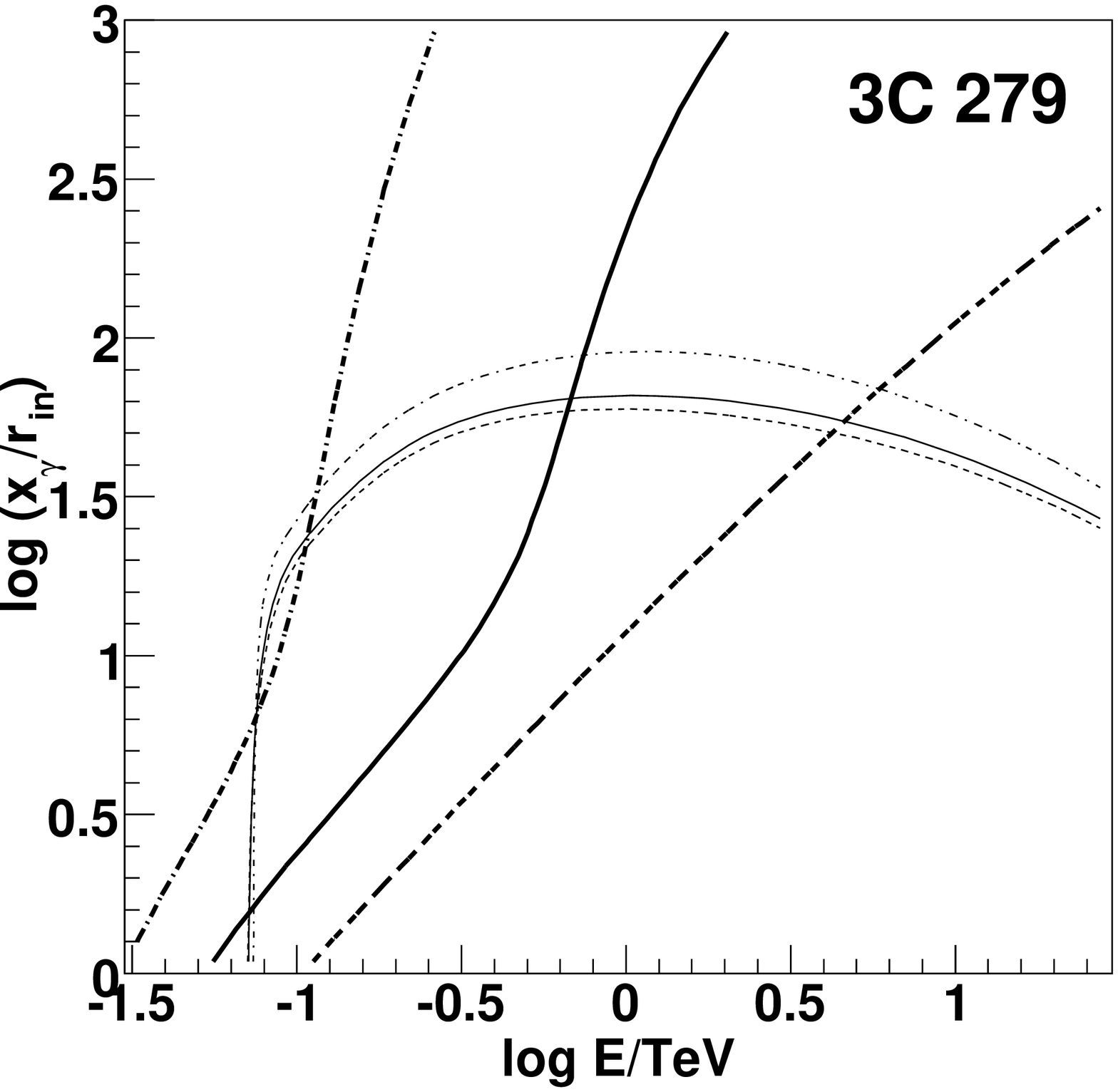}
\includegraphics[width=0.235\textwidth, height=0.235\textwidth, trim=26 0   0 4,clip]{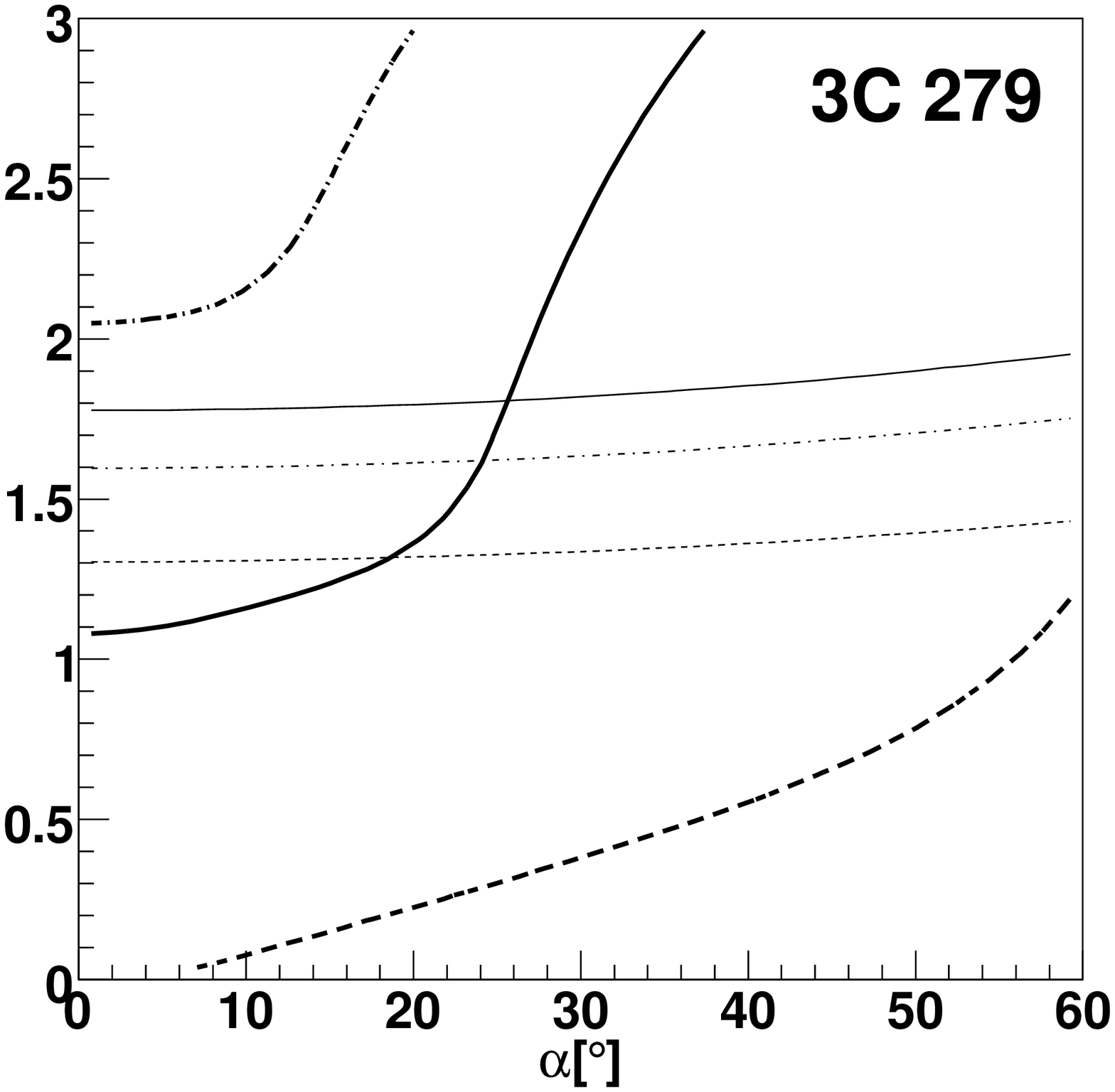}
\caption{$\gamma$-spheres for 3C~273 (upper panel) and 3C~279 (bottom) due to absorption of $\gamma$-rays in the disk radiation (thick curves), and in the radiation from the BLR (lines and continuous emission) (thin curves). The location of $\gamma$-spheres for different energies of injected $\gamma$-rays: $E_\gamma = 0.1$~TeV, (dashed curve); $1$~TeV, (solid); and $10$~TeV, (dot-dashed) are shown on the left and for selected injection angles ($\alpha = 0^\circ$, dashed curves; $30^\circ$, solid; and $60^\circ$, dot-dashed) on the right.}
\label{fig4}
\end{figure}
\subsection{Three-dimensional $\gamma$-spheres}

Based on the above calculations of the escape of $\gamma$-ray photons, we determine the three-dimensional surfaces around the central engine of AGN at which the optical depths for $\gamma$-rays with specific energies, $E_\gamma$, and injection angles, $\alpha$ (measured in respect to the direction perpendicular to the disk surface), are equal unity. The shape of this surface can be in general quite complicated, although it is called in the literature as a $\gamma$-sphere.
The evaluation of such a surface is very practical since $\gamma$-rays produced inside $\gamma$-sphere are strongly absorbed, while those ones produced outside $\gamma$-sphere can escape with negligible absorption. 

The observation of the superluminal motion in 3C~273, with apparent velocity of $\approx 5-10c$ (Unwin et al.~1985, Abraham et al.~1996), allows to put constraints on the jet viewing angle $\alpha_{\rm jet} < 15^{\rm o}$ and $\Gamma_{\rm jet} > 10$. On the other hand, modeling of the disk emission gives the best results for the inclination angle of the accretion disk close to $\alpha_{\rm d}\approx 60^{\rm o}$ (Kriss et al.~1999). This discrepancy suggests that in general jets do not need to propagate along the accretion disk axis. Then, the proper consideration of the escape of $\gamma$-rays in such a more complicated geometrical situation requires calculation of such full three-dimensional $\gamma$-spheres.

For the applied example values of the re-radiation factors, $\eta_{\rm l}$, and, $\eta_{\rm c}$ (mentioned above), the locations of the $\gamma$-spheres due to their absorption in the disk radiation and in the BLR radiation are of similar order provided that $\gamma$-rays are injected along the disk axis. The location of these specific $\gamma$-spheres are shown on Fig.~\ref{fig4} in the case of both OVV blazars (3C~273 and 3C~279), for the parameters of the radiation fields described above.
However, these two different radiation fields have different effects on the escape of $\gamma$-ray photons with the TeV energies. The $\gamma$-spheres in the BLR radiation are only weakly sensitive on energies of the $\gamma$-rays above 0.1 TeV and on their injection angles $\alpha$, since the BLR radiation field is quite isotropic. On the other hand, the $\gamma$-spheres due to the absorption in the disk radiation strongly depend on energies and injection angles of $\gamma$-rays. In general, $\gamma$-spheres broaden with energies of the TeV $\gamma$-rays and also with their injection angles. This is due to the relatively narrow peak in the spectrum of soft photons coming from the accretion disk. 

\begin{figure*}
\includegraphics[width=0.325\textwidth, height=0.32\textwidth, trim=  0 30 0 0,clip]{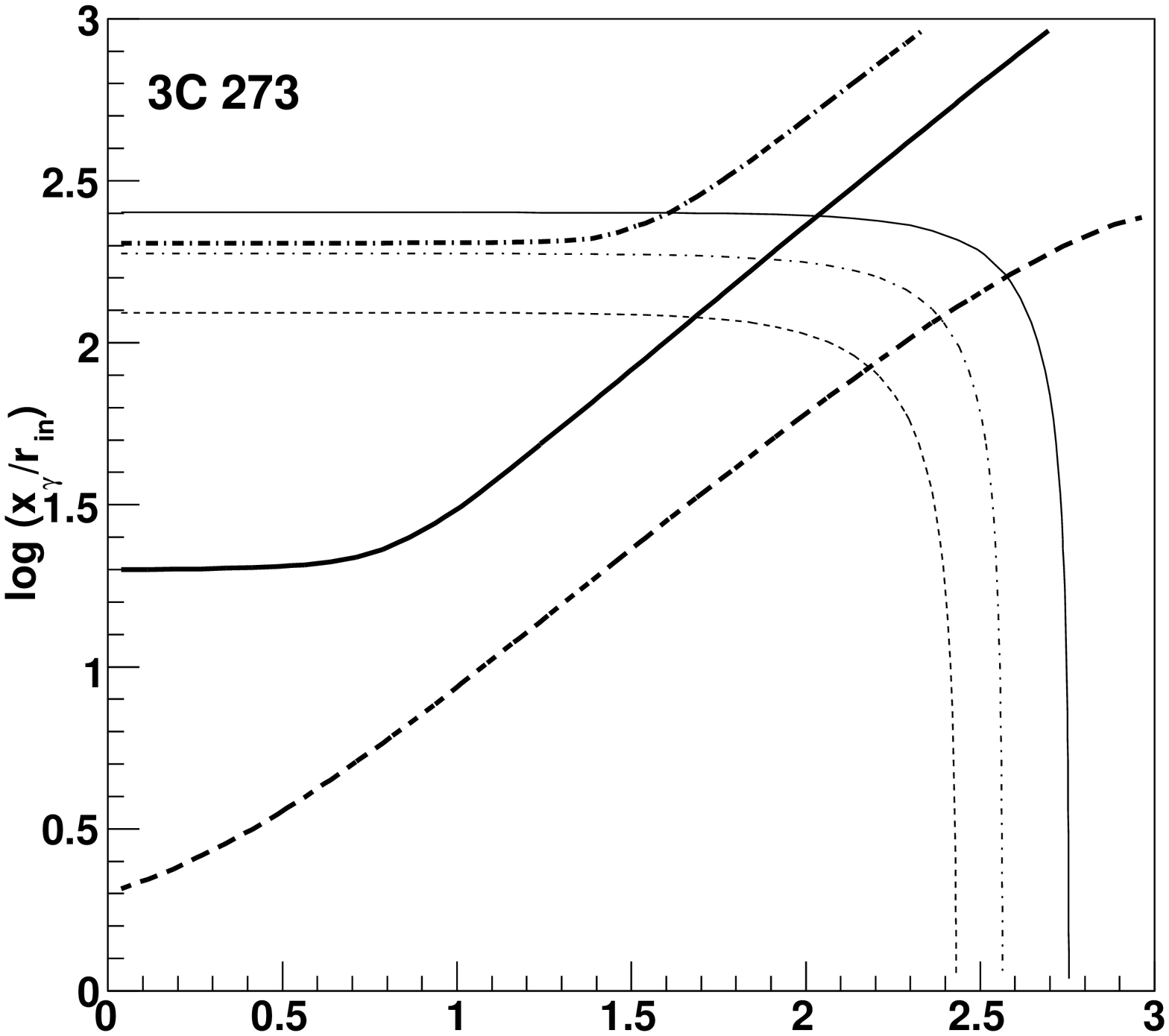}
\includegraphics[width=0.325\textwidth, height=0.32\textwidth, trim= 20 30 0 0,clip]{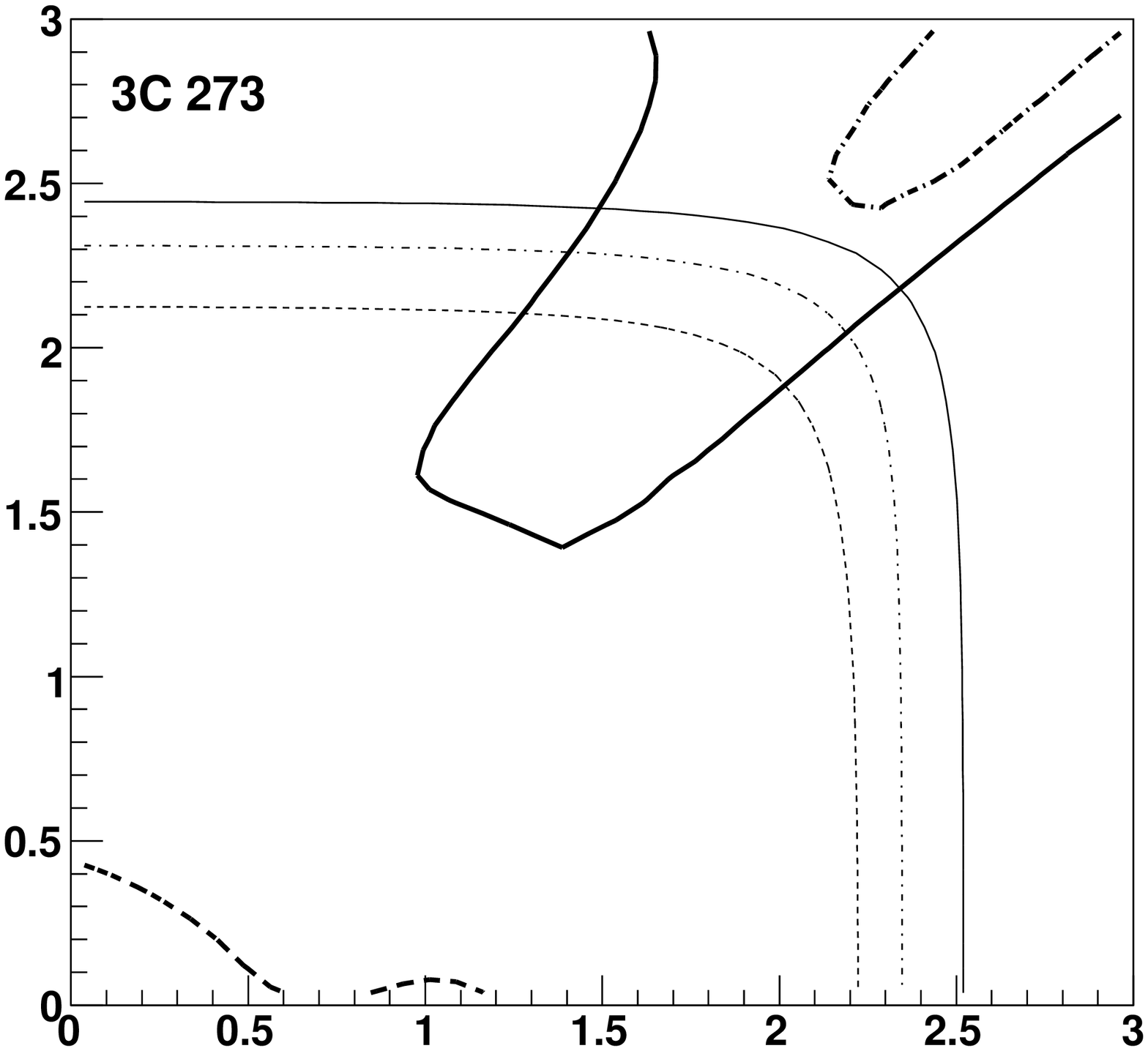}
\includegraphics[width=0.325\textwidth, height=0.32\textwidth, trim= 20 30 0 0,clip]{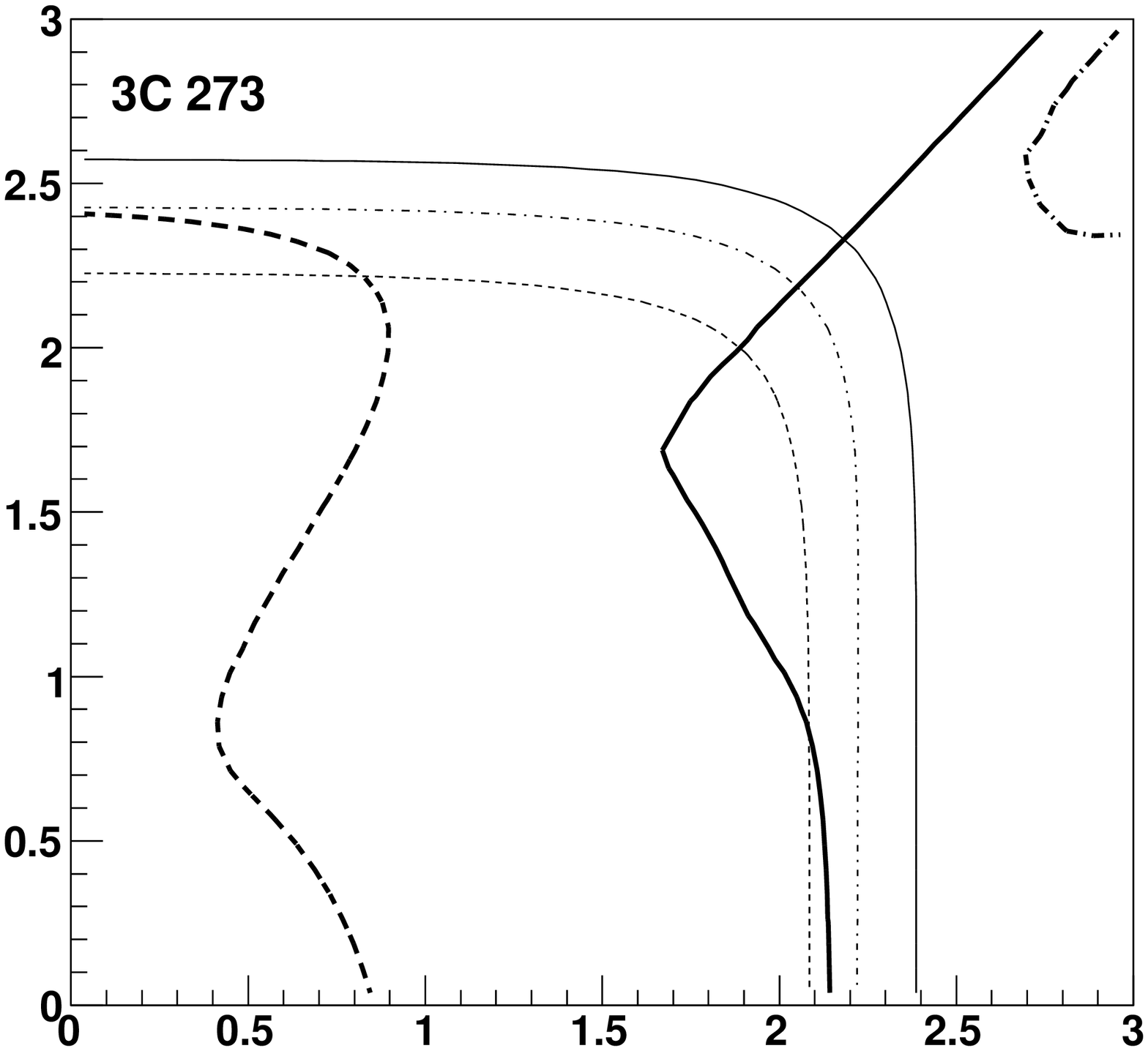}
\includegraphics[width=0.325\textwidth, height=0.32\textwidth, trim=  0 3 0 0,clip]{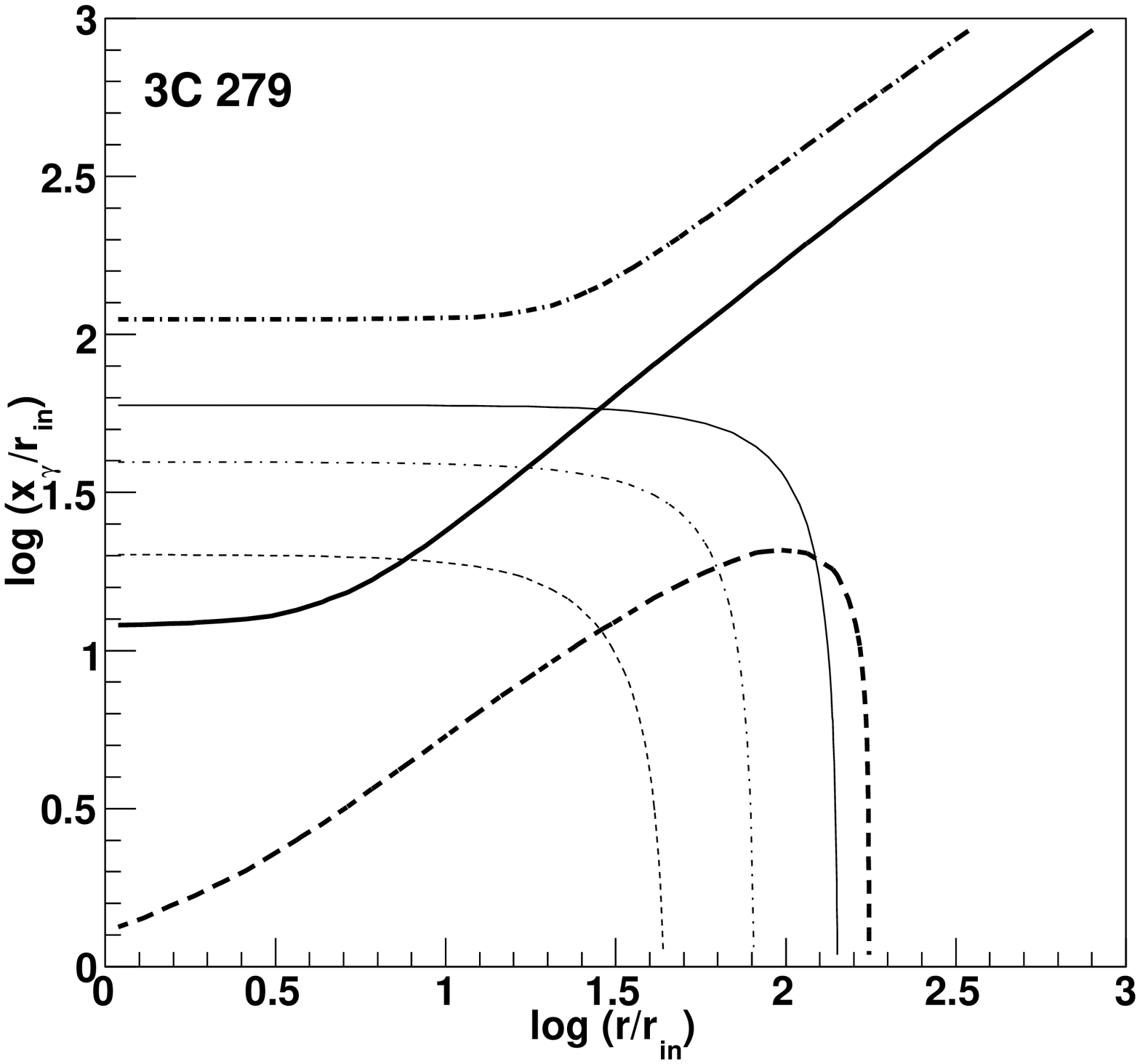}
\includegraphics[width=0.325\textwidth, height=0.32\textwidth, trim= 20 3 0 0,clip]{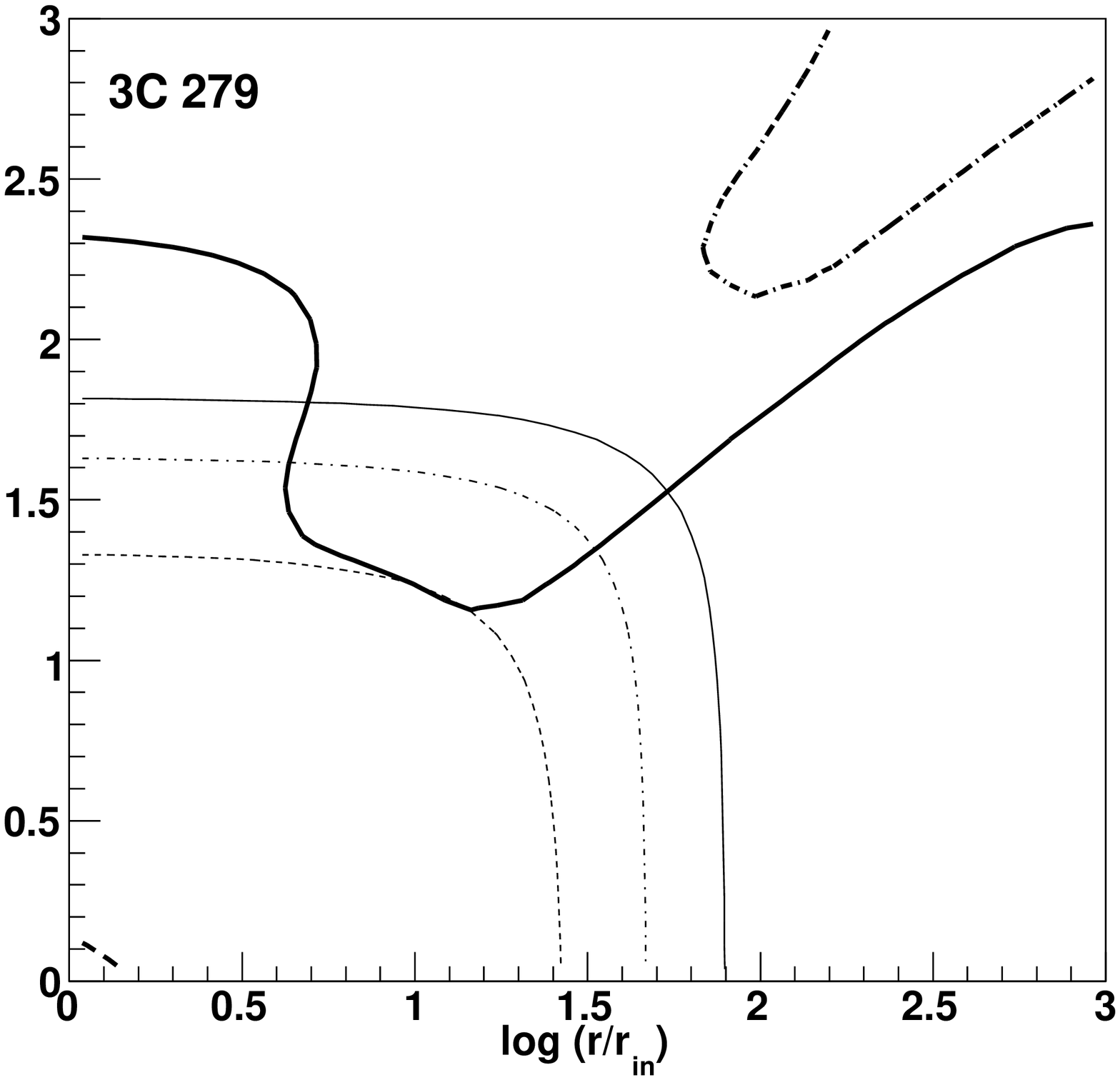}
\includegraphics[width=0.325\textwidth, height=0.32\textwidth, trim= 20 3 0 0,clip]{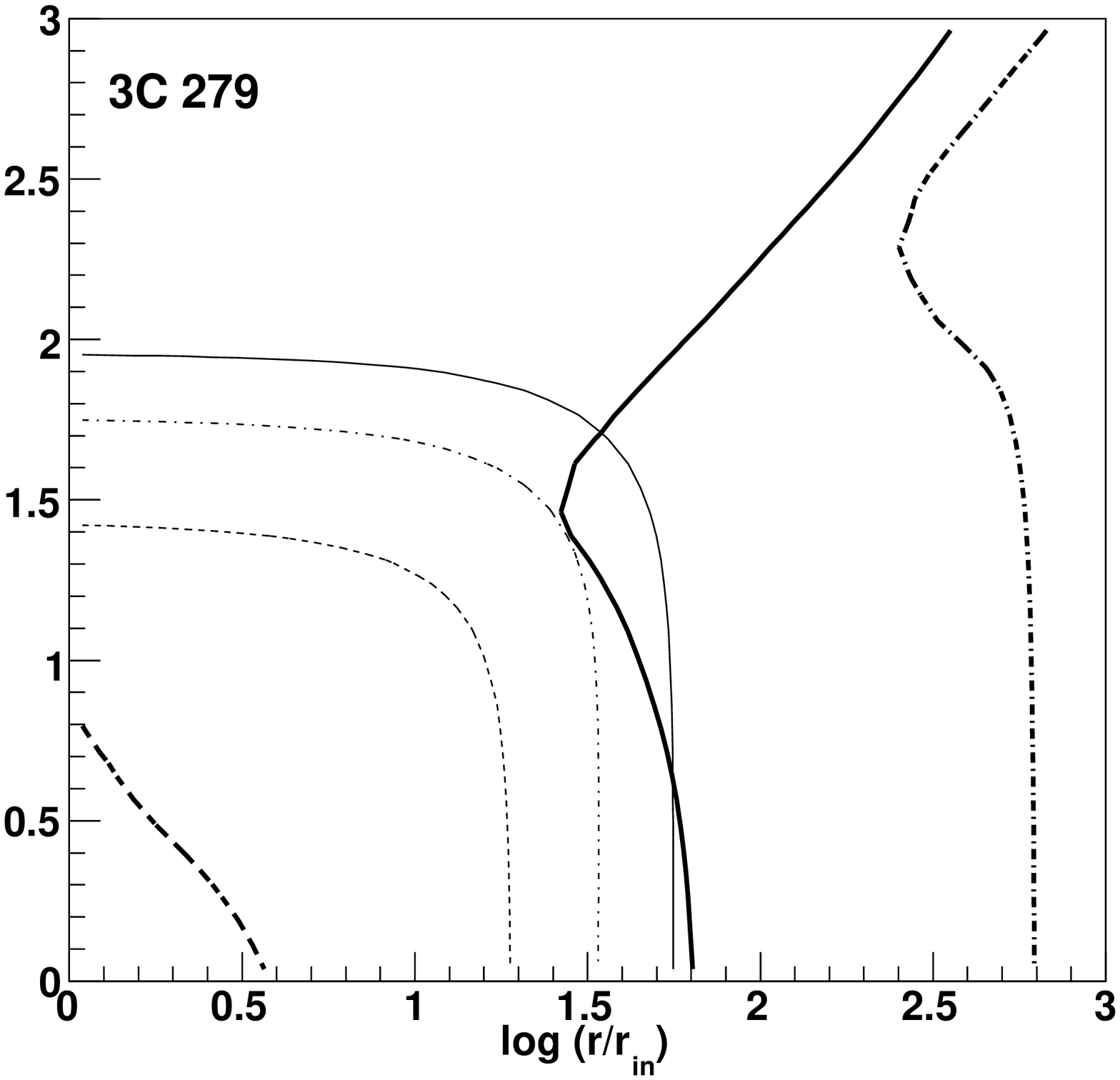}
\caption{Three-dimensional $\gamma$-spheres for 3C~273 (upper panel) and 3C~279 (bottom) due to the absorption of $\gamma$-rays in the disk radiation (thick curves) and in the radiation from the BLR (lines and continuous emission) (thin curves). 
The location of the $\gamma$-sphere (in coordinates: distance $r$ from the disk axis in units, $r_{\rm in}$, and distance $x$ from the disk surface in these same units) is shown for different energies of injected $\gamma$-rays: $E_\gamma = 0.1$~TeV (dashed curve), $1$~TeV (solid), and $10$~TeV (dot-dashed), and the injection angles $\alpha = 0^\circ$ (left figures), $30^\circ$ (middle), and $60^\circ$ (right). The parameters describing the radiation of the accretion disk and the BLR are mentioned in the text.}
\label{fig5}
\end{figure*}

The detailed structures of the three-dimensional $\gamma$-spheres for these two sources are shown in Fig.~\ref{fig5}. We investigate the location of the $\gamma$-sphere as a function of the injection angles, $\alpha$, and energies of $\gamma$-rays. In the case of their absorption in the BLR radiation, the 
$\gamma$-spheres depend only weakly on the injection angle due to high level of isotropy of
BLR radiation. However, the shapes of $\gamma$-spheres due to the absorption in the disk radiation are quite complicated. Note the curious shapes of the $\gamma$-spheres for
different injection angles of $\gamma$-ray photons. The $\gamma$-spheres at the axis of the accretion disk strongly depend on the injection angles of $\gamma$-rays. However,
they are at the closest distances to the central engines along directions corresponding to the values of the injection angles of $\gamma$-rays. Therefore, even in the case of highly inclined jet to the axis of the accretion disk, the $\gamma$-rays can escape efficiently.

According to our calculations, the $\gamma$-spheres around the central engine of 3C~273 are located at significantly larger distances than in the case of 3C~279. Therefore, in principle
TeV $\gamma$-rays originating closer to the central engine can escape from 3C~279.

\section{Modification of the injected gamma-ray spectra}

\begin{figure*}
\includegraphics[width=0.325\textwidth, height=0.325\textwidth, trim=  0 34 45 49,clip]{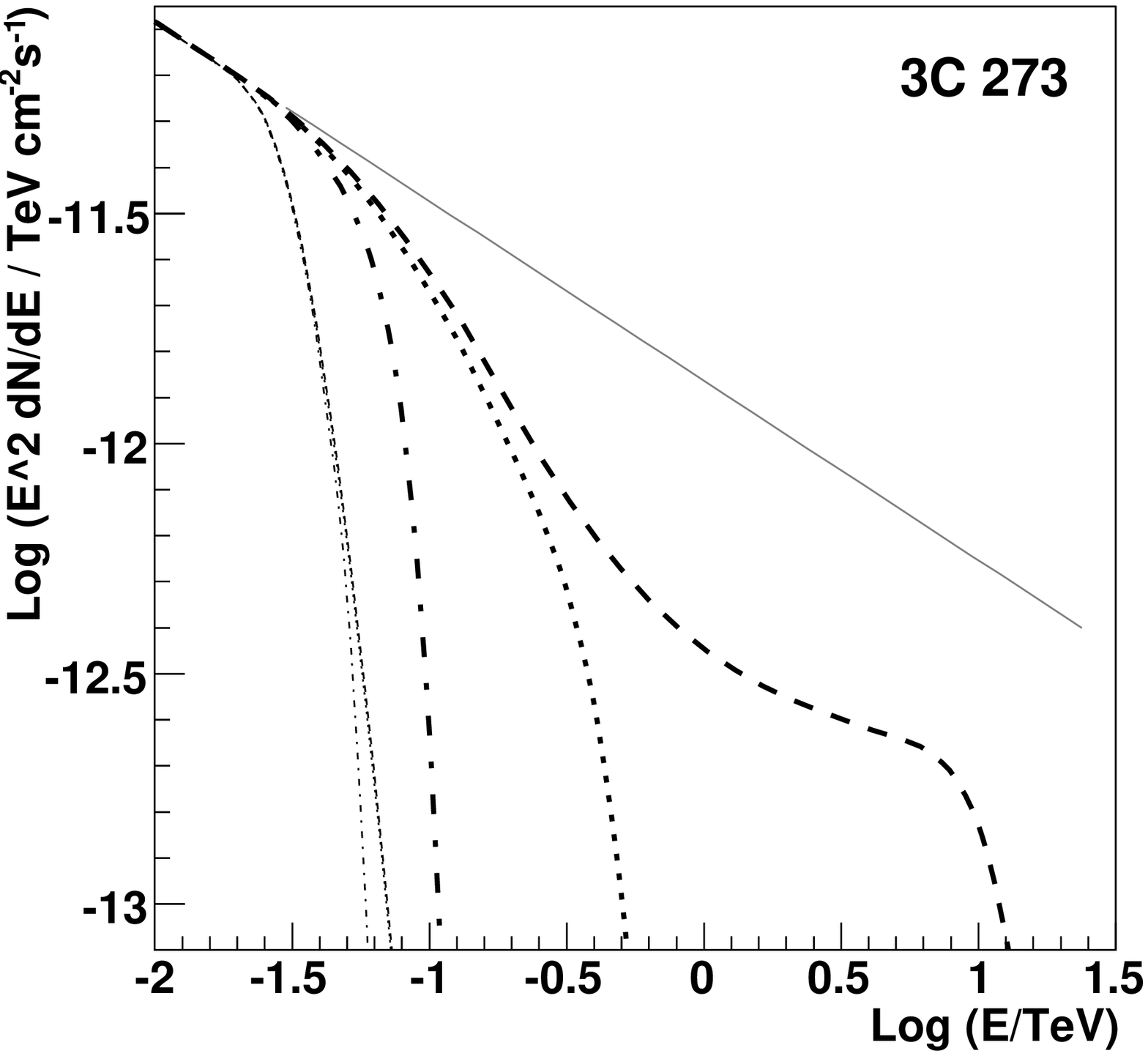}
\includegraphics[width=0.325\textwidth, height=0.325\textwidth, trim= 23 34 45 49,clip]{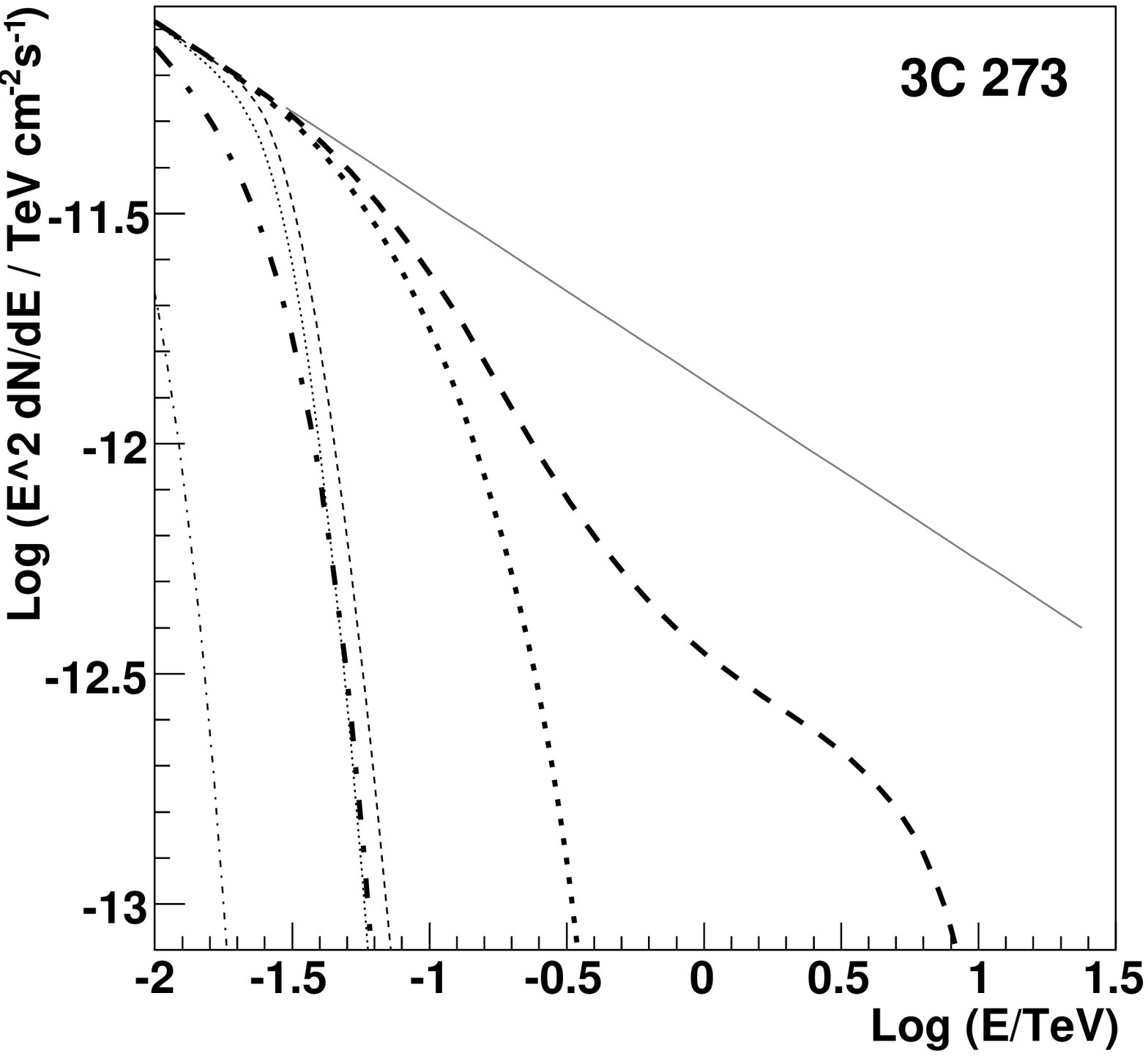}
\includegraphics[width=0.325\textwidth, height=0.325\textwidth, trim= 24 34 45 49,clip]{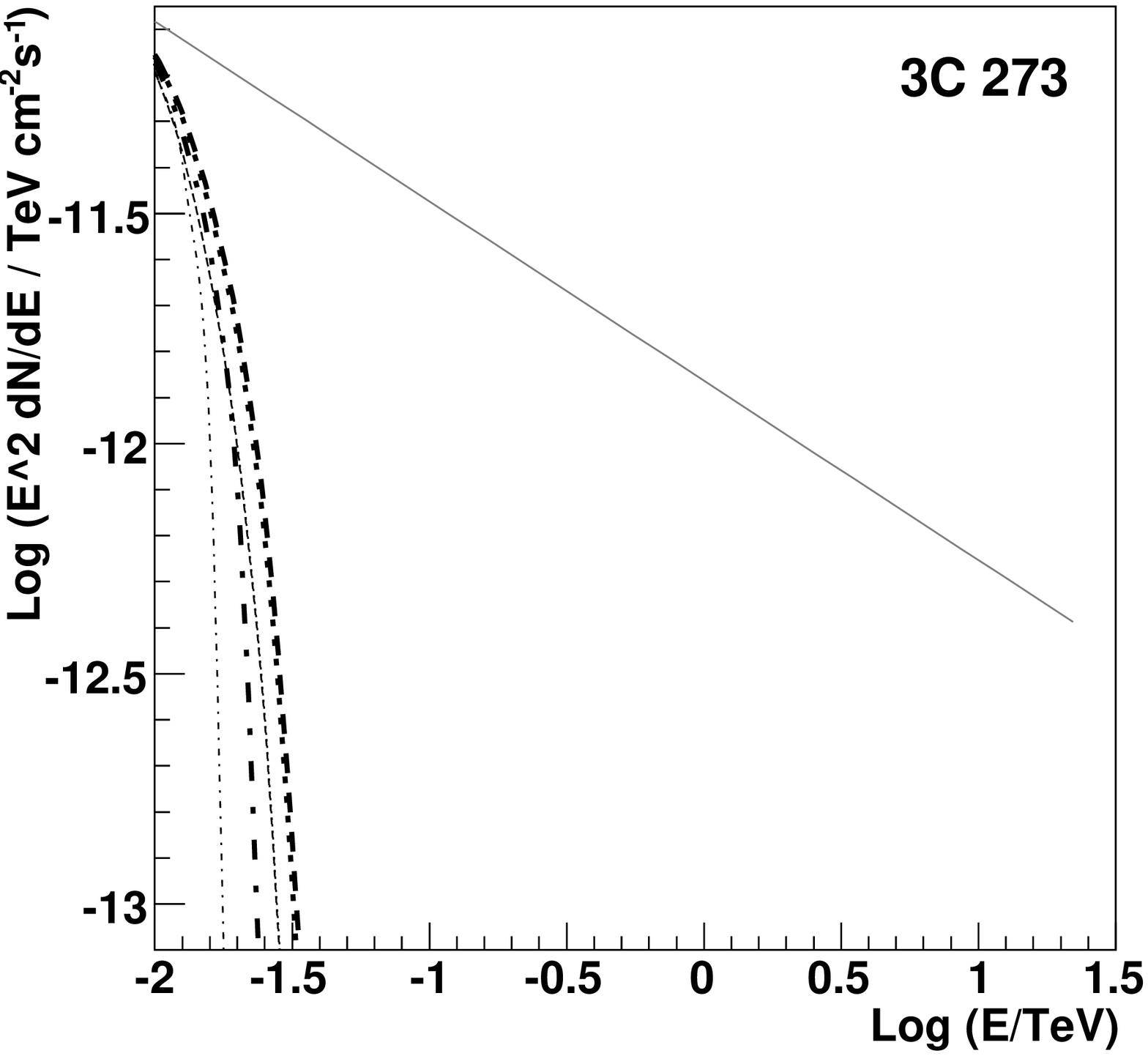}
\includegraphics[width=0.325\textwidth, height=0.325\textwidth, trim=  0  6 45 49,clip]{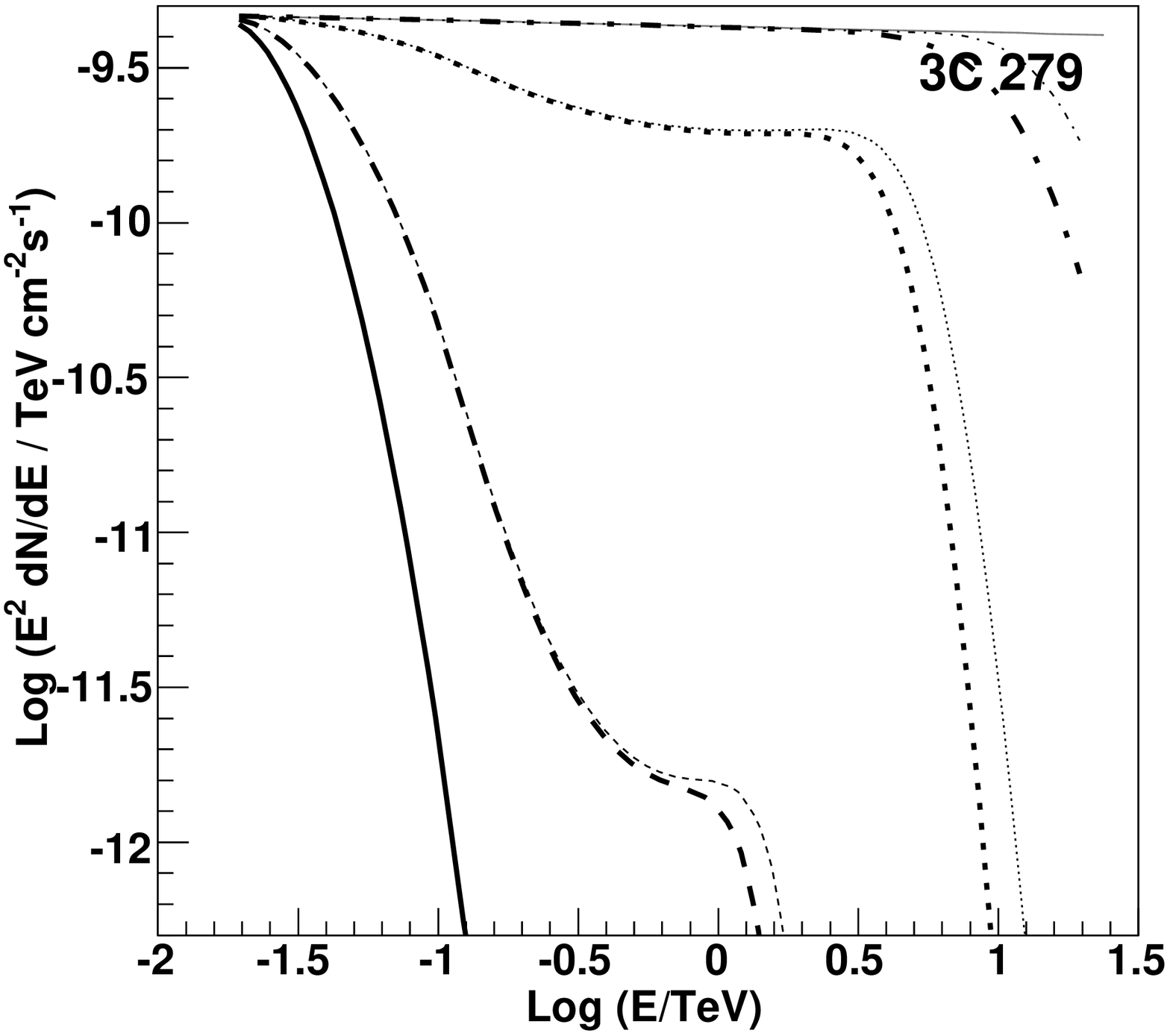}
\includegraphics[width=0.325\textwidth, height=0.325\textwidth, trim= 27  6 45 49,clip]{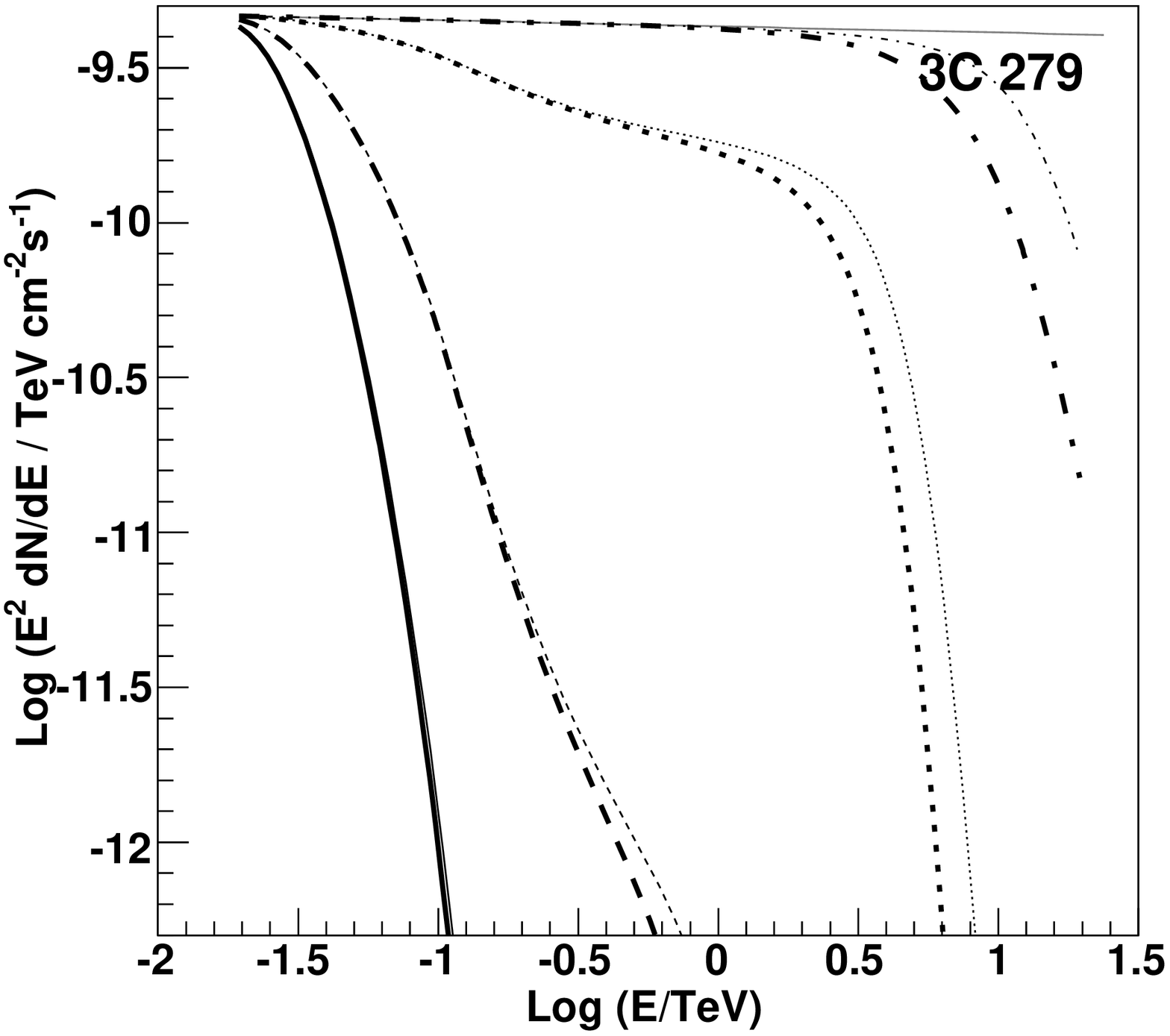}
\includegraphics[width=0.325\textwidth, height=0.325\textwidth, trim= 27  6 45 49,clip]{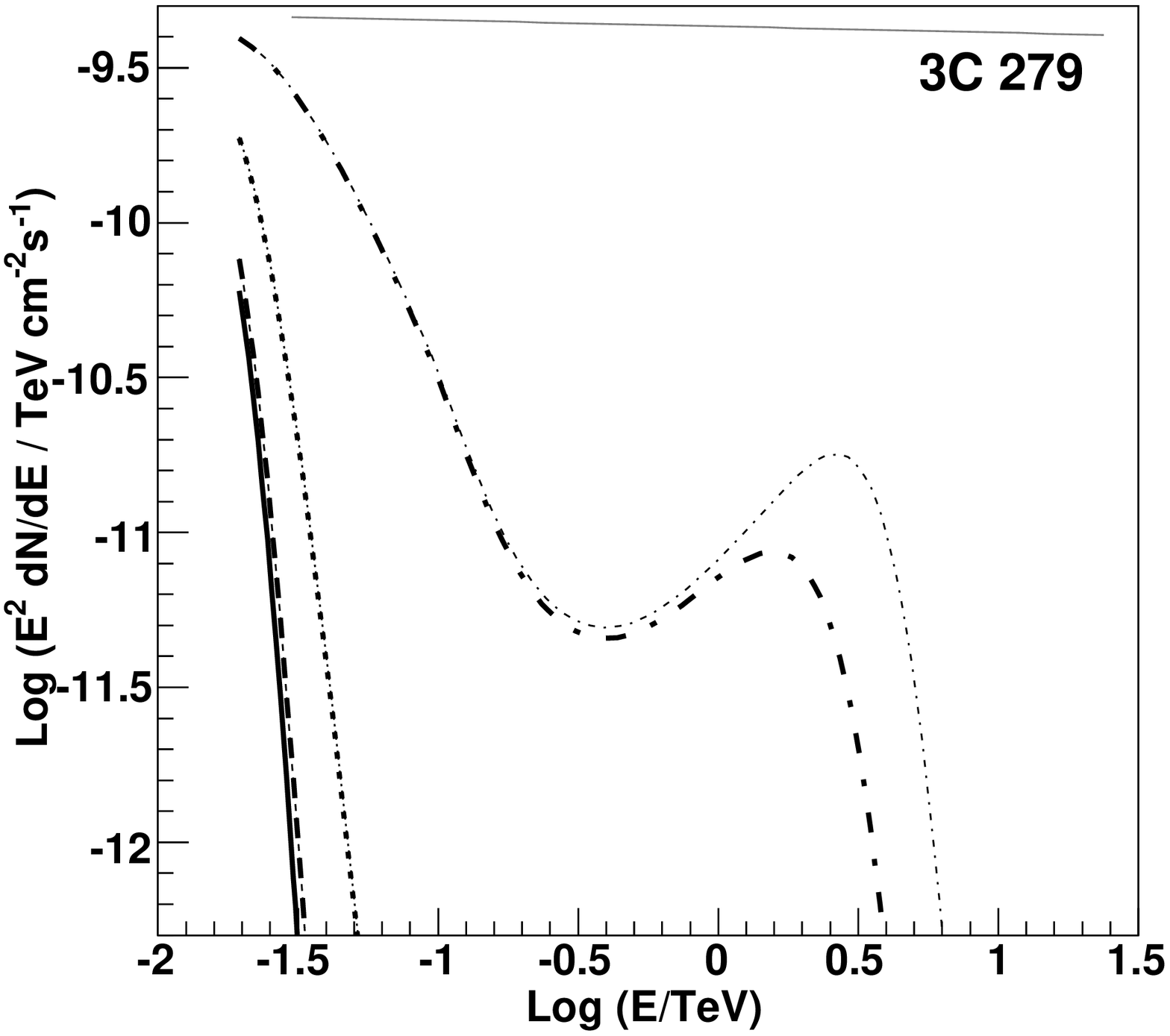}
\caption{Modification of the $\gamma$-ray spectrum extrapolated from the EGRET energy range to TeV energies, 3C~273 (upper panel) and 3C~279 (lower), by internal absorption in the disk and the BLR radiation for different models of radiation produced in the accretion disk: Shakura-Sunyaev disk model (model I, left figures), Shakura-Sunyaev model + power law high energy tail (model II, middle), high temperature disk model (model III, right). 
In the upper figures the injection angles of $\gamma$-rays are $\alpha=0^\circ$ (dashed), $30^\circ$ (dotted) or $60^\circ$ (dot-dashed curves) and distances from the accretion disk are: $x_\gamma = 30 r_{in}$ (thin curves) or $x_\gamma = 300 r_{in}$ (thick).
In the bottom figures the injection angle of $\gamma$-rays is $\alpha=0^\circ$ (thin curves) and $10^\circ$ (thick), and distances from the accretion disk are: $x_\gamma = 10 r_{in}$ (solid curves), $x_\gamma = 30 r_{in}$ (dashed), $x_\gamma = 100 r_{in}$ (dotted) and $x_\gamma = 300 r_{in}$ (dot-dashed).} 
\label{fig6}
\end{figure*}

As we have shown above, the $\gamma$-ray emission produced inside jets of blazars, but relatively close to the accretion disk and/or the BLR, may suffer significant absorption due to collisions with the soft radiation field. Therefore, the emerging $\gamma$-ray spectra from the source region can be strongly modified in respect to the injected $\gamma$-ray spectra in some specific radiation mechanism. Applying the above discussed models for the soft radiation field from the disk and around the jet, we calculate the $\gamma$-ray spectra which escaped from the source by assuming their simple modification according to the law: $F = F_{\rm 0}e^{-\tau_{\rm tot}}$, where $F$ and $F_0$ are the observed and intrinsic flux respectively, and $\tau_{\rm tot} = \tau_{\rm d} + \tau_{\rm BLR,l} + \tau_{\rm BLR,c}$ is the sum of the optical depths on different radiation fields. Such simple modification can be only considered in the case of complete isotropization of secondary $e^\pm$ pairs produced in $\gamma$-$\gamma$ absorption process. In this case, the next generation of $\gamma$-rays (which appear as a result of the Inverse Compton $e^\pm$ pair cascade process) is produced in different directions than the location of the observer. On the other hand $e^\pm$ pairs, produced by $\gamma$-rays propagating in other directions, can contribute to the $\gamma$-ray spectrum towards the observer. However, detailed calculations of 
$\gamma$-ray spectra, which also include such complicated cascade processes in the anisotropic radiation fields, are very difficult to perform realistically due to strong dependence of the final results on the unknown structure of the magnetic field around and inside the jet. We leave this very complicated problem for the future work.

Note, that since the optical depths for $\gamma$-rays depend on their energy, the spectrum of escaping $\gamma$-rays (in the case of simple absorption) can change significantly its spectral index. This is very important feature since in some cases the spectral index in specific energy range may become very flat, i.e. flatter than expected from any single radiation production process (see e.g. calculations by Bednarek~1997). Below we show the modified by absorption $\gamma$-ray spectra escaping from both considered here OVV quasars.

\subsection {3C~273}

The Optically violently variable (OVV) quasar 3C~273, at the redshift z=0.157, has been observed by the EGRET telescope on the board of the Compton GRO during several periods. A positive signal has been detected a few times (Lichti et al.~1995, von Montigny et al.~1997, Collmar et al.~2000). The $\gamma$-ray flux from 3C~273 shows strong variability during which the spectral index changes between  $3.2$ and $2.2$ (von Montigny et al.~1997). 
It has the form of flares characterized by the rise time of the order of  $\sim 2$ weeks and the fall time $\sim 1$ week (Collmar et al.~2000). The strongest outburst has been observed during 7 weeks monitoring in 1996-1997. Its spectrum in the high state can be described by a simple power law: 
$F=(3.0\pm1.7)\cdot 10^{-4} \cdot (E/\mathrm{MeV})^{-(2.39\pm0.13)}\mathrm{[ph\, cm^{-2}\, s^{-1}\, MeV^{-1}]}$ (Lichti et. al. 1995). At TeV energies only the upper limit has been reported up to now by the Whipple group (von Montigny et al.~1997). It lays clearly above the extrapolation of the EGRET spectrum. The multi-wavelength spectrum of 3C~273 (Lichti et al.~1995) shows two general peaks, first in the infrared (likely synchrotron origin) and the  second in the MeV energies. Additional strong peak at optical-UV range is usually interpreted as due to thermal emission from the accretion disk (e.g. Kriss et al.~1995). 

We calculate the $\gamma$-ray spectra expected from 3C~273 at TeV energies by including the absorption effects of primary injected $\gamma$-rays in the accretion disk and BLR radiation. It is assumed that the $\gamma$-ray spectrum, injected from the jet of 3C~273, extends to TeV energies with the spectral index and flux extrapolated from the EGRET energy range during the high state of activity. Note that, production of TeV $\gamma$-rays in OVV blazars has been postulated by some models (e.g. see Georganopoulos et al.~2006).
The $\gamma$-ray spectra emerging from the vicinity of the source are shown for different parameters of the considered models (see Fig.~\ref{fig6}). On the left Fig.~\ref{fig6}, we show the results for the simple Shakura-Sunyaev disk model. As expected, $\gamma$-rays injected close to the base of the jet are very strongly absorbed.  The observed superluminal motion in  3C~273 puts strong constraints on the viewing angle of the jet. However we consider a range of angles for the propagation of the $\gamma$-rays in respect to the surface of the accretion disk since as we note above the estimated inclination angle of the accretion disk in 3C 273 is quite large. Note that this situation might not be so surprising since the jet does not need to be perpendicular to the disk surface. Thus, the observation of the superluminal motion in this source does not exclude injection of $\gamma$-rays at large angles to the disk axis. 

In general, absorption effects strongly depend on the injection distance from the accretion disk and the injection angles $\alpha$. In the case of gamma-rays injected at the larger distances (see the case for $x_\gamma = 300r_{\rm in}$)
the shape of the spectrum depends strongly on the propagation angle $\alpha$. 
Only for small angles $\alpha$, $\gamma$-ray spectra clearly extends through the TeV energy range with only moderate absorption (see the case for $\alpha=0$ and $x_\gamma = 300r_{\rm in}$). Absorption of $\gamma$-rays in the BLR radiation produces clear flattening of the spectrum around $\sim 1$ TeV. Due to this feature the emerging spectrum is actually harder in the TeV energies in respect to the injected spectrum (see dashed thick curve in Fig.~\ref{fig6}).

The emerging $\gamma$-ray spectra calculated in the case of model II and III for the disk radiation field are shown in the middle and left Fig.~\ref{fig6}, respectively. 
The biggest difference in optical depths with and without corona occur at multi-GeV energies. Because of the sharp exponential cut-off in Planck spectrum emitted by the accretion disk, there is not enough UV photons for efficient absorption of $\gamma$-rays. 
In the case of additional power-law radiation from the accretion disk corona,  the number of those photons is much greater, so the optical depths in the case of the model with additional  disk corona are much larger comparing to the optical depths in the case of  only Shakura-Sunyaev disk radiation.
Nevertheless absolute values of optical depths are still quite small at those energies (compare with Fig.~\ref{fig3}). So then, differences in resulting spectra with and without corona are visible only at sub-TeV energies. 

The model with very hot disk (the temperature at the inner radius a factor of 3 higher) produces very strong absorption of the $\gamma$-ray spectra injected at distances $x_\gamma < 300 r_{\rm in}$ at energies above $\sim 30$ GeV.
Note that in this model, $r_{\rm in}$ is 4.3 times smaller than in the model I and II,. 
Therefore, not only the radiation field is much stronger, but also $\gamma$-rays are injected significantly closer to the central engine.

\subsection{3C~279}

The OVV blazar 3C~279, at the redshift z=0.538, has been observed by the EGRET telescope simultaneously with 3C~273 since both objects are at the angular distance of only a few degrees. A strong flare with a very flat spectrum (differential spectral index $1.89$) has been detected from 3C~279 already in June 1991 (Hartman et al.~1992). The flux increased during $\sim 1$ week and declined during 2 days (Kniffen et al.~1993). Even stronger $\gamma$-ray flare has been observed by the EGRET in 1996 with a similar time structure and spectral index (Wehrle et al.~1998). In spite of much larger distance than to 3C~273, 3C~279 has been positively detected by the EGRET telescope in all observation periods (Hartman et al.~2001a), showing a variety of flux and spectral index stages. The shortest variability time scales observed in these flares were below $\sim 1$ day (Wehrle et al.~1998, Hartman 2001b). 

\begin{figure*}
\includegraphics[width=0.325\textwidth, height=0.325\textwidth, trim=  0 34 46 49,clip]{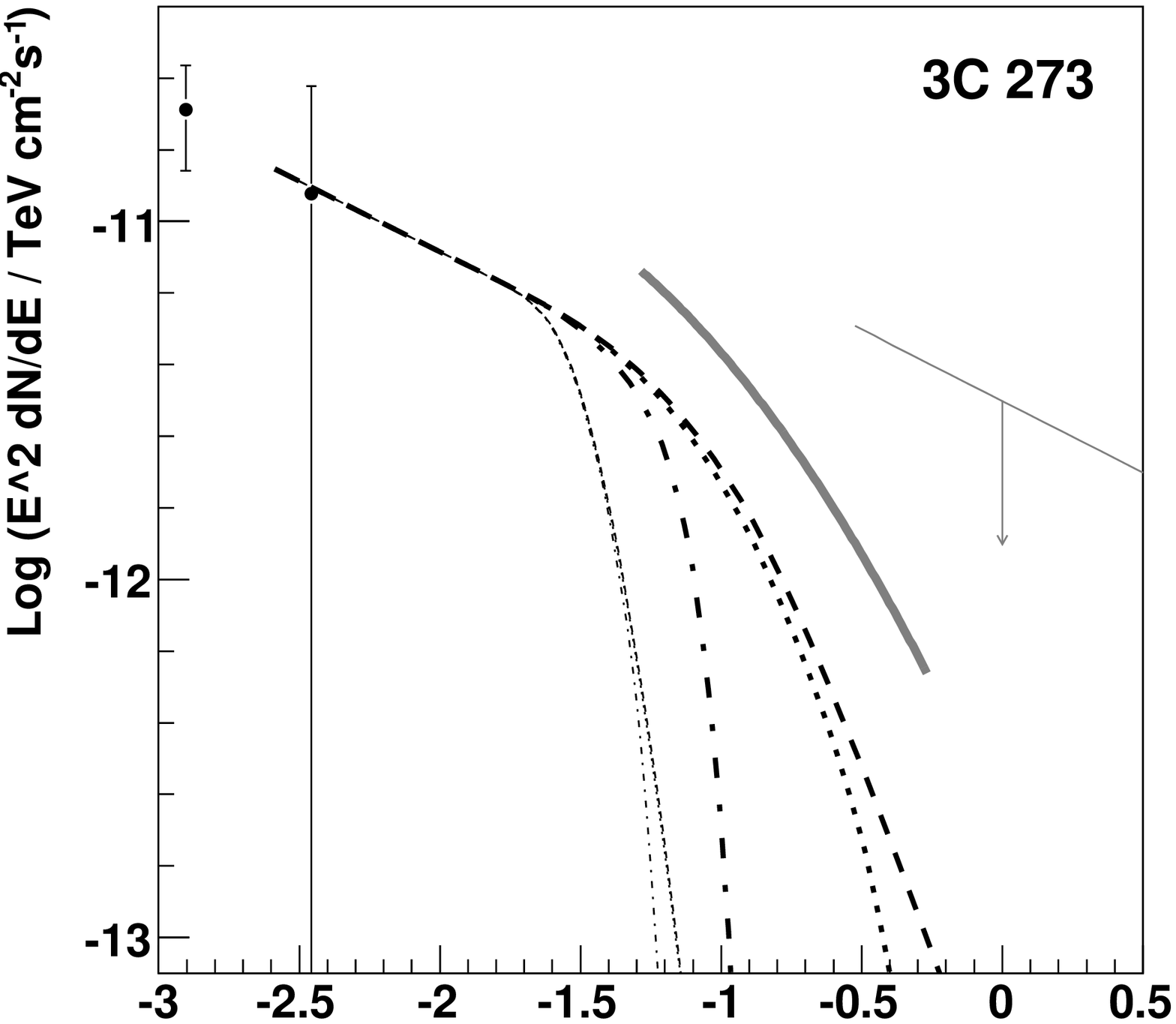}
\includegraphics[width=0.325\textwidth, height=0.325\textwidth, trim= 27 34 46 49,clip]{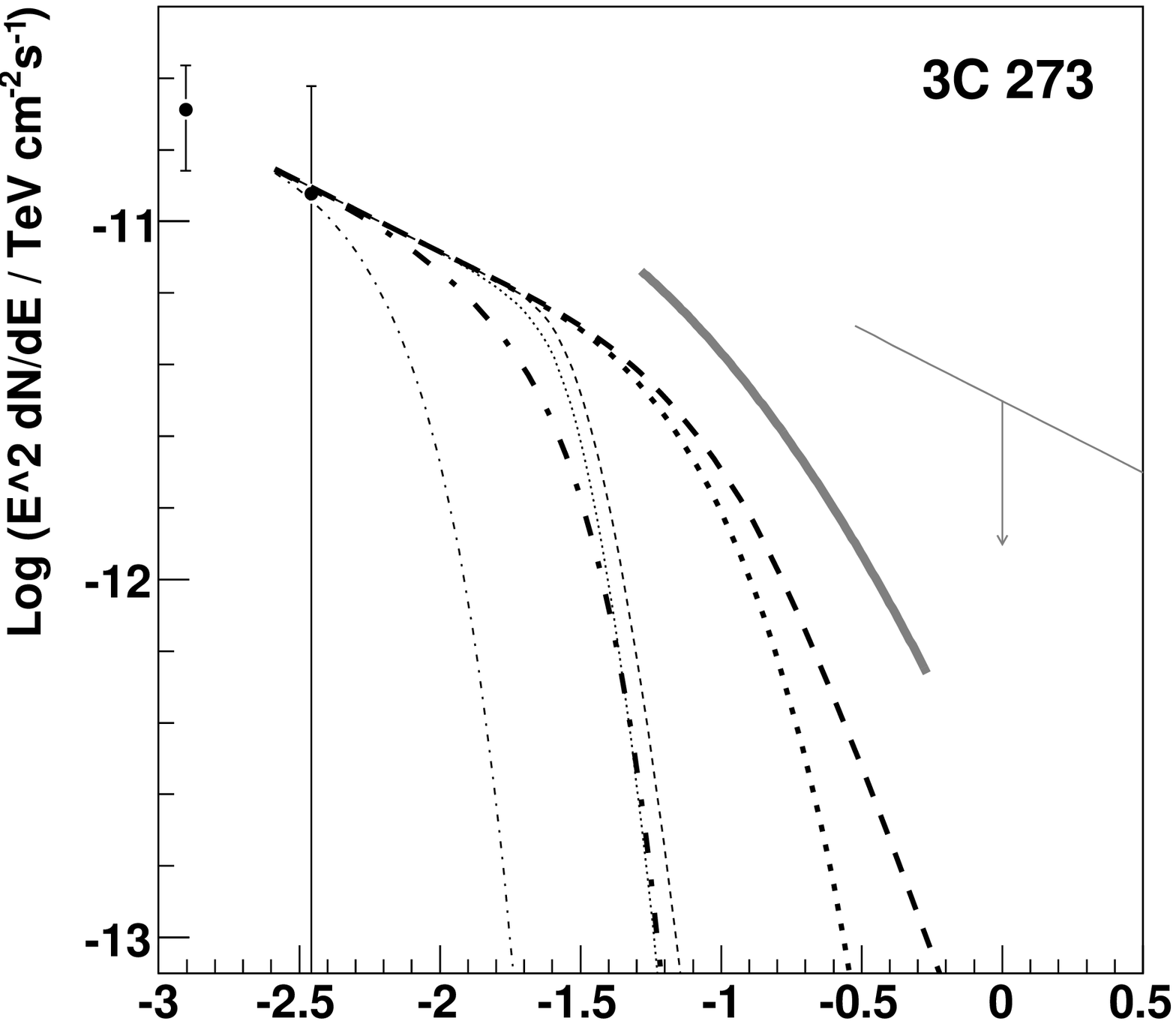}
\includegraphics[width=0.325\textwidth, height=0.325\textwidth, trim= 27 34 46 49,clip]{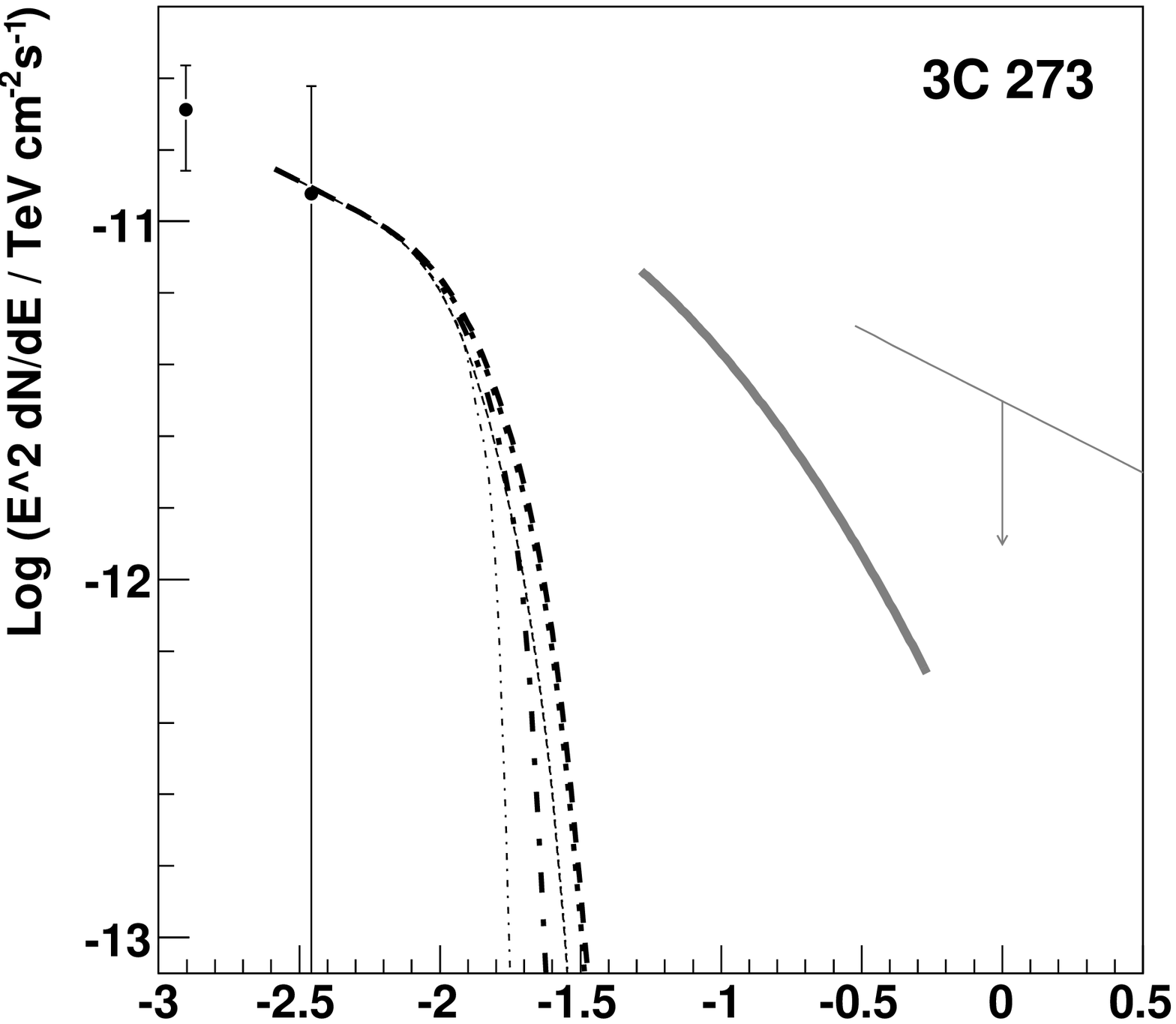}
\includegraphics[width=0.325\textwidth, height=0.325\textwidth, trim=  0  6 46 49,clip]{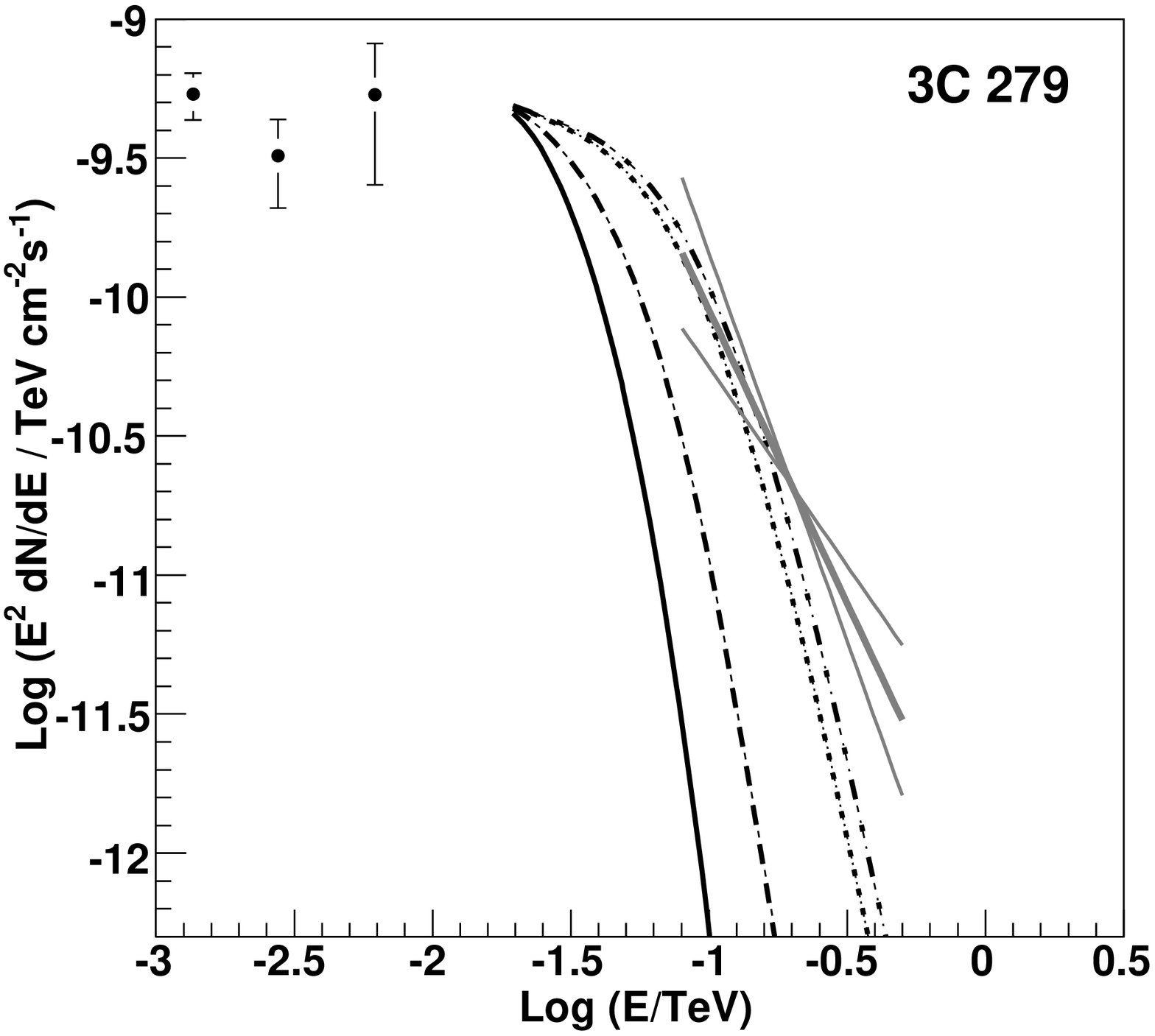}
\includegraphics[width=0.325\textwidth, height=0.325\textwidth, trim= 27  6 46 49,clip]{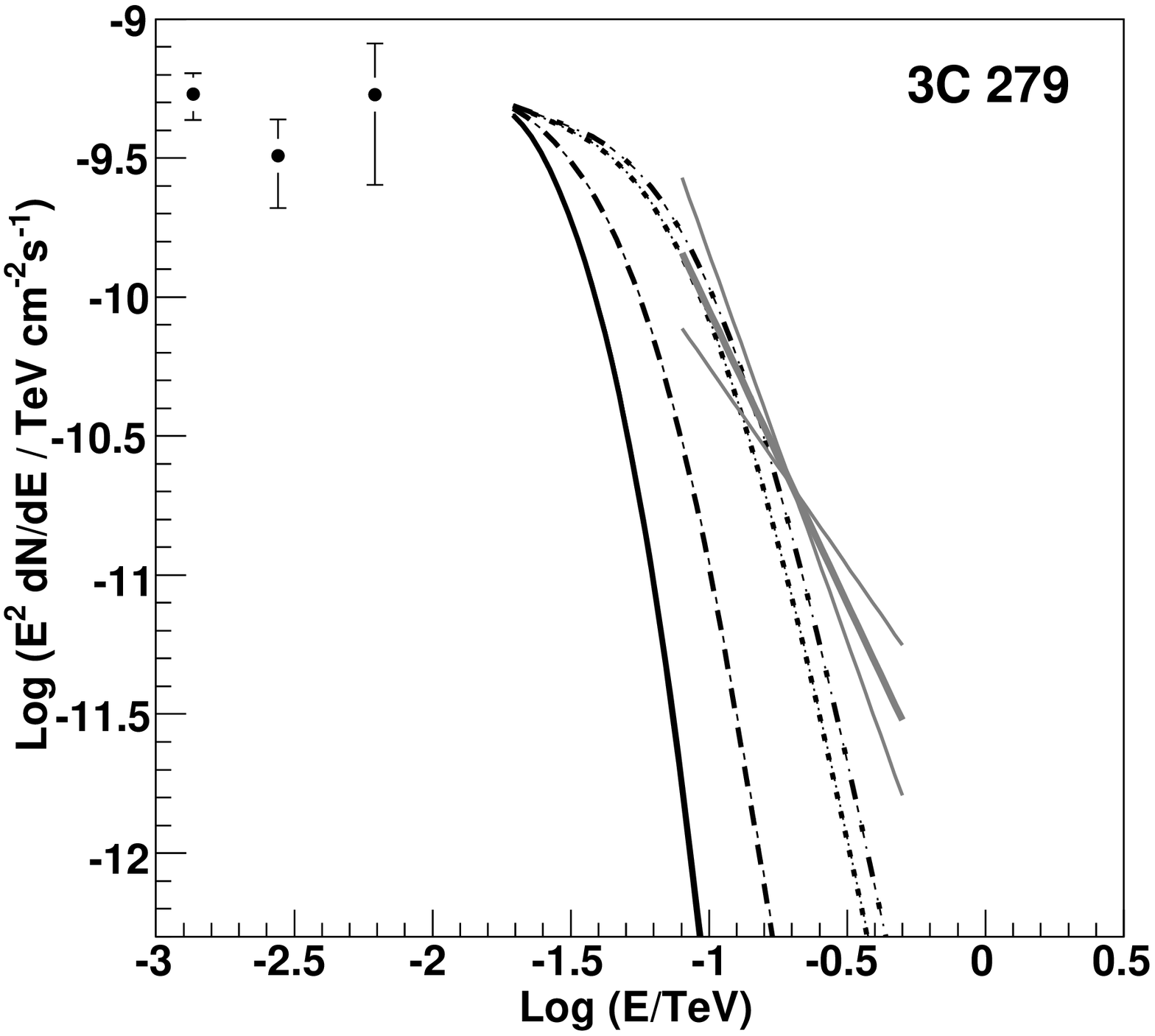}
\includegraphics[width=0.325\textwidth, height=0.325\textwidth, trim= 27  6 46 49,clip]{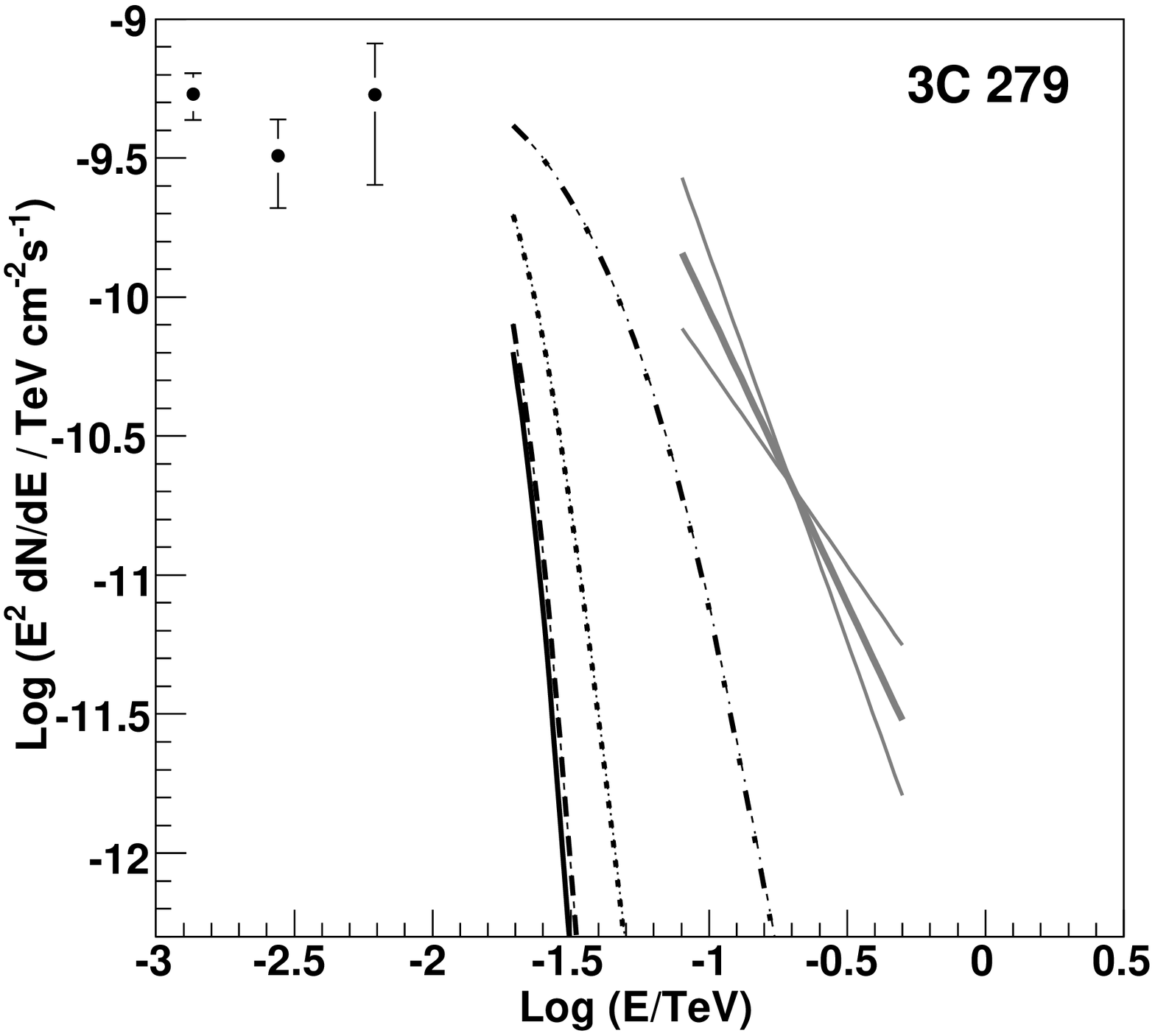}
\caption{$\gamma$-ray spectra at the observer from 3C~273 (upper panel) and 3C~279 (lower panel), modified by absorption in the intergalactic space on the EBL and by internal absorption in the disk and BLR radiation for different models of radiation produced in the accretion disk (as discussed in Fig.~\ref{fig5}). The absorption in the EBL is included according to Stecker et al.~(2006) model. The specific results marked by different curve styles correspond to the $\gamma$-ray spectra at the source calculated in Fig.~\ref{fig5}. The thin grey line (with an arrow) in figures for 3C 273 is the limit obtained by the Whipple Observatory (see von Montigny et al.~1997), and the thick grey curve is the MAGIC telescope sensitivity. The spectrum detected by the MAGIC telescope form 3C~279 is marked 
by the thick grey line with the error boxes marked by the thin lines.}
\label{fig7}
\end{figure*}

As in the case of 3C~273, we assume that the injection spectrum of 3C~279 extends from the GeV energy range through the TeV energy range with similar spectral index. As an example, the EGRET $\gamma$-ray spectrum observed during the strong flares have been considered (i.e. spectral index close to 2). 
In the case of the flare in 1996, the differential $\gamma$-ray flux can be fitted with the simple power law, $4.3\times 10^{-10} (E/\mathrm{\rm TeV})^{-2.02} \mathrm{\rm TeV\, cm^{-2}\, s^{-1}}$ (Hartman et al. (2001). We calculate the $\gamma$-ray spectra emerging from the source by including absorption of $\gamma$-rays in the disk and BLR radiations and by applying the above described models for these radiation fields (see Section~2). However, in the case of 3C~279, there is no information on the inclination angle of the accretion disk to the observer's line of sight. We limit our calculations only to small values of the angle $\alpha$. The $\gamma$-ray spectra modified by absorption are shown for different injection distances of primary $\gamma$-rays from the accretion disk (see Fig.~\ref{fig6}). It is clear that emerging $\gamma$-ray spectra can be very strongly modified provided that the injection distances are $x_\gamma < 100 r_{\rm in}$. The dependence on the injection angle below $\sim 10^{\rm o}$ is small.   

Due to generally weaker disk radiation in 3C 279 (in respect to 3C 273), a significant part of the TeV $\gamma$-rays is able to emerge from distances of the order of a few hundred inner disk radii
even in the case of the model III in which the inner disk temperature is 3 times larger. However, the $\gamma$-ray spectrum emerging in this last case is strongly modified. Strong absorption dip is seen at $\sim 300$ GeV with a very flat spectrum above this energy.

\section{Gamma-ray spectra after propagation in the EBL} 

\begin{figure*}
\includegraphics[width=0.325\textwidth, height=0.325\textwidth, trim=  0 34 46 49,clip]{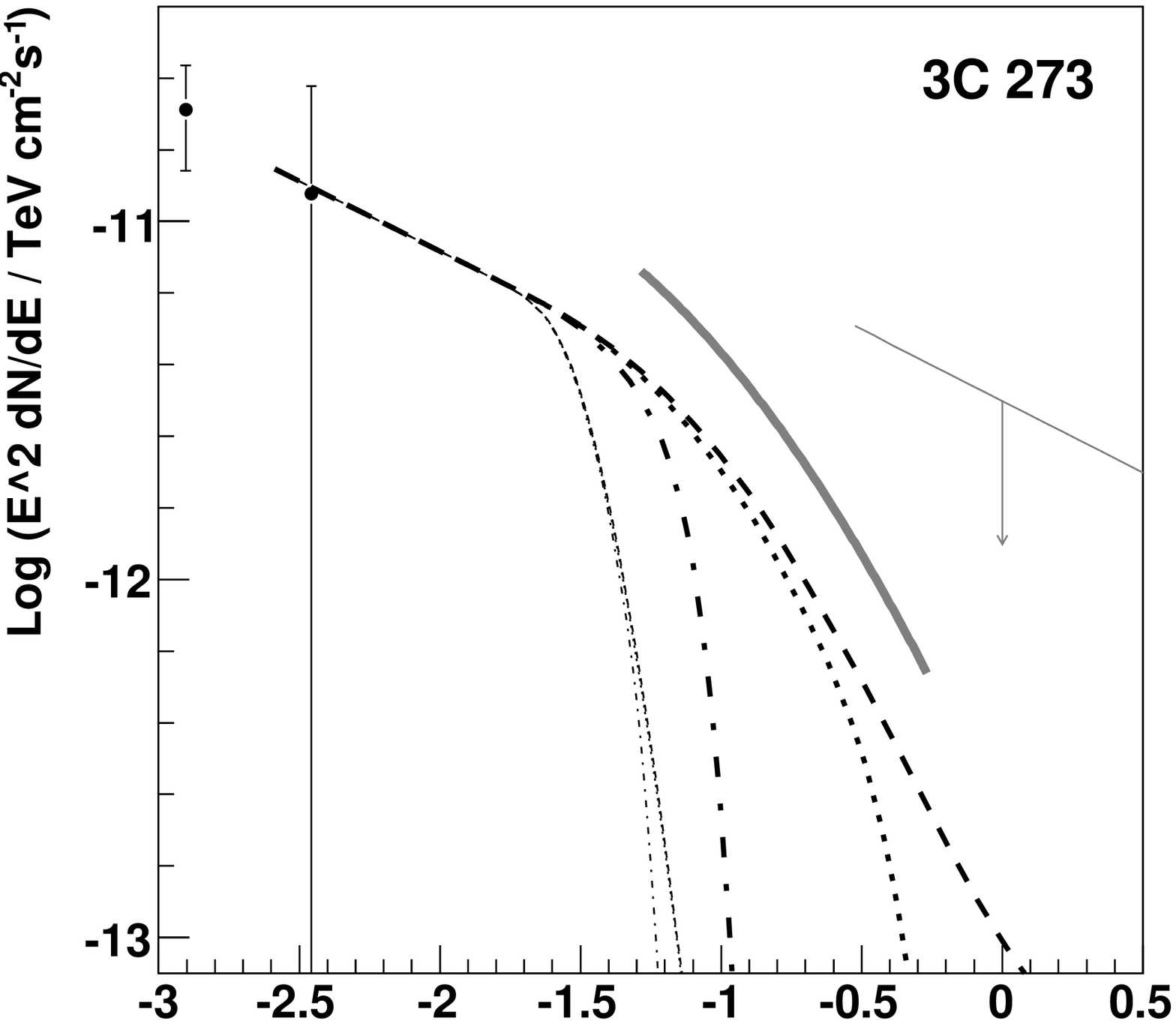}
\includegraphics[width=0.325\textwidth, height=0.325\textwidth, trim= 27 34 46 49,clip]{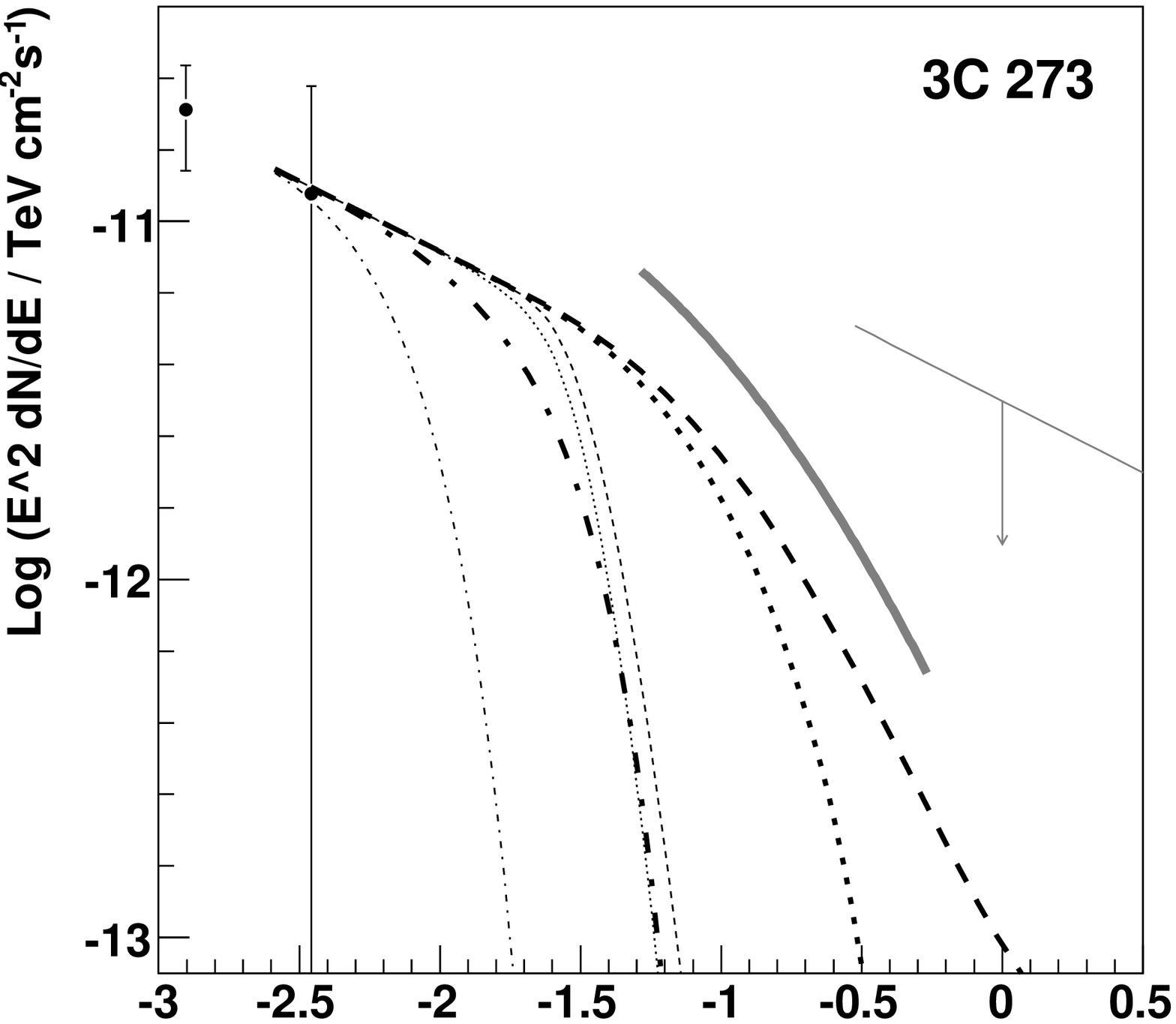}
\includegraphics[width=0.325\textwidth, height=0.325\textwidth, trim= 27 34 46 49,clip]{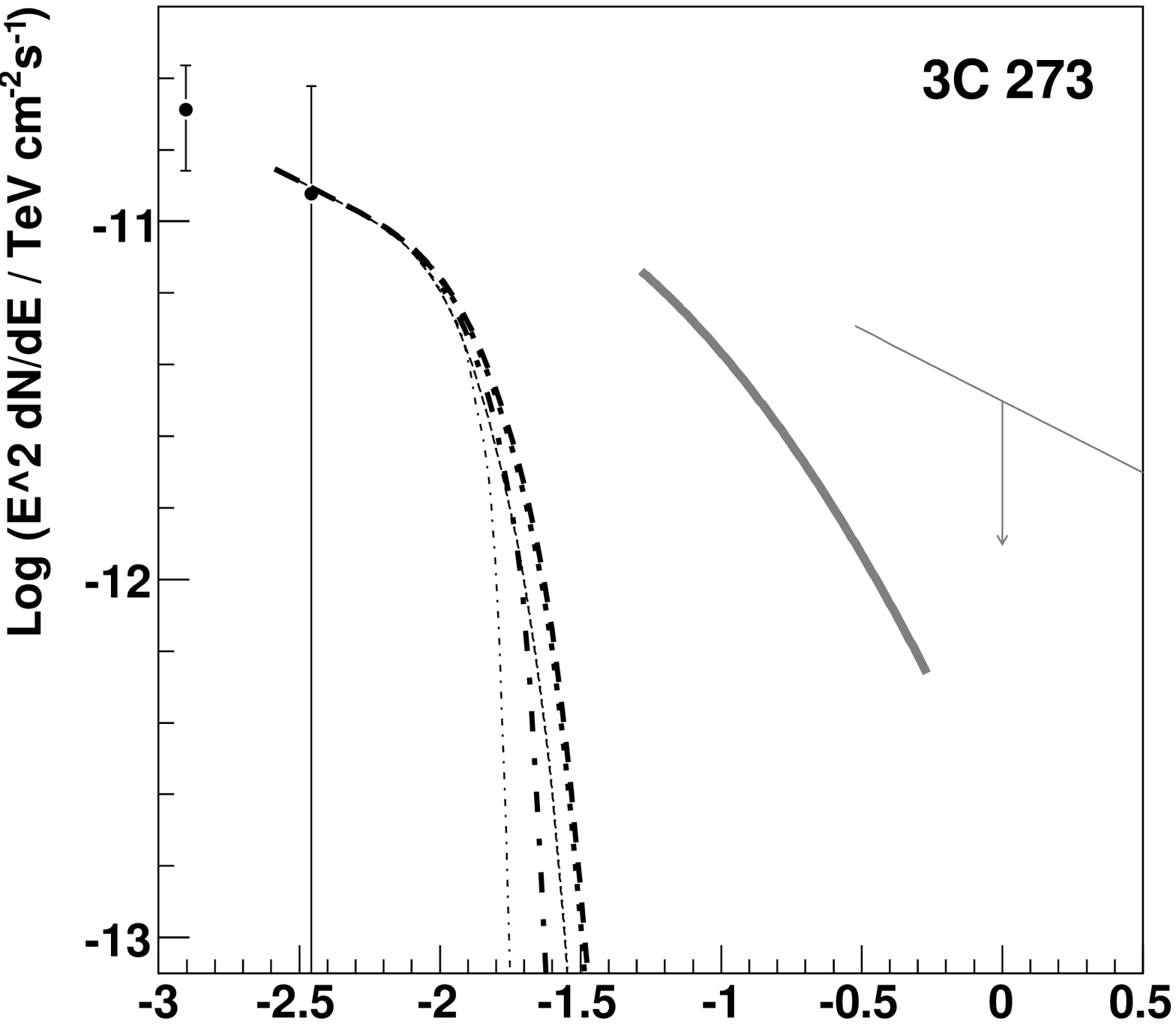}
\includegraphics[width=0.325\textwidth, height=0.325\textwidth, trim=  0  6 46 49,clip]{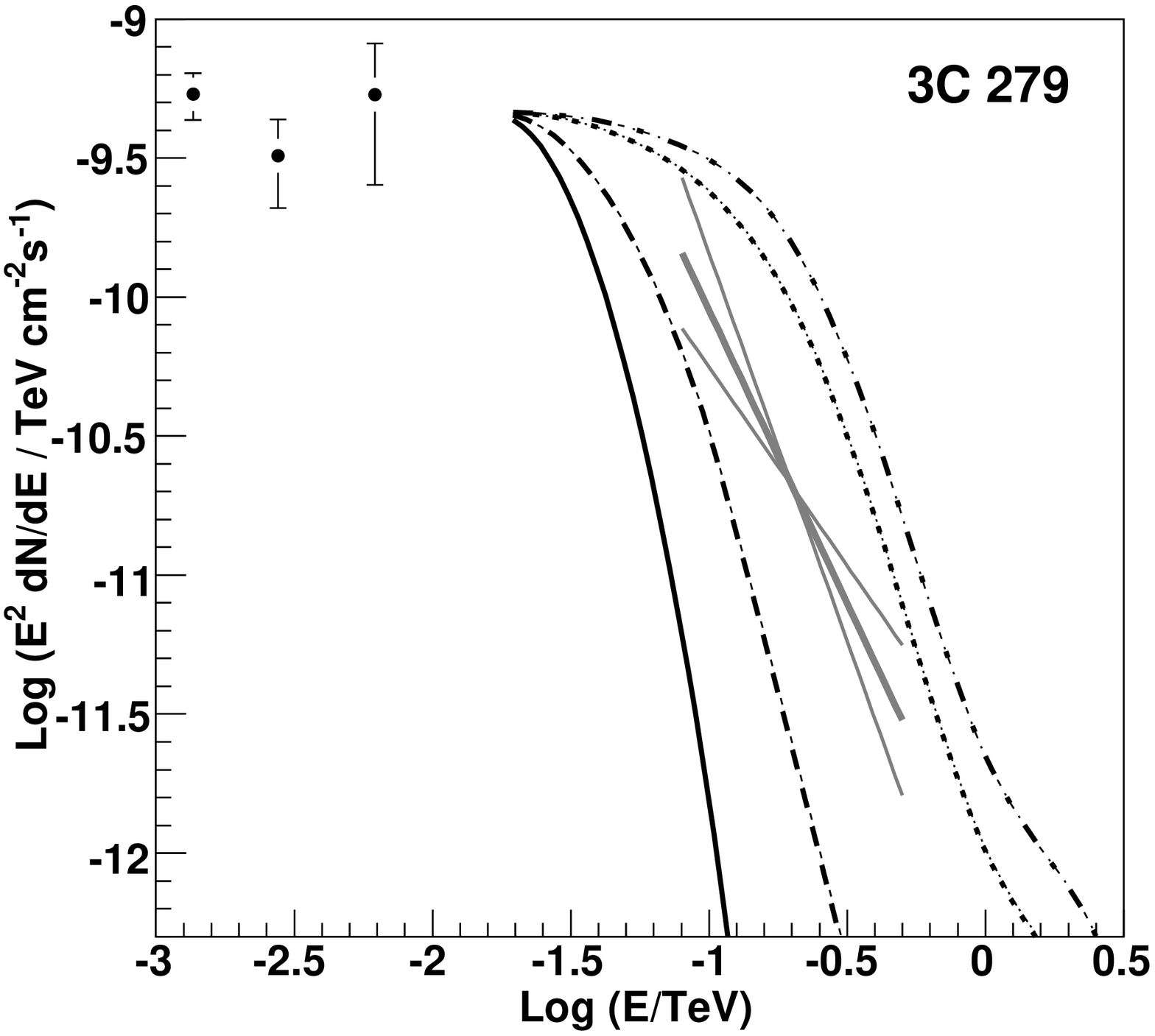}
\includegraphics[width=0.325\textwidth, height=0.325\textwidth, trim= 27  6 46 49,clip]{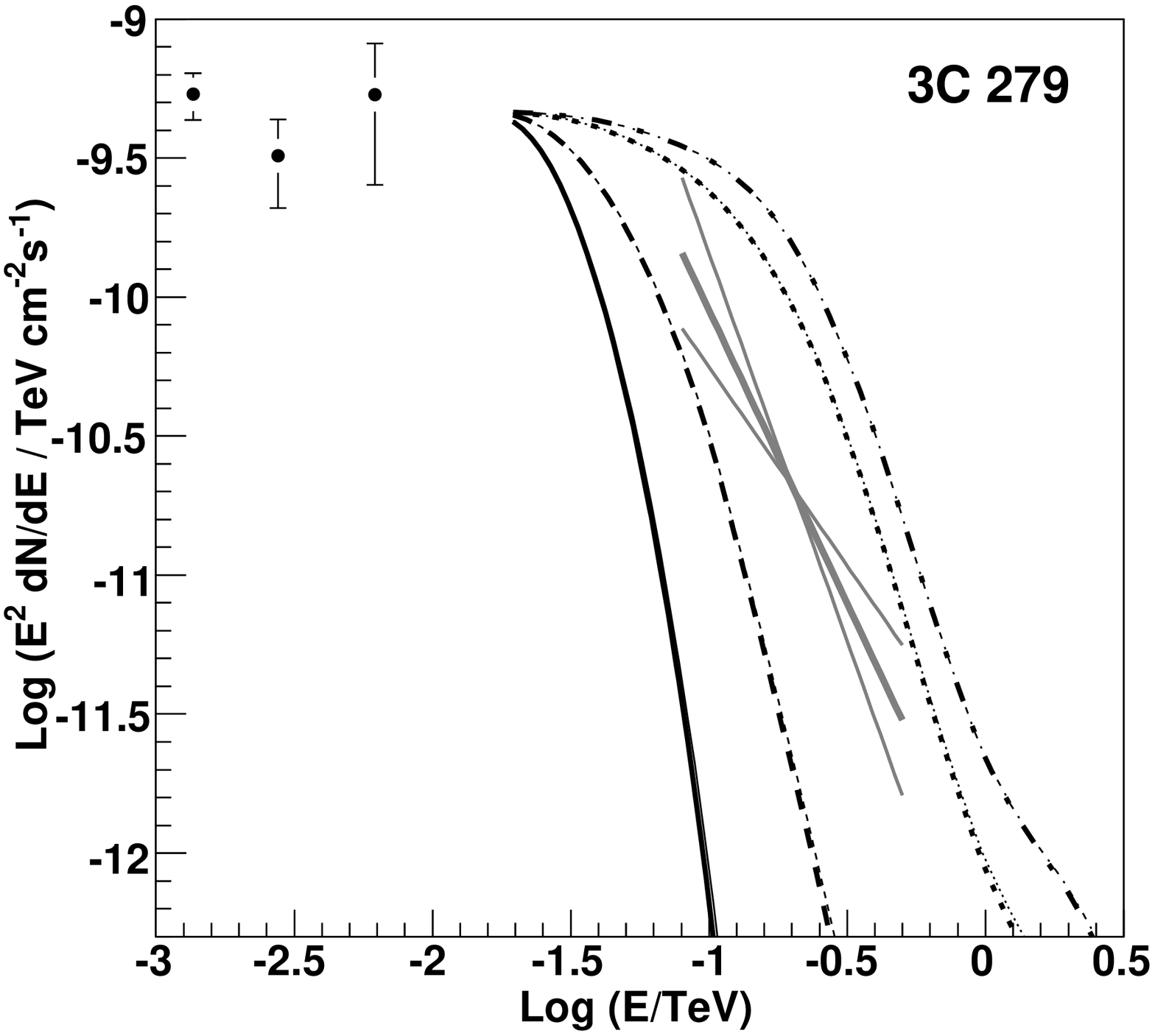}
\includegraphics[width=0.325\textwidth, height=0.325\textwidth, trim= 27  6 46 49,clip]{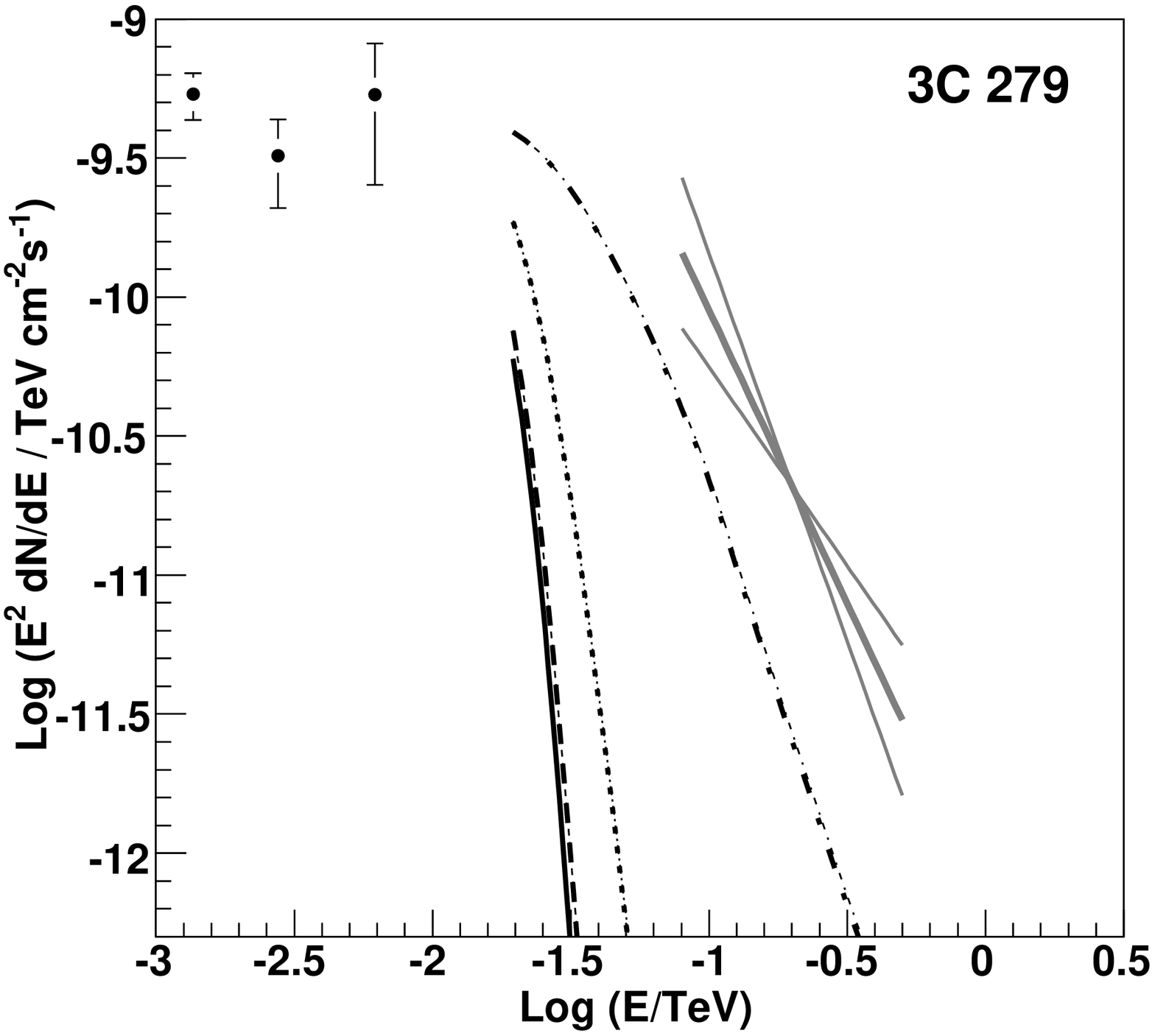}
\caption{As in Fig.~\ref{fig7}, but for the Primack et al. (2005) model for the EBL.}
\label{fig8}
\end{figure*}

In the case of both discussed quasars, the absorption of TeV $\gamma$-rays during their propagation through the intergalactic space have to be taken into account. However the radiation field filling the intergalactic space (EBL) is not precisely known. Two limiting estimates are usually considered in the literature. In the model by Stecker et al. (2006), the EBL is relatively strong and the absorption effects on the TeV $\gamma$-ray spectra even from close sources are strong.
In this model the EBL spectra are calculated from so called ``backward-evolution'' scheme.
In this approach luminosities of galaxies at higher redshifts are predicted using present galaxy SEDs, luminosity functions and redshift evolution of luminosity function.
Spectral energy distribution of galaxy is deduced from it relation to luminosity in one IR band. The luminosity function of galaxies used in this model is analytic fit of smoothed broken power law to the observational data.  
Redshift evolution of galaxies adopted in this model is a pure luminosity evolution, which is more important that galaxies number density evolution. 
There are two posibilities of redshift evolution of galaxies consider in Stecker model: so called baseline and fast evolution models. The former one predicts slightly lower values of optical depths then the latter one. In this work we use baseline Stecker model.
Stecker EBL model requires very flat injection spectra of TeV $\gamma$-rays even from the observed BL Lacs. Such production spectra are not widely accepted. 
However, they are a viable alternative (Katarzy\'nski et al.~2006; Stecker, Baring \& 
Summerlin~2007).
Although, in general they are possible in the case of strong internal absorption as postulated by e.g. some cascade models for $\gamma$-ray production in BL Lacs (see e.g. Bednarek~1997).
In the Primack et al.~(2005) model, the absorption of TeV $\gamma$-rays from the observed BL Lacs is rather moderate. This model follows the ``forward-evolution'' approach. 
The emission of galaxies is predicted from general theory of cosmology and galaxy formation. 
Note however, that this model is not viable lower limit to the EBL because it lies below the solid observational lower limits from galaxy counts at infrared wavelengths as shown in Albert et al.~2008).
In the case of the second model, for quasars at the distance of 3C~279, the $\gamma$-ray absorption becomes sever at energies above a few hundred GeV. Below, we show how the $\gamma$-ray spectra are modified in the case of both EBL models for the injection spectra discussed above in which we also take the internal absorption at the source into account (as shown in example Fig.~\ref{fig6}).

\subsection{$\gamma$-rays injected at specific places in the jet}

The $\gamma$-ray spectra, calculated for specific parameters of our models (considered in Fig.~\ref{fig6}), are shown in Figs.~\ref{fig7} and~\ref{fig8}, after including the effects of propagation in the EBL estimated according to the Stecker and Primack models. 
Let's us first consider the spectra obtained with the Stecker model.
The $\gamma$-ray emission above $\sim 0.4$ TeV (in the case of 3C~273) and above $\sim 1$ TeV
(in the case of 3C~279) are severely attenuated in all considered models for the soft radiation field of the accretion disk and BLR. Note that the spectra can extend to higher energies in the case of 3C~279 which is due to significantly flatter injection spectrum of $\gamma$-rays inside the jet (spectral index $\sim 2$ in 3C~279 in respect to $\sim 2.4$ in 3C~273). In Fig.~\ref{fig7} we also show the measured $\gamma$-ray spectrum from 3C~279 by the MAGIC telescope (Albert et al.~2008) and the available upper limit on the spectrum from 3C~273 by the Whipple telescope (von Montigny et al.~1997). The observed spectrum from 3C~279 can be only consistent with the calculated spectra provided that the maximum rate of the injection of primary $\gamma$-rays occurs at distances significantly larger than $x_\gamma\sim 100r_{\rm in}$ from the base of the jet. However, in the case of the Primack EBL model the maximum of the flare might occur already at distances as close as a few tens of $r_{\rm in}$. The injection places of primary $\gamma$-rays have to be located at much larger distances from the base of the jet in the case of the radiation model with 3 times larger temperature at the inner radius. 

The $\gamma$-ray spectra expected from 3C 273 steepens significantly above a few hundred GeV even for larger distances from the base of the jet. For the considered parameter range, they are also below the present sensitivity of the MAGIC I telescope (differential $5\sigma$ in 50 hrs sensitivity re-calculated for the spectrum of $\gamma$-rays with the index 2.6). Therefore we conclude that detection of 3C 273 by the present Cherenkov telescopes is unlikely if the injection region is located within the jet at distances considered on these figures. However, such emission should be within the sensitivity limits of the planned 
Cherenkov Telescope Array (CTA) whose sensi\-ti\-vi\-ty will be an order of magnitude better
(Hermann~2007).

\subsection{Model for injection of primary $\gamma$-rays} 

The strong $\gamma$-ray flares observed from these two OVV quasars in the GeV energy range  develop typically on a time scale of a few days to two weeks (the rise time, $\tau_{\rm r}$) and fall on a time scale of a day up to a few days (the fall time, $\tau_{\rm f}$) in the case of 3C 279 and 3C 273, respectively. These time scales can be related to the distances at which production of $\gamma$-rays occurs inside the jet. However, these distance scales are also related to the Doppler factors ($D$) of the emission regions and at a smaller extent on the distance to the quasar. The characteristic distance scales on which the flares occur can be estimated from,
\begin{eqnarray}
L\cong D^2c(\tau_{\rm r} + \tau_{\rm f})/(1 + z),
\label{eq2}
\end{eqnarray}
\noindent
where $c$ is the velocity of light, and $z$ is the redshift of the source. For mentioned above durations of the observed $\gamma$-ray flares and estimated Doppler factors 
(typically of the order of $\sim 10$), $\gamma$-ray injection has to occur along quite a large region extending along the jet if compared to the characteristic dimension of the source scaled by the inner radius of the accretion disk. The absorption of the TeV $\gamma$-rays throughout such extended region can change drastically as we have shown in the previous Section~(5.1). Here, we calculate the emerging spectra in the TeV energy range by taking into account the absorption effects of primary $\gamma$-rays in the case of specific model 
for the development of the flare inside the jet.

Let us consider a model for the injection of $\gamma$-rays in which emission region in the jet (the blob) is very compact, i.e. its dimension is much smaller than the characteristic distance scale along the jet on which injection of $\gamma$-rays occurs. 
As we noted above, the shortest variability detected from 3C~279 are below $\tau_{\rm s}\sim 1$ day. This allows us to put the upper limit on the dimension of the blob,
\begin{eqnarray}
R_{\rm b} < Dc\tau_{\rm s}/(1 + z).
\label{eq3}
\end{eqnarray}
\noindent  
The blob  moves along the jet with the Doppler factor which is independent on time. 
We assume that $\gamma$-rays are injected from the blob with the rate corresponding to the observed $\gamma$-ray light curve measured by the EGRET telescope, i.e. with the initial, slow increase and the final fast decrease. The basic assumption of our calculations is that the injection spectra can be extrapolated from the EGRET energy range without a break and extend up to at least $\sim 1$ TeV. Note, that our assumption is less rigorous than recent
interpretation of the TeV $\gamma$-ray observations of some BL Lacs in which  the upper limits on the EBL are derived based on the assumption on the spectral indexes of $\gamma$-ray spectra at the source (spectral indexes $\propto E^{-1.5}$ or even $\propto E^{-2/3}$, Katarzy\'nski et al.~2006 ). 
We assume that GeV $\gamma$-rays are able to escape freely from the central region without absorption. Therefore, our hypothesis of the common origin of the GeV and TeV emission in blazars allows us to postulate that also the rate with which the TeV $\gamma$-rays 
are injected has similar time structure as observed by the EGRET telescope. 

Below, we investigate the effects of the spectral changes due to the internal absorption at the source in the above described model for injection of primary $\gamma$-rays in the case of both sources: 3C~273 and 3C~279. We calculate the expected spectra at the observer, i.e. after propagation in the EBL, in terms of both considered EBL models and next compare them with the observations of the TeV $\gamma$-rays from these two objects and/or with the sensitivities of the present and future Cherenkov telescopes.

\begin{figure*}
\includegraphics[width=0.325\textwidth, height=0.325\textwidth, trim=  0 34 46 49,clip]{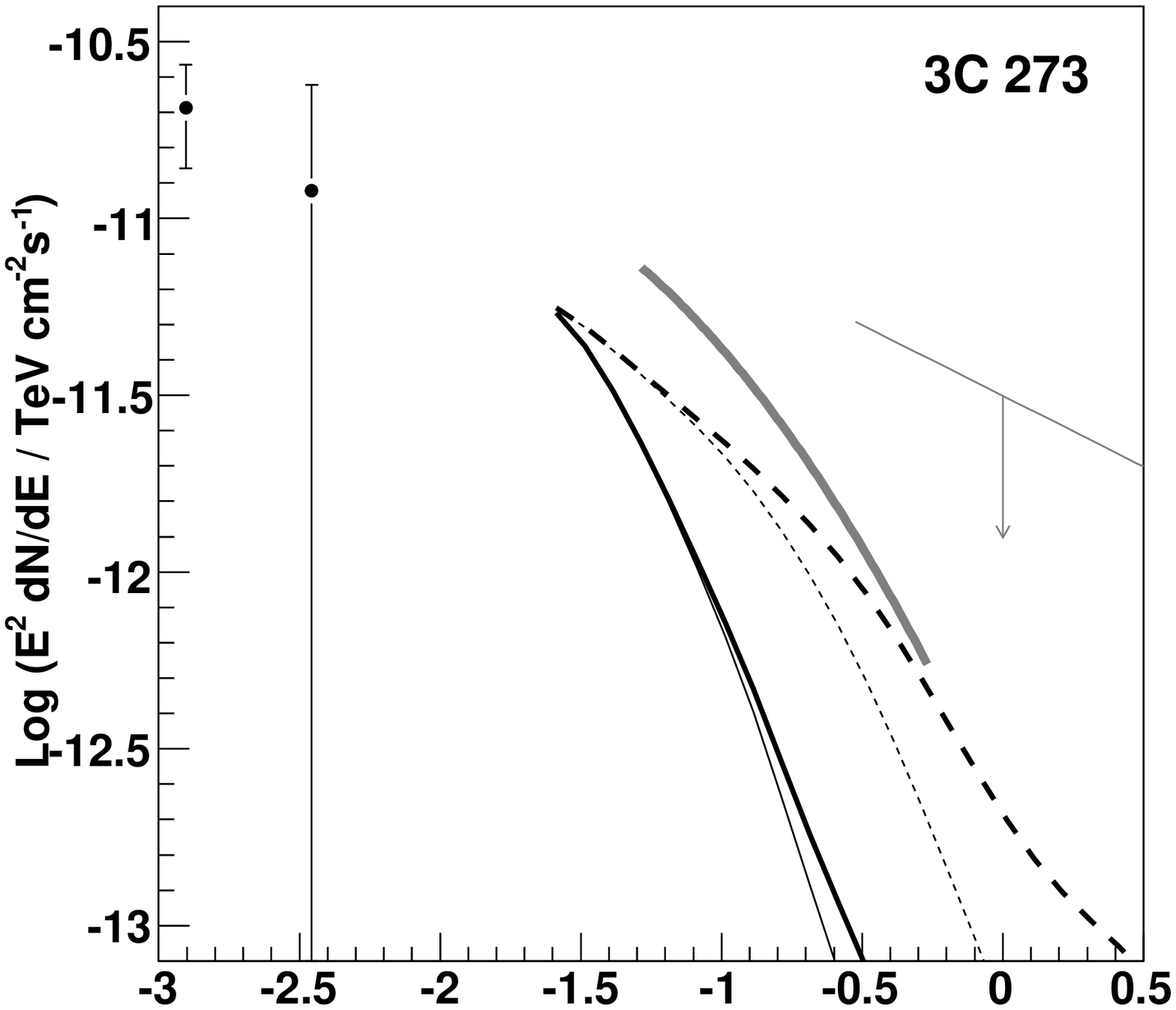}
\includegraphics[width=0.325\textwidth, height=0.325\textwidth, trim= 26 34 46 49,clip]{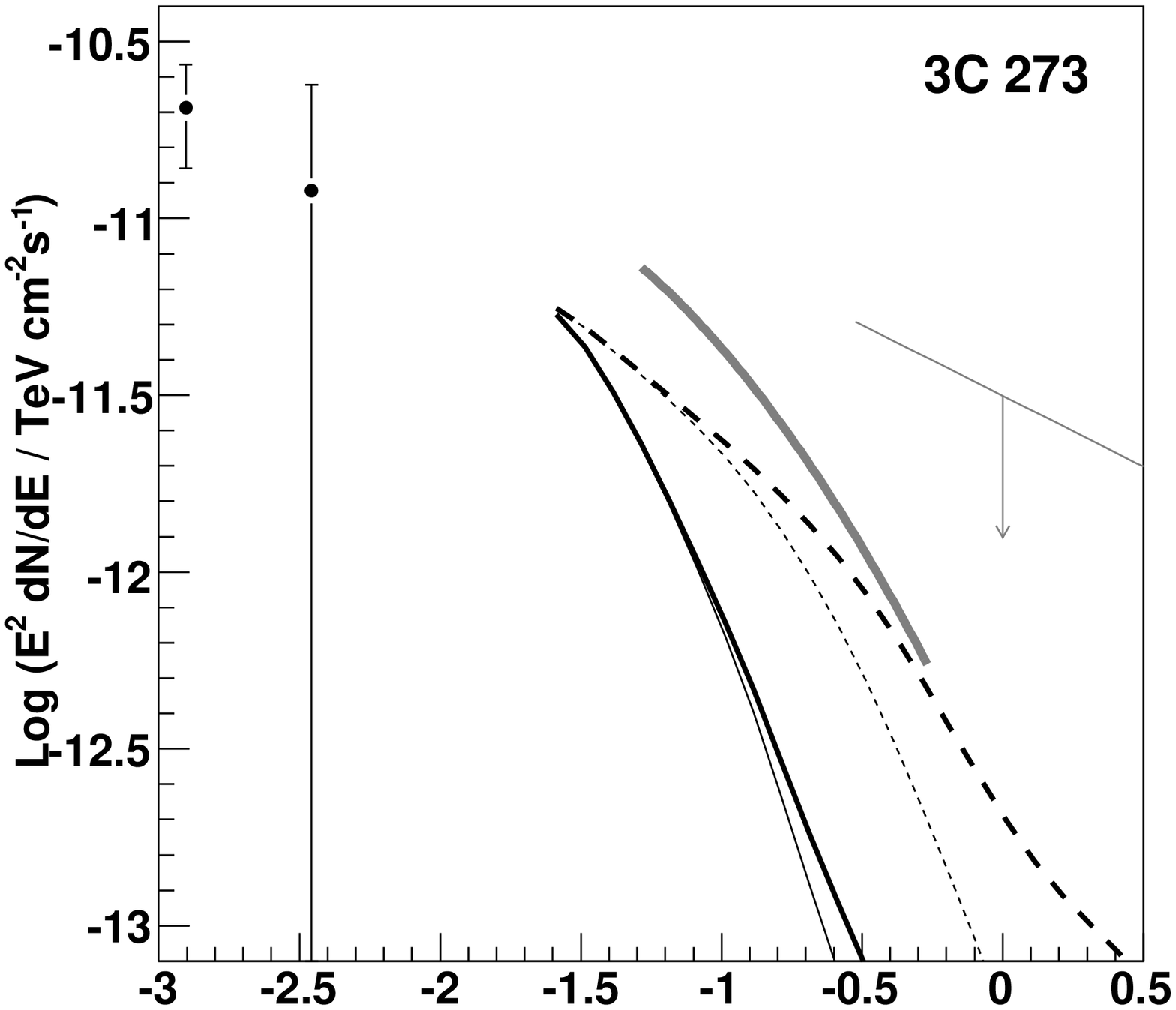}
\includegraphics[width=0.325\textwidth, height=0.325\textwidth, trim= 27 34 46 49,clip]{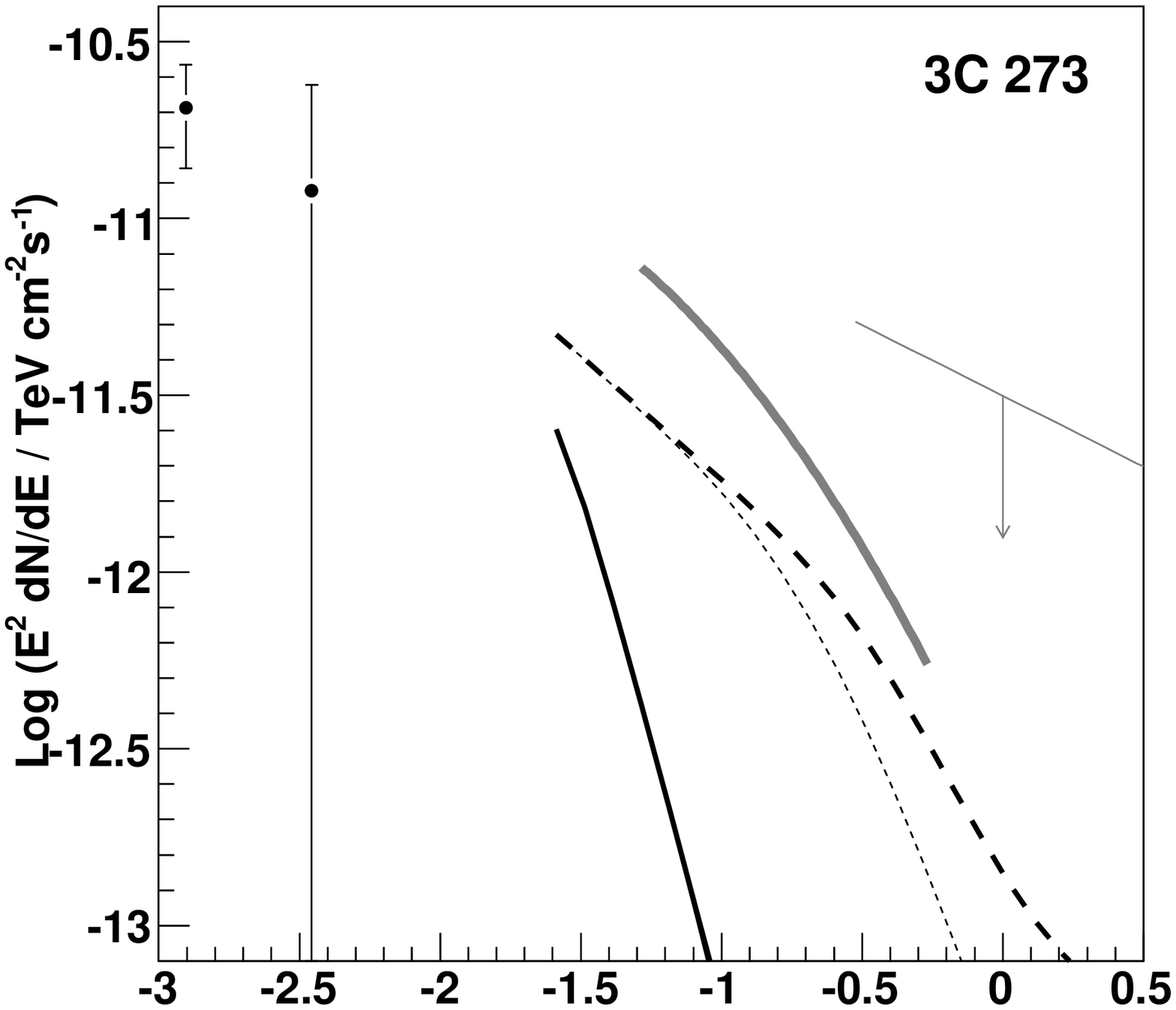}
\includegraphics[width=0.325\textwidth, height=0.325\textwidth, trim=  0  6 46 49,clip]{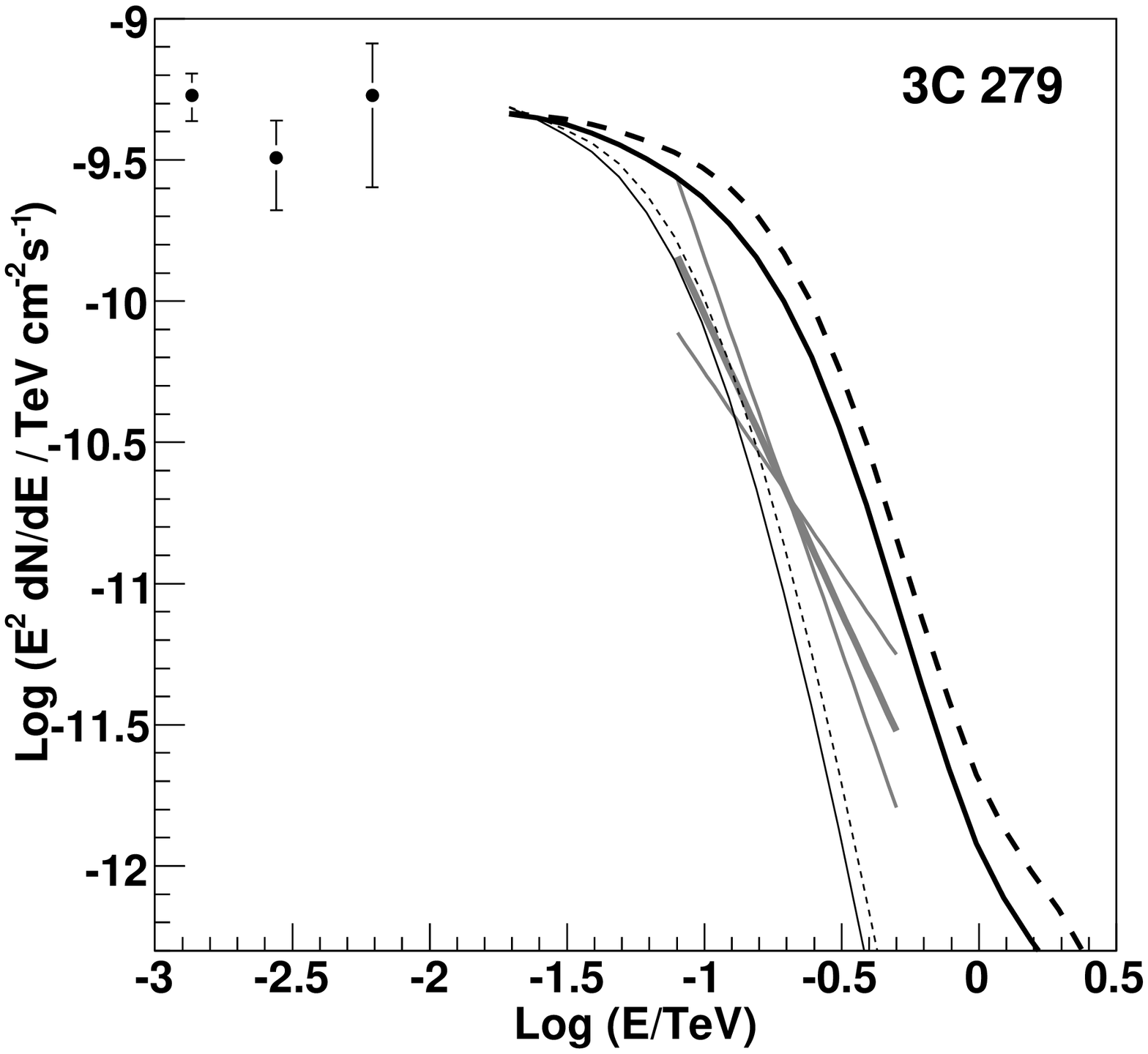}
\includegraphics[width=0.325\textwidth, height=0.325\textwidth, trim= 26  6 46 49,clip]{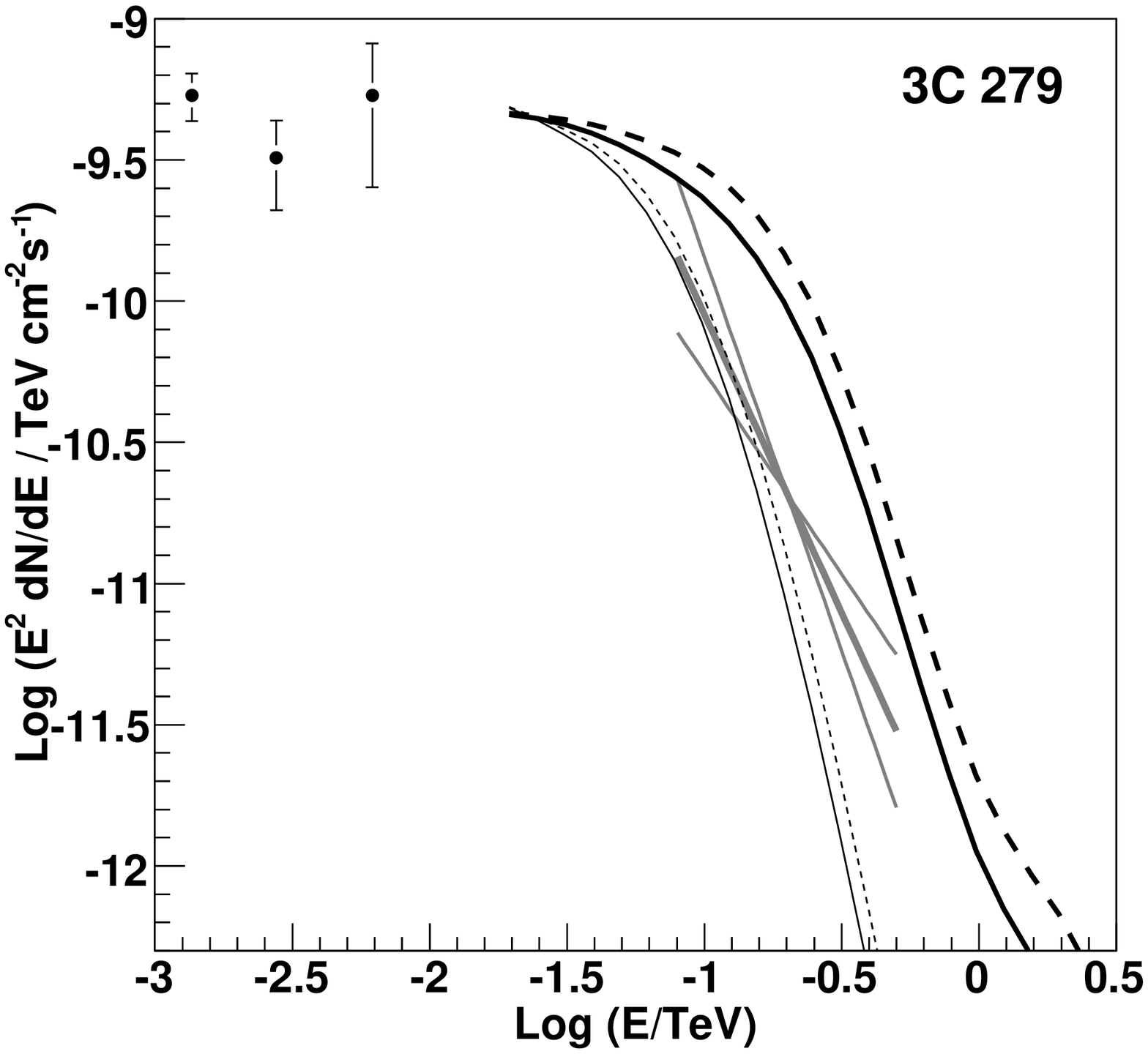}
\includegraphics[width=0.325\textwidth, height=0.325\textwidth, trim= 27  6 46 49,clip]{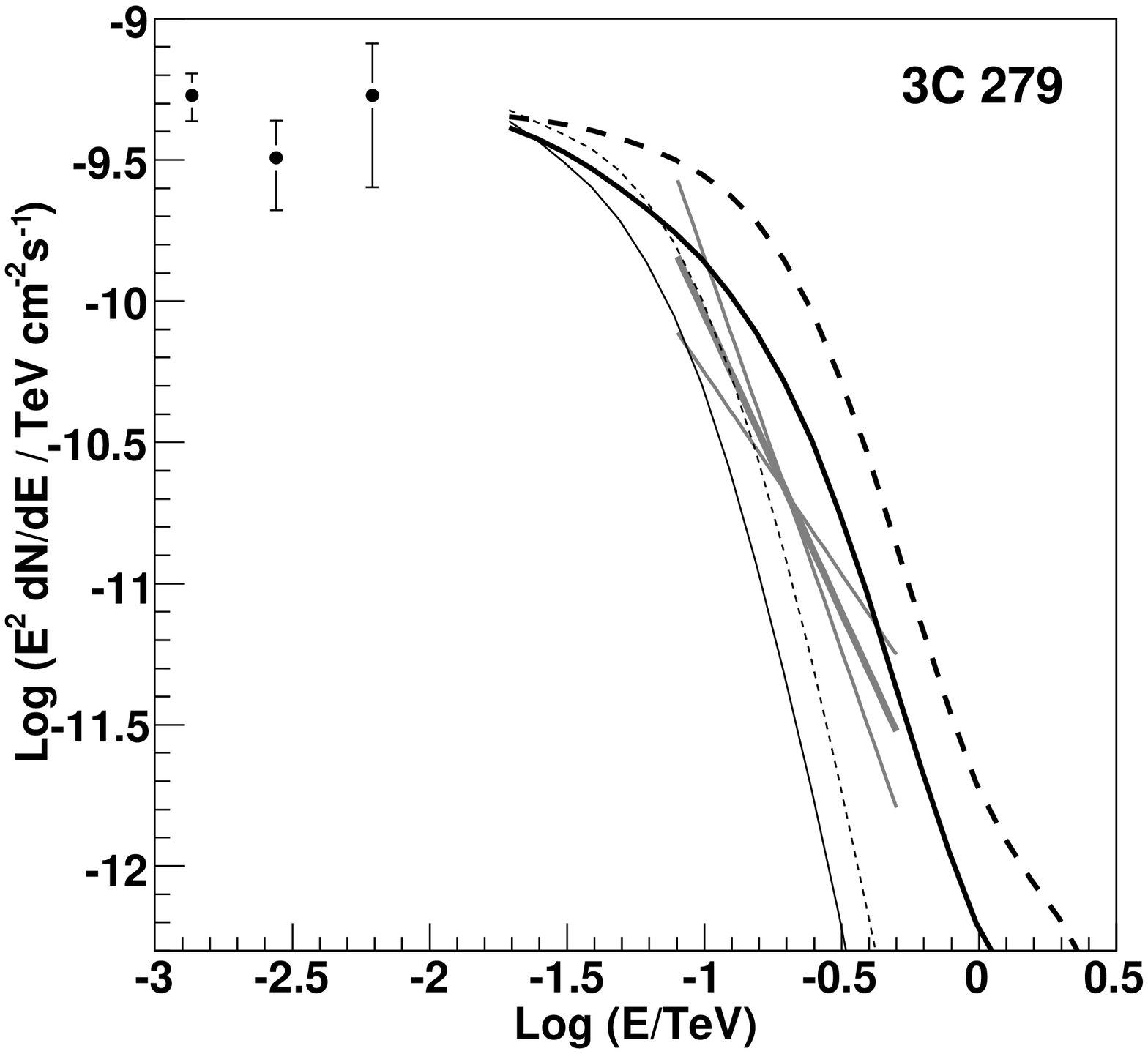}

\caption{Gamma-ray spectra at the observer calculated for injection of primary $\gamma$-rays into the jet of 3C~273 (upper panel) and 3C~279 (bottom), for different models of the radiation field created by the accretion disk and BLR: Shakura-Sunyaev disk model (model I, left figures), Shakura-Sunyaev + power law tail from the disk corona (model II, middle),  high temperature disk profile model (model III, right). The $\gamma$-ray spectra are 
shown by taking into account the absorption in the EBL according to the 
Stecker et al. model (thin curves) and the Primack et al. model (thick). 
The injection rate of $\gamma$-rays into the jet has been estimated based on the observations in the GeV energy range by the EGRET telescope (see Collmar et al.~2000 and Wehrle et al.~1998). They are characterized by the rise times of the flares, equal to 2 weeks in 3C~273 and 7 days in 3C~279, and the fall times of the flares,  equal to 1 week  (3C~273) and 2 days (3C~279). The spectra are integrated over a part of the jet corresponding to the mentioned above duration of the flares (see Eq.~\ref{eq3}). The Doppler factor of the emission region is equal to $D = 7$ (full curves) and $D = 11$
(dashed). The grey curves mark the positive detection, the upper limit, and the sensitivity of the Cherenkov telescopes as in Fig.~\ref{fig7}.}
\label{fig9}
\end{figure*}
\subsection{$\gamma$-ray spectra at the observer}

Applying the model for the development of the flare inside the jets of these two OVV blazars
(defined above), we calculate the $\gamma$-ray spectra at the observer by integrating over the duration of the flares (see Fig.~\ref{fig9}). In these spectra all the effects of the internal absorption and absorption in the EBL (with both, Stecker and Primack models) are taken into account. It is assumed that the flare starts to develop from the base of the jet.
The example calculations are shown for two values of the Doppler factors of the emission region $D = 7$ and $11$ and three models for the radiation field around the jet.
There is already clear difference between the spectra obtained for these two Doppler factors
since the extend of the emission region depends on $\propto D^2$. So then, the internal 
absorption effects are significantly lower for the $\gamma$-rays produced at the distance twice larger from the base of the jet. 
Interestingly, the spectra at the observer do not depend strongly on the applied model for
the soft radiation field close to the jet (disk emission) even in the case with 3 times
higher temperature. Note, however, that the total power in these spectra has been normalized in such a way in order to be consistent with the observations of the UV bump in their soft thermal spectrum. The emission from the disk corona has only important effect very close to
the base of the jet from which only a part of the observed $\gamma$-rays is produced.

The comparison of these calculations with the sensitivity of the MAGIC I telescope show that 
detection of the $\sim 2$ week flare from 3C 273 is rather unlikely. Only the future CTA telescopes, with the sensitivity of the order of magnitude better should detect this source.
The situation is much better with 3C 279 in which case already sub-TeV emission has been detected (Albert et al.~2008). The spectrum observed from 3C 279 by the MAGIC telescope is generally consistent with the considered here model for the $\gamma$-ray production in the inner part of the jet. Moreover, with the slight modification of the Stecker EBL model (smaller absorption at lower energies), the high estimate of the EBL also give correct description of the sub-TeV $\gamma$-ray spectrum. So then, positive detection of 3C 279 by the MAGIC telescope does not need any extraordinary assumptions on the production spectrum at the source (e.g. very flat spectra) but already simple extrapolation from the 
EGRET energy range is enough. In the case of low estimate of the EBL (Primack et al. model), the $\gamma$-ray spectrum above several GeV should have a clear break in order not to overcome the reported flux level and to be consistent in the sub-TeV energies with the MAGIC observations. Note, however that the above comparisons concern the $\gamma$-ray spectrum integrated over the whole duration of the
flare but the MAGIC observations are only limited to two days with the most of the signal detected within a few hours. Therefore, in the next figure (Fig.~\ref{fig10}, we predict the evolution of observed $\gamma$-ray spectra from these two objects at different stages of the 
development of their flares. According to our calculations, the power in the $\gamma$-ray spectra should increase in time (as expected from the normalization to the EGRET energy range) but also the cut-off in the spectrum should shift to higher energies. Therefore, we predict specific behavior of the $\gamma$-ray emission from OVV blazars which could be tested by the future sub-TeV observations even with the second phase of the present Cherenkov telescopes (e.g. MAGIC II, HESS II) or the GLAST telescope at $\sim 100$ GeV energies. As discussed above, the Stecker model for the EBL, is consistent with the detected
sub-TeV $\gamma$-ray emission from 3C 279 during the peak of the flare. In the case of the Primack model, significant part of a one week flare might be observed in the sub-TeV energies (provided that the spectrum extrapolated from he GeV energies does not break) or the significant break in the $\gamma$-ray spectrum should appear at several GeV. 

However, it looks that there is a chance to detect the sub-TeV $\gamma$-rays from 3C 273 during the peak emission with the MAGIC II stereo system (at least twice better
sensitivity). Note that marked on these figures differential sensitivity of the MAGIC I telescope concerns the $5\sigma$ detection within 50 hrs observation period. The peak of the flare detected by the EGRET from 3C 273 lasts for a few days. With the MAGIC II sensitivity,
and 15 hrs low zenith angle observations within a few days, detection of the peak of the flare from 3C 273 on the level of $\sim 5\sigma$ seems reasonable.

\begin{figure}
\includegraphics[width=0.235\textwidth, height=0.235\textwidth, trim=  0 35 46 45,clip]{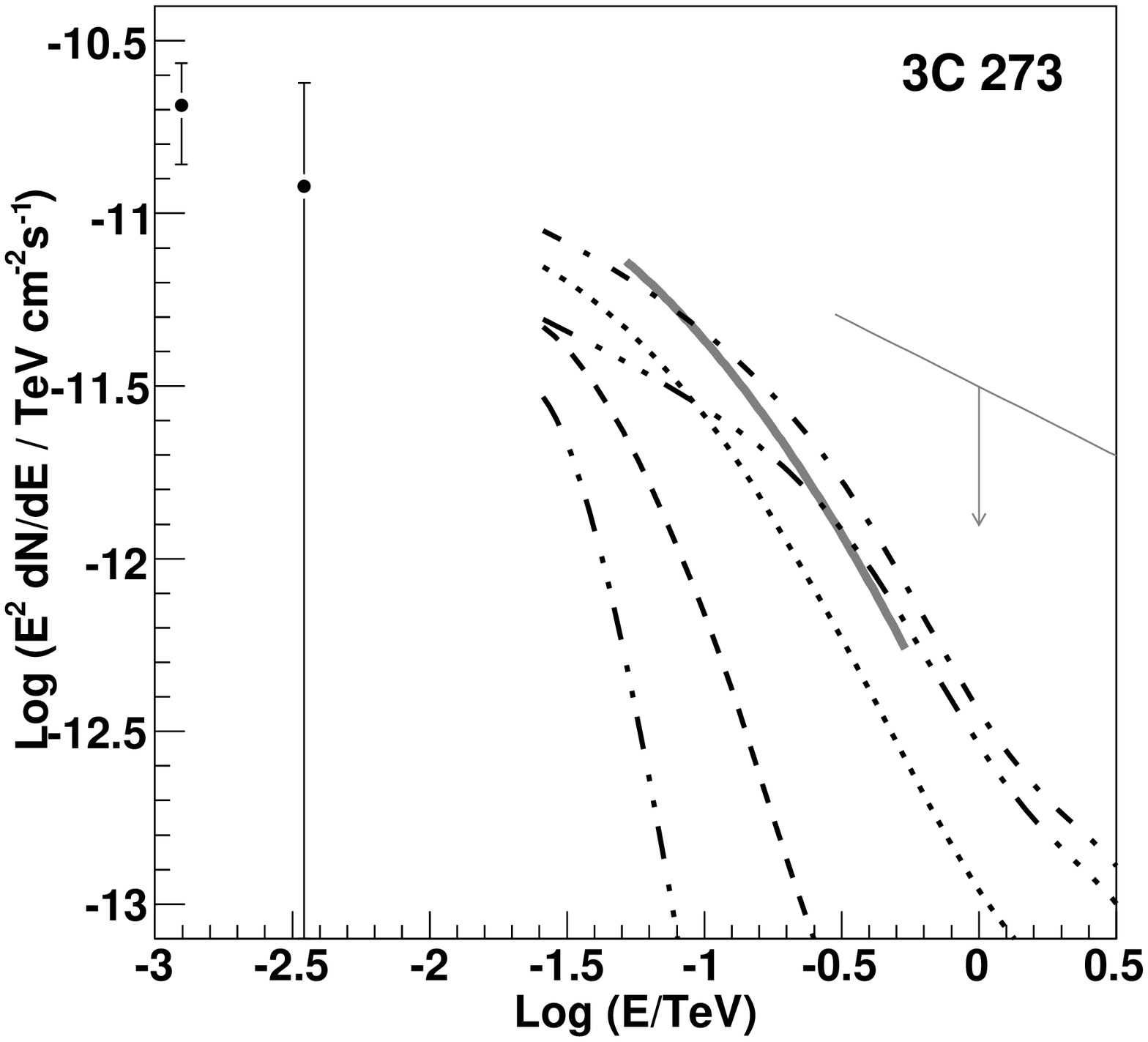}
\includegraphics[width=0.235\textwidth, height=0.235\textwidth, trim= 27 35 46 45,clip]{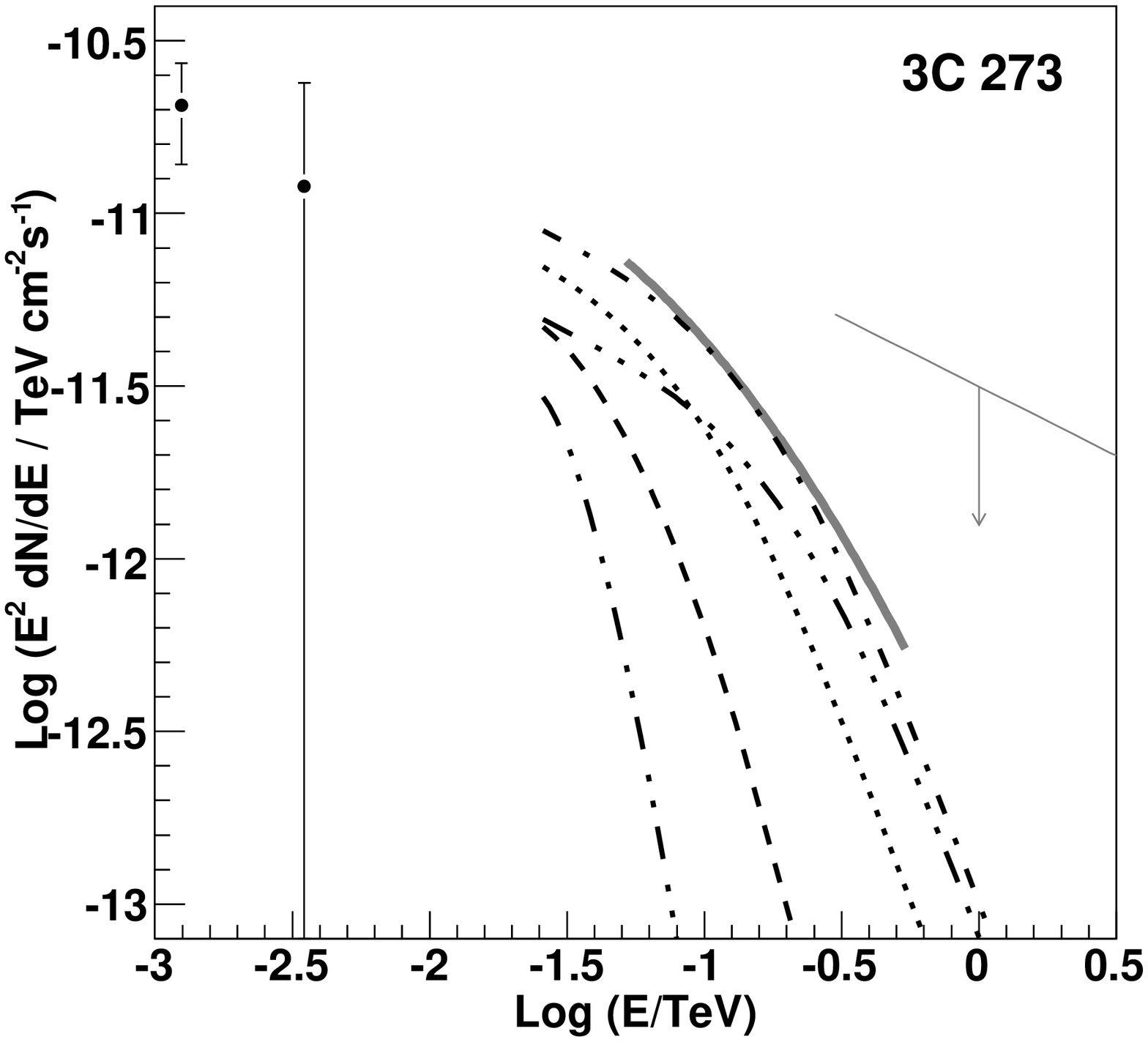}
\includegraphics[width=0.235\textwidth, height=0.235\textwidth, trim=  0  5 46 45,clip]{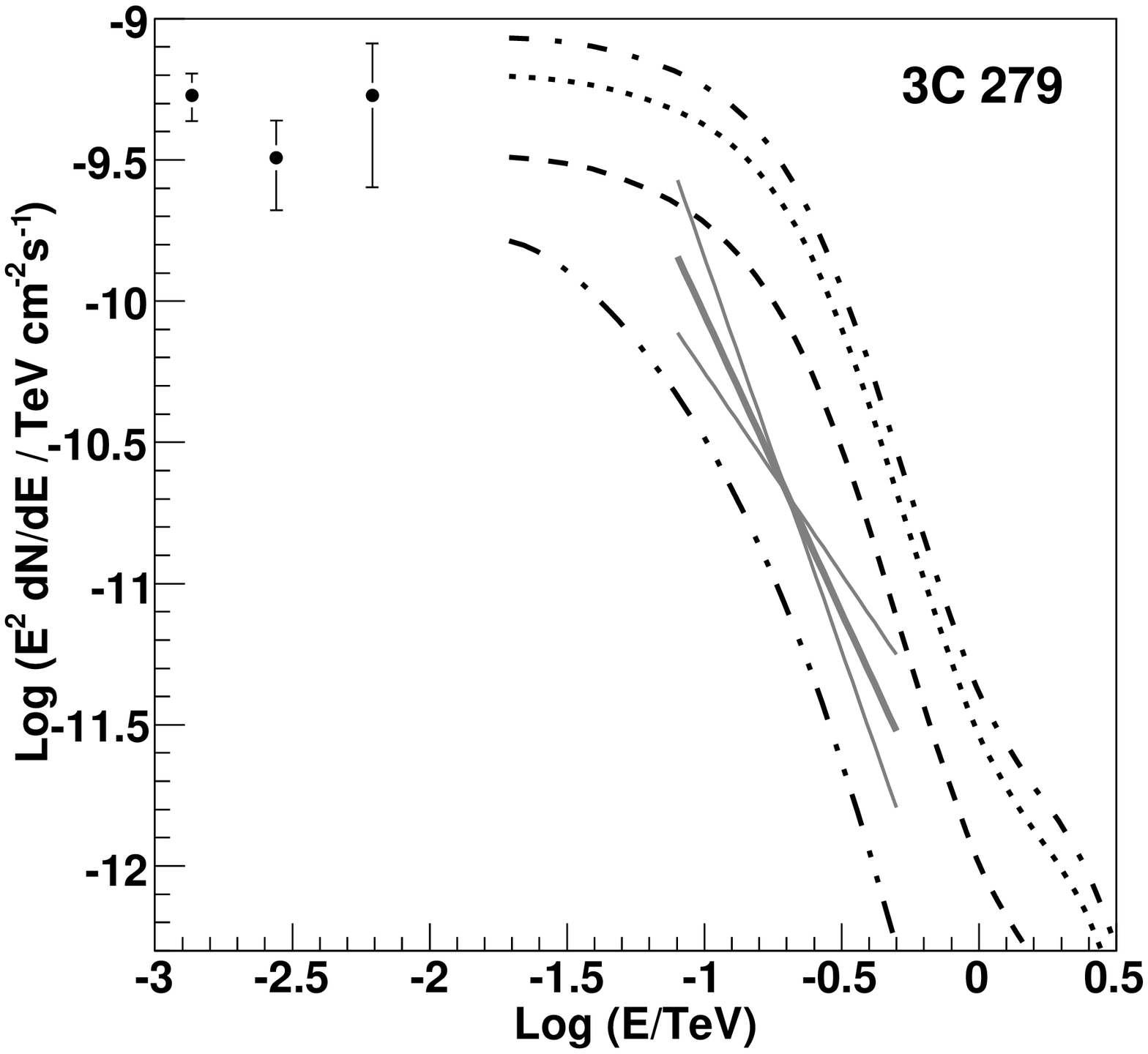}
\includegraphics[width=0.235\textwidth, height=0.235\textwidth, trim= 27  5 46 45,clip]{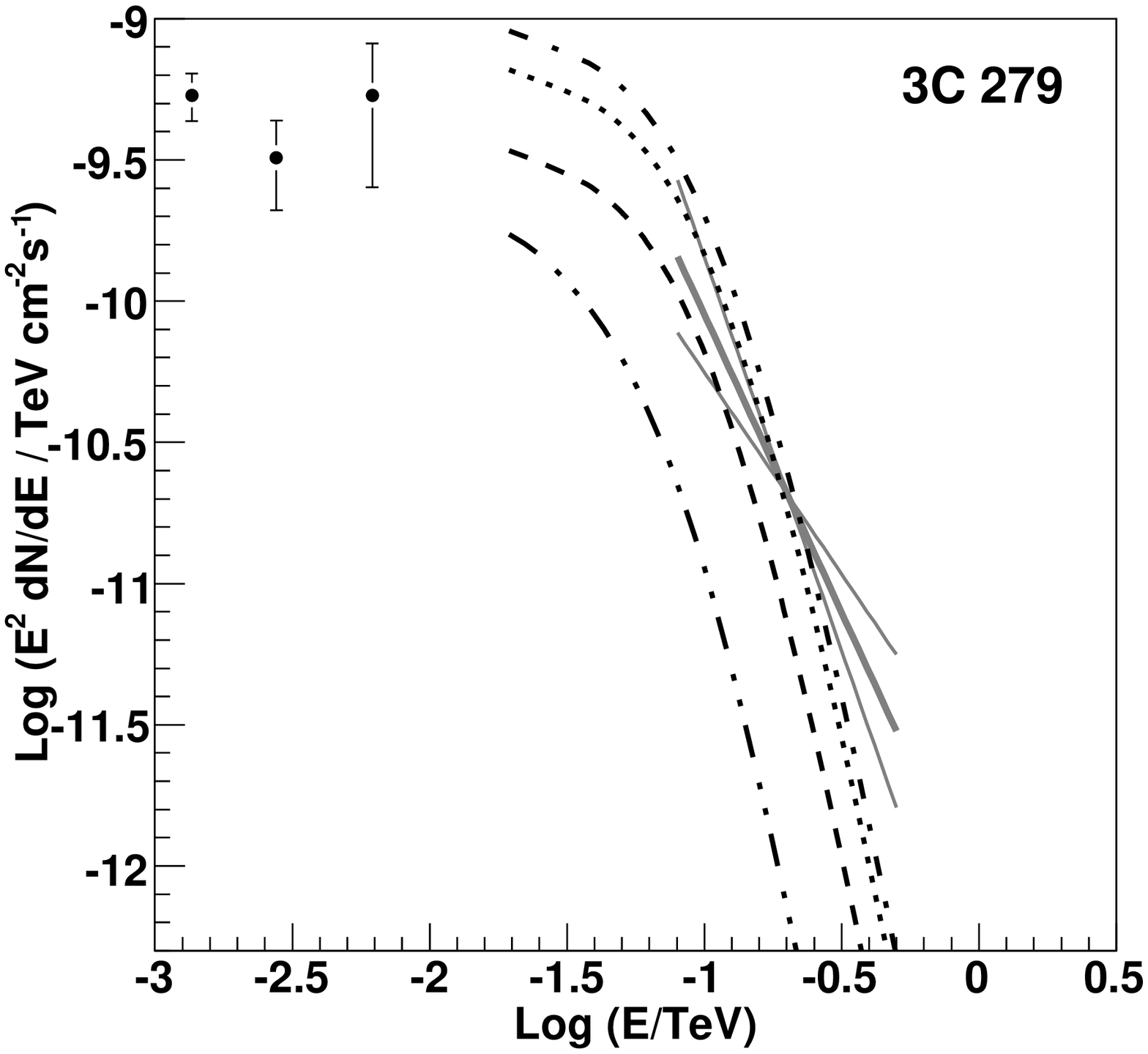}
\caption{The example time evolution of the $\gamma$-ray spectra during the development of the $\gamma$-rays flares from 3C 273 and 3C 279 for the Doppler factor $D=10$.
Specific spectra have been calculated in the specific time windows ($\Delta t = 4$ days in the case of 3C 273 and 2 days in the case of 3C 279). It is assumed that injection of 
$\gamma$-rays starts at the base of the jet. Specific curves are obtained for:
3C 273: 0-4 days (dot-dot-dashed curve), 4-8 (dashed), 8-12 (dotted), 12-16 (dot-dashed), 16-20 (dot-dot-dot dashed) in the case of 3C 273, and 0-2 days (dot-dot-dashed curve), 
2-4 (dashed), 4-6 (dotted), 6-8 (dot-dashed) in the case of 3C 279. 
The absorption effects in the EBL according to the Stecker et al. model (right figures) and Primack et al. model (left figures) is included. 
For the description of the grey curves see Fig.~\ref{fig7}.}
\label{fig10}
\end{figure}
\section{Conclusions} 

We discuss in detail the effects of internal absorption of $\gamma$-rays produced in jets of 
the OVV type blazars concentrating on two famous sources: 3C 279 and 3C 273. 
The first source has been recently detected in the sub-TeV energies by the MAGIC telescope (Albert et al.~2008), but only the upper limit exists on the TeV emission from the second source (von Montigny et al.~1997).

It is shown that the model for $\gamma$-ray production in the jet of 3C 279, in which the spectrum is extrapolated from the EGRET energy range (observed up to several GeV) 
to the $\sim 1$ TeV energy range, is quite consistent with reported by the MAGIC collaboration $\gamma$-ray spectrum, even in the case of the upper estimate on the extragalactic background light derived by Stecker et al.~(2006). Note that this is the furthest TeV $\gamma$-ray source ($z = 0.536$) detected up to now. Therefore, we conclude that the observation of this most distant sub-TeV source does not introduce any new important constraints on the EBL. Such constraints are used to be derived in the case of the population of the closer BL Lac type objects in which case the TeV emission extends to much higher energies. However, if the EBL is closer to the estimate by Primack et al.~(2005) model, then the $\gamma$-ray spectrum injected from the considered active region in the jet has to show a clear break above several GeV. At present it is difficult to conclude only on the theoretical grounds if the 
$\gamma$-ray spectrum in OVV blazars continue to higher energies or have significant break. 

We also show that detection of 3C 273 with the present MAGIC I telescope is rather
problematic even if the source is captured in the highest up to now observed emission state by 
the EGRET telescope and the EBL is well described by the Primack et al. model. 
However, there is a real chance to detect the peak of a few day flare from this 
source with the MAGIC II stereo system (which should have at least 2 times 
better sensitivity), provided that the EBL is closer to the Primack et al. model.

We have only shown the results of calculations for the two values of the 
Doppler factor of the emission region in the jet ($D = 7$ and 11). In the case of 
emission regions moving with larger Doppler factors, the internal absorption of $\gamma$-rays in the radiation field around the jet should be reduced due to the larger distance traveled by the 
emission region and related to this on the average more distant injection of $\gamma$-rays from the accretion  disk and BLR clouds. We also assumed that the flare in the jet starts to develop 
already from its base. However, farther away from the accretion disk, the internal absorption is 
certainly weaker. Therefore, we consider the most pessimistic case for the escape of 
TeV $\gamma$-rays from these OVV blazars. Note that location of the injection region 
of $\gamma$-rays inside the jet can be in the future constrained by the more sensitive
observations which allow detection of much shorter variability time scales of the 
$\gamma$-ray emission. Discovery of a very short time scale variability in blazars 
of the BL Lac type (of the order of a few minutes, see Mrk 501 (Albert et al.~2007), and 
PKS 2155-304 (Aharonian et al.~2007), and also radio galaxy M87 (Aharonian et al.~2006)),
strongly suggest that the $\gamma$-ray emission region is rather located  very close to 
the base of the jet, i.e. to the surface of the accretion disk.

\section*{Acknowledgments}
This work is supported by the Polish MNiSzW grant N N203 390834.


\label{lastpage}
\end{document}